\documentclass[useAMS,usenatbib,twocolumn]{mn2e}

\usepackage{graphicx,amsmath,amssymb}
\usepackage{natbib,graphicx,graphics,amssymb,amsmath,pdflscape,textcomp,epigraph}
\usepackage{setspace,subfig,rotating,multirow,psfrag,color,longtable,gensymb}
\usepackage{longtable,enumitem}

\pdfoutput=1

\newcommand\araa{{ARA\&A}}%
\newcommand\apj{{ApJ}}%
\newcommand\apjl{{ApJ}}%
\newcommand\apjs{{ApJS}}%
\newcommand\aap{{A\&A}}%
 \newcommand\actaa{{Acta Astronomica}}%
\newcommand\mnras{{MNRAS}}%
\newcommand\pasp{{PASP}}%
\newcommand\nat{{Nature}}%
\newcommand\PLslope{$d(\log f)/d(\log q)=-0.50\pm0.17$} 
\newcommand\BNslope{$d(\log f)/d(\log q)=0.32\pm0.38$} 

\title[The frequency of snowline-region planets]
{\vspace*{-0.9 truecm} The frequency of snowline-region planets from four-years of OGLE-MOA-Wise second-generation microlensing\vspace*{-0.6 truecm}}
\author[Shvartzvald et al.]
{Y.~Shvartzvald$^{1,w}$\thanks{E-mail: yossi@wise.tau.ac.il},
D.~Maoz$^{1,w}$,
A. Udalski$^{2,o}$,
T.~Sumi$^{3,m}$,
M.~Friedmann$^{1,w}$,
S.~Kaspi$^{1,w}$,
\and
R. Poleski$^{4,o}$,
M.\,K. Szyma{\'n}ski$^{2,o}$,
J. Skowron$^{2,o}$
S. Koz{\l}owski$^{2,o}$,
{\L}. Wyrzykowski$^{2,o}$,
\and
P. Mr{\'o}z$^{2,o}$,
P. Pietrukowicz$^{2,o}$,
G. Pietrzy{\'n}ski$^{2,o}$,
I. Soszy{\'n}ski$^{2,o}$,
K. Ulaczyk$^{5,o}$,
F.~Abe$^{6,m}$,
\and
R.~K.~Barry$^{7,m}$,
D.~P.~Bennett$^{8,m}$,
A.~Bhattacharya$^{8,m}$,
I.A.~Bond$^{9,m}$,
M.~Freeman$^{10,m}$,
\and
K.~Inayama$^{11,m}$,
Y.~Itow$^{6,m}$,
N.~Koshimoto$^{3,m}$,
C.H.~Ling$^{9,m}$,
K.~Masuda$^{6,m}$,
A.~Fukui$^{12,m}$,
\and
Y.~Matsubara$^{6,m}$,
Y.~Muraki$^{6,m}$,
K.~Ohnishi$^{13,m}$,
N.~J.~Rattenbury$^{10,m}$,
To.~Saito$^{14,m}$,
\and
D.J.~Sullivan$^{15,m}$,
D.~Suzuki$^{8,m}$,
P.~J.~Tristram$^{16,m}$,
Y.~Wakiyama$^{6,m}$,
A.~Yonehara$^{11,m}$\\
\\
$^1$ School of Physics and Astronomy, Tel-Aviv University, Tel-Aviv 69978, Israel\\
$^2$ Warsaw University Observatory, Al.~Ujazdowskie~4, 00-478~Warszawa, Poland\\
$^3$ Department of Earth and Space Science, Osaka University, Osaka 560-0043, Japan\\
$^4$ Department of Astronomy, Ohio State University, 140 W. 18th Ave., Columbus, OH 43210, USA\\
$^5$ Department of Physics, University of Warwick, Gibbet Hill Road, Coventry, CV4 7AL, UK\\
$^6$ Solar-Terrestrial Environment Laboratory, Nagoya University, Nagoya, 464-8601, Japan\\
$^7$ Astrophysics Science Division, NASA Goddard Space Flight Center, Greenbelt, MD 20771, USA\\
$^8$ University of Notre Dame, Department of Physics, 225 Nieuwland Science Hall, Notre Dame, IN 46556-5670, USA\\
$^9$ Institute of Information and Mathematical Sciences, Massey University, Private Bag 102-904, North Shore Mail Centre, Auckland, New Zealand\\
$^{10}$ Department of Physics, University of Auckland, Private Bag 92-019, Auckland 1001, New Zealand\\
$^{11}$ Department of Physics, Faculty of Science, Kyoto Sangyo University, 603-8555 Kyoto, Japan\\
$^{12}$ Okayama Astrophysical Observatory, National Astronomical Observatory of Japan, Asakuchi, Okayama 719-0232, Japan\\
$^{13}$ Nagano National College of Technology, Nagano 381-8550, Japan\\
$^{14}$ Tokyo Metropolitan College of Aeronautics, Tokyo 116-8523, Japan\\
$^{15}$ School of Chemical and Physical Sciences, Victoria University, Wellington, New Zealand\\
$^{16}$ Mt. John University Observatory, P.O. Box 56, Lake Tekapo 8770, New Zealand\\
$^{w}$ The Wise Observatory Group\\
$^{o}$ Optical Gravitational Lens Experiment (OGLE) Collaboration\\
$^{m}$ Microlensing Observations in Astrophysics (MOA) Collaboration
\vspace*{-0.5 truecm}
}
\date{Accepted 2016 Januray 20; in original form 2015 October 14 \vspace*{-0.5 truecm}}

\begin{document}
\maketitle
\begin{abstract}
 \noindent
We present a statistical analysis of the first four seasons from a ``second-generation'' microlensing survey for extrasolar planets,
consisting of near-continuous time coverage of 8 deg$^2$ of the Galactic bulge by the OGLE, MOA, and Wise microlensing surveys. 
During this period, 224 microlensing events were observed by all three groups. Over 12\% of the events showed a deviation from single-lens microlensing,
and for $\sim$1/3 of those the anomaly is likely caused by a planetary companion.
For each of the 224 events we have performed numerical ray-tracing simulations to calculate the detection efficiency of possible companions as a function of
companion-to-host mass ratio and separation.
Accounting for the detection efficiency, we find that $55^{+34}_{-22}\%$ of microlensed stars host a snowline planet.
Moreover, we find that Neptunes-mass planets are $\sim10$ times more common than Jupiter-mass planets.
The companion-to-host mass ratio distribution shows a deficit at $q\sim10^{-2}$, separating the distribution into two companion populations,
analogous to the stellar-companion and planet populations, seen in radial-velocity surveys around solar-like stars.
Our survey, however, which probes mainly lower-mass stars, suggests a minimum in the distribution in the super-Jupiter mass range,
and a relatively high occurrence of brown-dwarf companions.
\end{abstract}


\section{INTRODUCTION}
\label{sec:intro}

Over the last 20 years, our knowledge about planetary systems has increased dramatically, from one example with eight planets (our own Solar system) to over 1100 planetary systems hosting more than
1800 planets\footnote{\tt http://exoplanet.eu}. The various planet--detection techniques are sensitive to different regions of the planetary parameter space, but the overall emerging picture
from combining their results is that planetary systems are very common. The occurrence rate found by radial velocity and transits surveys, which have detected the majority of the known exoplanets,
and are most sensitive to close-in planets, shows that, on average, about half of Sun-like stars host small planets (Earth to Neptune) within 1 AU,
10\% have giant planets within a few AU, and about 10\% among those giants are within a few hundredths of an AU (\citealt{Winn.2015.A}).
Moreover, systems with multiple close-in planets are not rare.
The existence of close-in massive planets (``hot Jupiters'') requires a migration mechanism for those planets from beyond the ''snowline``, where they likely form according
to planet formation scenarios (\citealt{Ida.2005.A}).

Complementing these discovery methods, gravitational microlensing, while having detected only tens of planets, offers a number of unique advantages.
Microlensing is sensitive to planets at projected host separations of 1 to 10 AU, orbiting stars throughout a large volume of the Galaxy,
among both the bulge and the disk populations (\citealt{Gaudi.2012.A}). Initial statistical analyses of microlensing surveys have suggested that, on average,
every star in the Galaxy hosts at least one snowline planet
(\citealt{Cassan.2012.A}), and that the majority of them might be multiple-planet systems (\citealt{Han.2013.A}).
Microlensing surveys are also the only technique sensitive, in principle, to unbound planets. \cite{Sumi.2011.A} argue that such free-floating planets are at least as common as
main-sequence stars.
However, due to the current small total number of microlensing-detected planets
and the difficulties in turning these
numbers into a statistical occurrence rate, the abundance of Solar-like planetary systems is still poorly known.

The traditional ``first generation'' strategy for exoplanet searches by the microlensing community has been to focus on specific microlensing events discovered by the OGLE (\citealt{Udalski.2009.A})
and the MOA (\citealt{Sumi.2003.A}) microlensing surveys.
These events have generally been bright and of high-magnification. Following alerts announced by the surveys, events that promised to be of high magnification ($A\gtrsim 100$) were
monitored quasi-continuously by global follow-up networks, in order to search for planetary anomalies. The light curve of a bright and highly magnified event has a high signal-to-noise
ratio (SNR), permitting observations with small (and even amateur) telescopes, which are a significant part of the follow up networks. The motivation for this strategy was that
high-magnification events have the highest sensitivity to planets, since the source star image is distorted into a nearly full Einstein ring, and hence a planet anywhere in the Einstein ring
vicinity will perturb the light curve and will be detected (\citealt{Gould.1992.A,Griest.1998.A}).
Although this strategy was important for discovering the first microlensing planets, it also led to the
main limitations of the first generation: (a) high-magnification events are rare ($\sim$1\%) and hence overall a small number of microlensing planets have been discovered with this approach;
(b) the alert and follow up process involves a complex social decision and communication process, sometimes after an event has already hinted at the possibility of an anomaly.
This process is difficult to account for in a statistical analysis.

Exploiting the full planet discovery potential of microlensing requires a ``second generation'' (hereafter - genII) microlensing survey (\citealt{Gaudi.2009.A}),
in which a large fraction of all ongoing microlensing events toward the Galactic bulge -- of both low and high magnification --
are monitored continuously with a cadence that is high enough to detect planetary perturbations in the light curves.
Apart from the larger number of expected planet discoveries, the ``controlled experiment'' nature of a genII survey allows for safer statistical inferences on the planet population.
The requirements are straightforward -- a network of 1m-class telescopes situated so as to allow 24-hour coverage of the bulge, equipped with degree-scale imagers.

There have been three main works to constrain snowline planet properties and frequencies based on first-generation surveys.
\citet{Sumi.2010.A} used the first ten microlensing-discovered planets to constrain the mass-ratio distribution of snowline region exoplanets and their hosts, and concluded
that Neptune-mass planets are at least three times as common as Jupiters in this region.
\citet{Gould.2010.B} analyzed six planets discovered in high-magnification microlensing events that were intensively followed-up by the $\mu$FUN network (\citealt{Gould.2008.A}),
and found that $\sim1/3$ of microlensed stars host a snowline planet in the planet-to-star mass-ratio interval $-4.5<{\rm log}~q<-2$, corresponding to the range of ice to gas giants. Moreover, since one of the systems
had two planets with the same mass and distance ratios as Jupiter and Saturn to the Sun, they argued that the frequency of Solar-like systems around lens stars is about 1/6.
\citet{Gould.2010.B} argued that the chaotic nature
of the ``alert and follow-up'' process results, in the end, in a randomization that
minimizes selection effects and although it cannot be claimed to be a controlled experiment, it gives a high level of completeness.
Finally, \citet{Cassan.2012.A} combined the previous results and three additional planets that were discovered by the PLANET follow-up network (\citealt{Albrow.1998.A}), and deduced a
frequency of more than unity for snowline-region planets.

In June-July 2010, we carried out a 6-week pilot of a genII experiment.
In 2011, we initiated the first full genII microlensing experiment, involving OGLE, MOA and the Wise Observatory microlensing group (\citealt{Shvartzvald.2012.A}).
During four observing seasons, we have monitored all events in a specific, 8 deg$^2$, field toward the Galactic bulge.
In this paper we present a statistical analysis of these first four observing seasons (2011-2014) of our genII survey.
The sample includes 224 microlensing events, monitored by all three groups.
Over 12\% of the events show clear anomalies in their light curves indicating a companion to the lens star.
To assess the nature of the companions we perform a coarse grid-based search of the microlensing model
parameter space for each event.
Comparing the actual number of light curves having planetary anomalies to the total number of events, after accounting for the detection efficiency of each event,
allows to estimate, for the first time, the frequency of planetary systems from a controlled microlensing experiment.

The paper is arranged as follows. In Section \ref{sec:obs_stat}, we describe the observations and reductions of our sample of lensing events.
In Section \ref{sec:detections}, we model each light curve in our sample with a single-lens model,
describe our anomaly detection filter, and present the anomalous events among our sample.
In Section \ref{sec:frequency} we derive the frequency and the mass-ratio distribution of snowline planets implied by our survey,
and discuss the results in the context of current knowledge in Section \ref{sec:Discussion_stat}.

\section{Observations}
\label{sec:obs_stat}

\subsection{Observational data and reduction}
\label{sec:data}

Our genII survey network is a collaboration between three groups: OGLE, MOA, and Wise.
The OGLE and MOA groups regularly monitor a large region of the Galactic bulge, and routinely identify and monitor microlensing events.
The Wise group monitors a field of 8 $\rm deg^2$, within the observational footprints of both OGLE and MOA, having the highest event rates based on previous years' observations
(see \citealt{Shvartzvald.2012.A}).

The OGLE group operates the 1.3m Warsaw University telescope at Las Campanas Observatory in Chile, with the 1.4 $\rm deg^2$ field of view OGLE-IV camera.
The genII survey's 8 $\rm deg^2$ footprint includes three extremely high-cadence OGLE-IV fields observed every $\sim$15 minutes and five high-cadence
OGLE-IV fields observed once per $\sim$45 minutes. Standard OGLE observations are through an $I$ filter.
OGLE photometry used here was extracted by their standard difference image analysis (DIA) procedure (\citealt{Udalski.2003.A}).
These OGLE data are the preliminary measurements posted on the Early Warning System website\footnote{http://ogle.astrouw.edu.pl/ogle4/ews/ews.html}, which are not suitable for detailed modeling
of the events, but sufficient for the anomaly-detection search and rough modeling that we perform here. 

The MOA group operates the 1.8m MOA-II telescope in New Zealand with the MOA-cam3 camera, with a 2.2 $\rm deg^2$ field of view (\citealt{Sako.2008.A}). 
MOA-II monitors six extremely high-cadence fields (every $\sim$15 minutes) which cover most of the genII 8 $\rm deg^2$ field, 
and an additional six high-cadence fields (every $\sim$45 minutes) which complement the network's footprint.
MOA uses a wide $R/I$ filter (``MOA-Red''). The MOA data were reduced using their routine DIA procedure (\citealt{Bond.2001.A}).

The Wise group's main setup is the 1m telescope at Wise Observatory in Israel equipped with the LAIWO camera, with a 1 $\rm deg^2$ field of view.
The cadence for each of the eight Wise pointings (which define the survey footprint) is $\sim$30 minutes. 
Observations are in the $I$ band.
In the first half of the 2012 season, the LAIWO camera had readout electronics problems and 
the Wise group used, as an alternative, the Wise C18 0.46m telescope with an $I$ filter, with a 1 $\rm deg^2$ field of view.
The cadence for the eight C18 fields (which overlap with, but differ from the eight LAIWO fields) was $\sim$1 hour.
All Wise data were reduced using the pySis DIA software (\citealt{Albrow.2009.A}).

\subsection{Sample of lensing events}
\label{sec:sample_stat}

The sample of microlensing events analysed here consists of 224 events from the 2011-2014 bulge seasons, observed by all three groups,
and with each group having data near the peak of the event. Without the last criterion, events with a long timescale that peak before the beginning of a season might enter the sample,
while such events with a short timescale would not.
Being the northernmost node of the network, with the shortest bulge observing season,
Wise sets the time limits for the genII survey, from mid-April to mid-September of each season.

With the above criteria, and considering the limiting magnitudes of the various groups (see definition in Section \ref{sec:model_stat}),
the initial sample has a relatively small number of events (230).
After identifying anomalous events in our sample (see Section \ref{sec:anomlous}), we noted six events that passed the limiting magnitude criteria only
because of the anomaly
itself (due to a caustic crossing), and otherwise would not meet the criteria.
Including these events would introduce a bias toward anomalous events, and therefore we exclude them from the sample.
Table~\ref{table:data} lists the final sample of 224 events.  For completeness, the excluded six events are listed separately at the end of the Table.

\section{Companion detections}
\label{sec:detections}

\subsection{Single-lens modeling of all events}
\label{sec:model_stat}

As a first stage in our analysis,
we model each of the 224 events with a single point-lens model, including the possibility of parallax and finite source effects. Such a model is necessary in order to subsequently detect
those events that have lensing anomalies (Section \ref{sec:anomlous}), as well as to gauge the planet detection efficiency of each event (Section \ref{sec:efficiency}). 
Both of these elements of the analysis then enter our planet frequency calculation (Section \ref{sec:frequency}).

The characteristic angular scale of a lensing phenomenon is the angular Einstein radius,
\begin{equation}
\theta_{\rm E}^2 = \kappa M \pi_{\rm rel},~\kappa = \frac{4G}{c^2{\rm AU}}\simeq 8.14{{\rm mas}\over M_\odot},
\end{equation}
where $M$ is the mass of the lens star, and the relative parallax is $\pi_{\rm rel}={\rm AU}(D_L^{-1}-D_S^{-1})$,
where $D_S$ and $D_L$ are the distances to the source and lens stars, respectively.
A standard point-source, point-lens, microlensing model requires three ``Paczy{\'n}ski parameters'': $t_{\rm E}$---the event time scale,
over which the source crosses one angular Einstein radius;
$u_0$---the impact parameter, the minimal separation between the source star and lens star in angular Einstein radius units;
and $t_0$---the time of minimal separation between the source star and the lens star.
There are two higher-order effects, which can change the light curve even in the absence of a companion to the lens.
First, the finite source size needs to be taken into account when the scale in the source plane over which the magnification varies is
of the order of the source size. This effect is parametrized by the ratio between the angular size of the source, $\theta_*$, and the angular Einstein radius,
$\rho= \theta_*/\theta_{\rm E}$.
Second, the orbital microlens parallax, $\pi_{\rm E}$, changes the relative proper motion between the source star and the lens during an event, due to the Earth's orbit.
The effect is parametrized by the size of the angular Einstein radius, projected onto the observer plane in astronomical units (AU), and can be represented by:
$\boldsymbol{\pi_{\rm E}}= (\pi_{\rm rel} / \theta_{\rm E})(\boldsymbol{\mu}/\mu$), where $\boldsymbol{\mu}$ is the relative proper motion between the source and the lens.
The microlens parallax is divided into a north and east component with respect to the Galactic coordinate system, $\pi_{E,N}$ and $\pi_{E,E}$.
The orbital parallax involves a well known $u_0 \leftrightarrow -u_0$ ecliptic degeneracy for sources near the ecliptic (\citealt{Jiang.2004.A,Poindexter.2005.A}),
and we explore both options in our modeling. 

During a microlensing event, an additional constant flux from stars along the line of sight to the source,
that do not get magnified due to the lens (in particular, the lens star itself), contributes to the light curve.
Thus the total flux, $f(t)$, is the sum of the source flux, $f_{\rm s}$, times the magnification, $A(t)$, and the background flux, $f_{\rm b}$.
Due to the different filters and seeing conditions of each survey group, this is calculated separately for each dataset, $i$,
\begin{equation}
f_i(t)=f_{\rm{s}\it{,i}}\cdot A(t)+f_{\rm{b}\it{,i}}.
\end{equation}

When exploring the lens-model parameter space, we limit the ranges of some of the parameters, to avoid biases and unphysical solutions.
First, we include only events with $|u_0|<1$, i.e. with peak magnifications $A>1.34$.
In addition, we limit the crossing time scale to values $t_{\rm E}<500$ days.
Finally, we limit each component of $\pi_{\rm E}$ to be less than 1.5.
Events that exceed these limits  on $t_{\rm E}$ and $\pi_{\rm E}$ can, in principle, occur for extreme configurations. However, their probability is negligible given the number of events in our sample.

\begin{figure}
\centering
\begin{tabular}{c}
\includegraphics[width=0.5\textwidth]{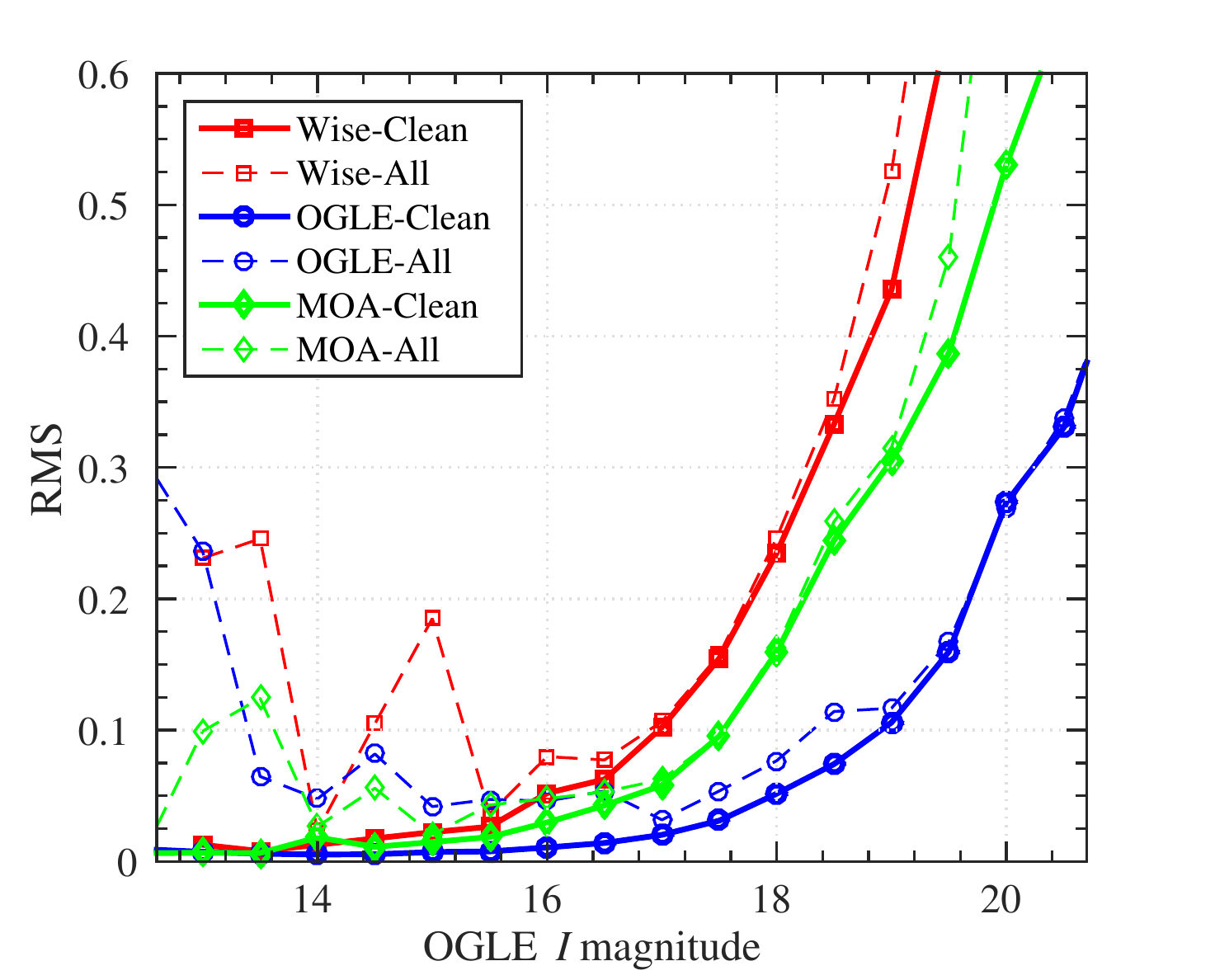}\\
\includegraphics[width=0.5\textwidth]{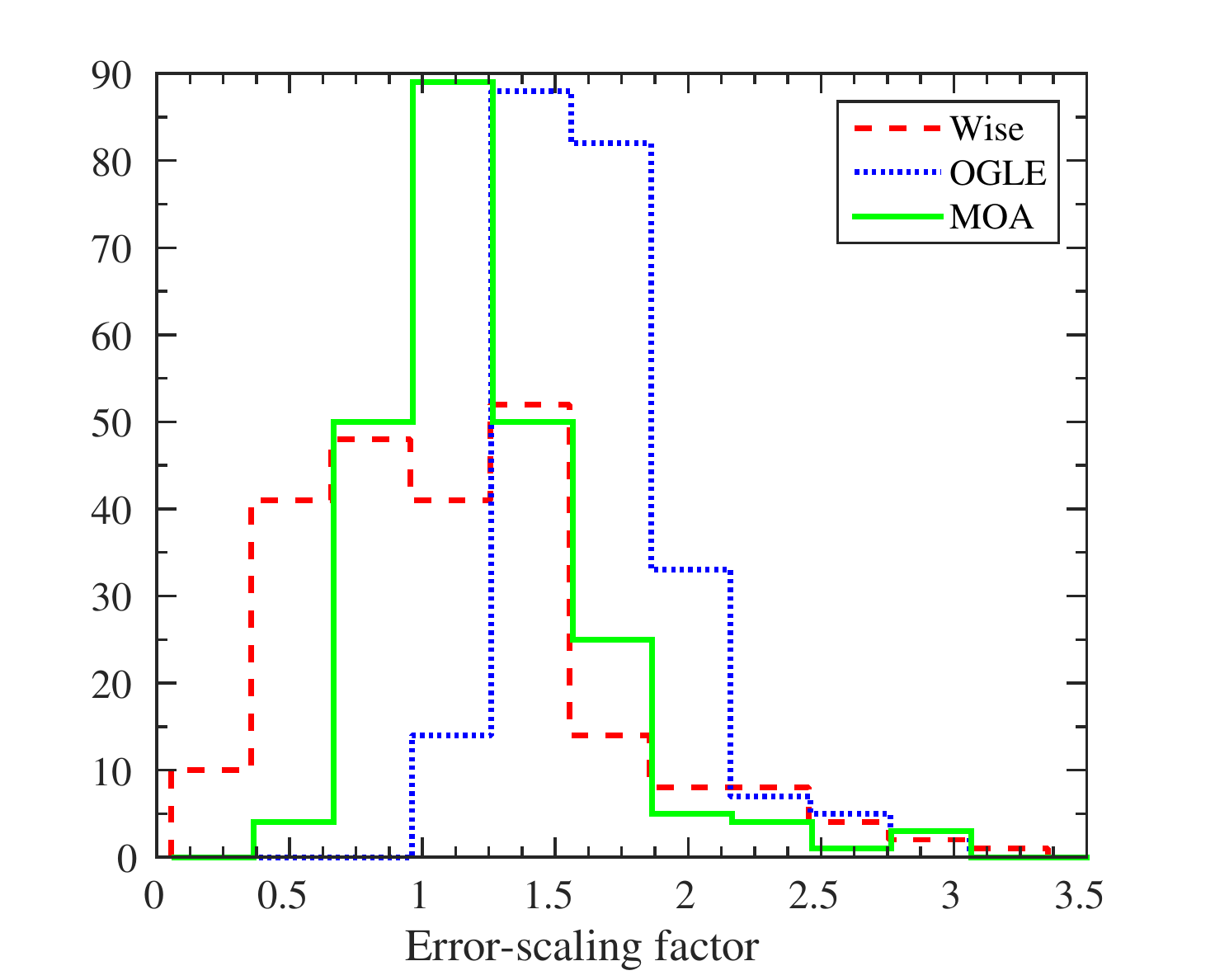}
\end{tabular}
\caption{{\it Top}: root mean square (RMS) of residuals of best fits of single-lens microlensing models to all observed light curves,
including all data (dashed line) and after excluding outliers and anomalies (solid line),
for each group as a function of OGLE $I$-band magnitude at a given point in the light curve (OGLE - blue circles, MOA - green diamonds, Wise - red squares). The differences between groups reflect both the instrumental and site quality
(collecting area, seeing conditions, CCD sensitivity) and the DIA pipeline for extracting the photometry.
{\it Bottom}: error-scaling factor distributions for the 224 events, and for each group. Most of the pipeline-reported errors underestimate the real uncertainties. The tails shown by each distribution
at large factors reflect events with long-duration anomalies.\label{fig:RMS}}
\end{figure}

The MOA and Wise fluxes were aligned to the OGLE $I$-band magnitude scale, and inter-calibrated to the single-lens microlensing model.
In order to obtain the different photometric precision for each group, we construct an observed distribution of the residuals from the model
for the entire dataset (from all events), and calculate its root mean square (RMS) as a function of OGLE $I$-band magnitude.
This accounts for differences in telescope aperture, the different seeing conditions at each site,
and overall systematics in the reduction pipelines.
Figure~\ref{fig:RMS} shows the RMS of the residuals for each group.

\cite{Yee.2012.A} have studied the effect of faint sources on microlens parameters for the lensing event MOA-2011-BLG-293, and found that when the measured flux errors are
comparable to, or larger than the source flux, particularly near baseline, this leads to biases in the measured Einstein timescale.
Based on their conclusions and by analyzing the precision at each site, found above,
we exclude from each light curve in our sample data points for which the observed residual distribution corresponding to
the underlying event's $I$-magnitude (as suggested by the single lens model) has a root-mean square (RMS) greater than 0.33 mag (SNR$\sim$3, 
below that the microlensing signal cannot be reliably detected).
This corresponds to events with Wise points with baseline magnitudes fainter than 18.5 mag,
and MOA baseline points fainter than 19.5 mag. For OGLE we include all points.

\begin{figure*}
\begin{minipage}{\textwidth}
\centering
\begin{tabular}{c}
\includegraphics[width=0.8\textwidth]{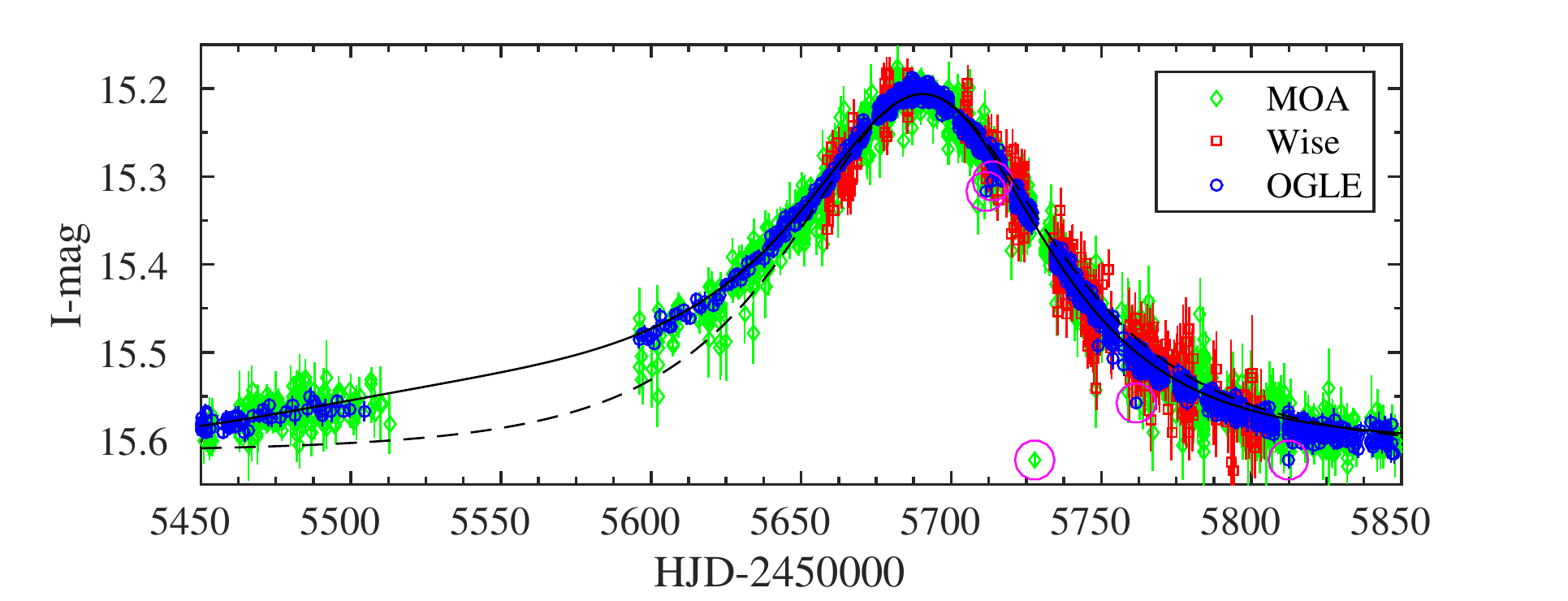}
\end{tabular}
\caption{Example of detection and exclusion of outlier points from the single-lens model in
event OGLE-11-0022/MOA-11-025, which has a strong microlens parallax signal (but no detected anomaly from a companion). The detected outlier points are circled.
The solid line is the best-fit microlensing model including parallax, and the dash-dot line is one without parallax.
The light curve is clearly asymmetric, and thus without including parallax it would have been flagged as anomalous.\label{fig:outliers}}
\end{minipage}
\end{figure*}

\begin{figure*}
\begin{minipage}{\textwidth}
\centering
\begin{tabular}{cc}
\includegraphics[width=0.5\textwidth]{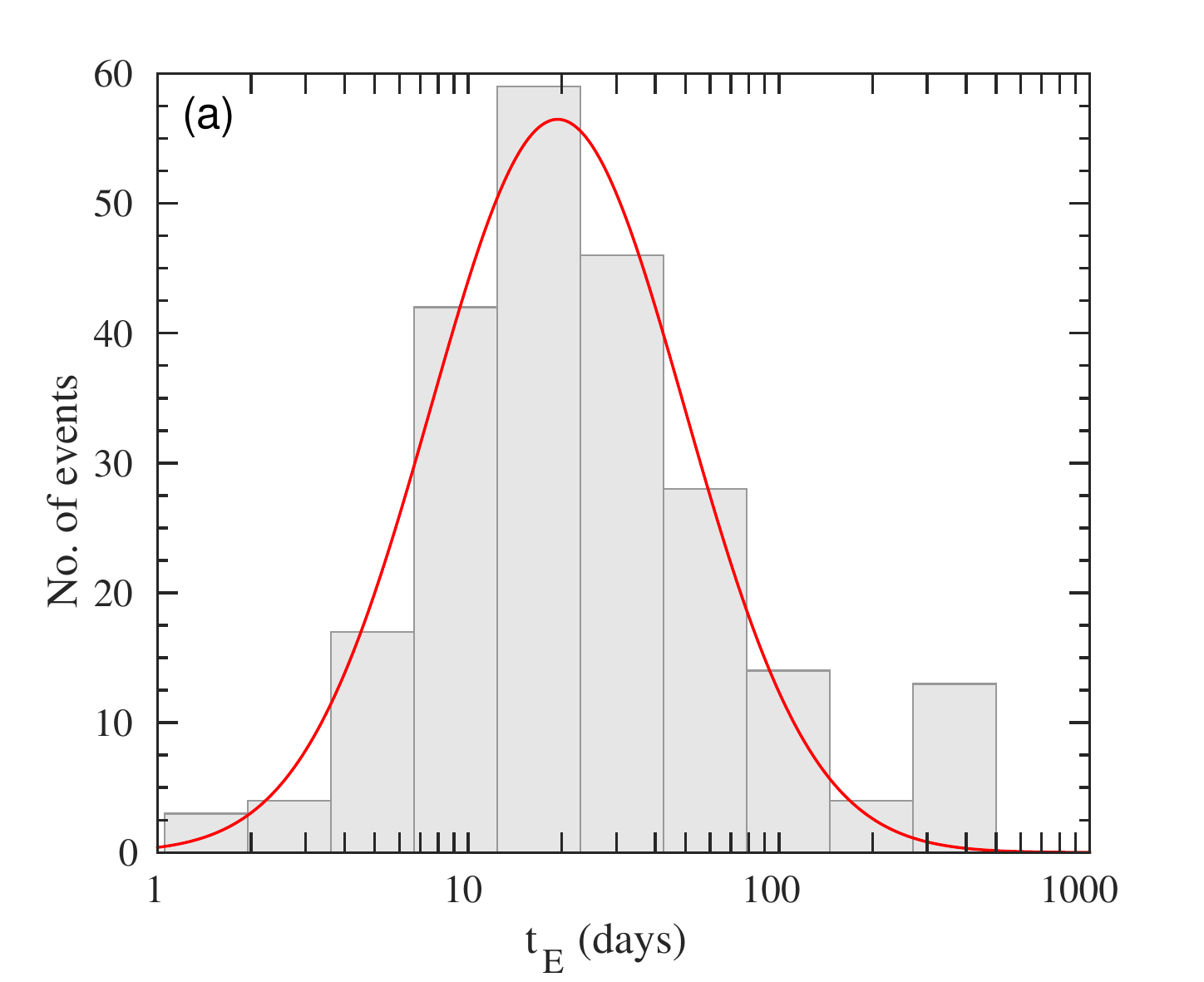}&
\includegraphics[width=0.5\textwidth]{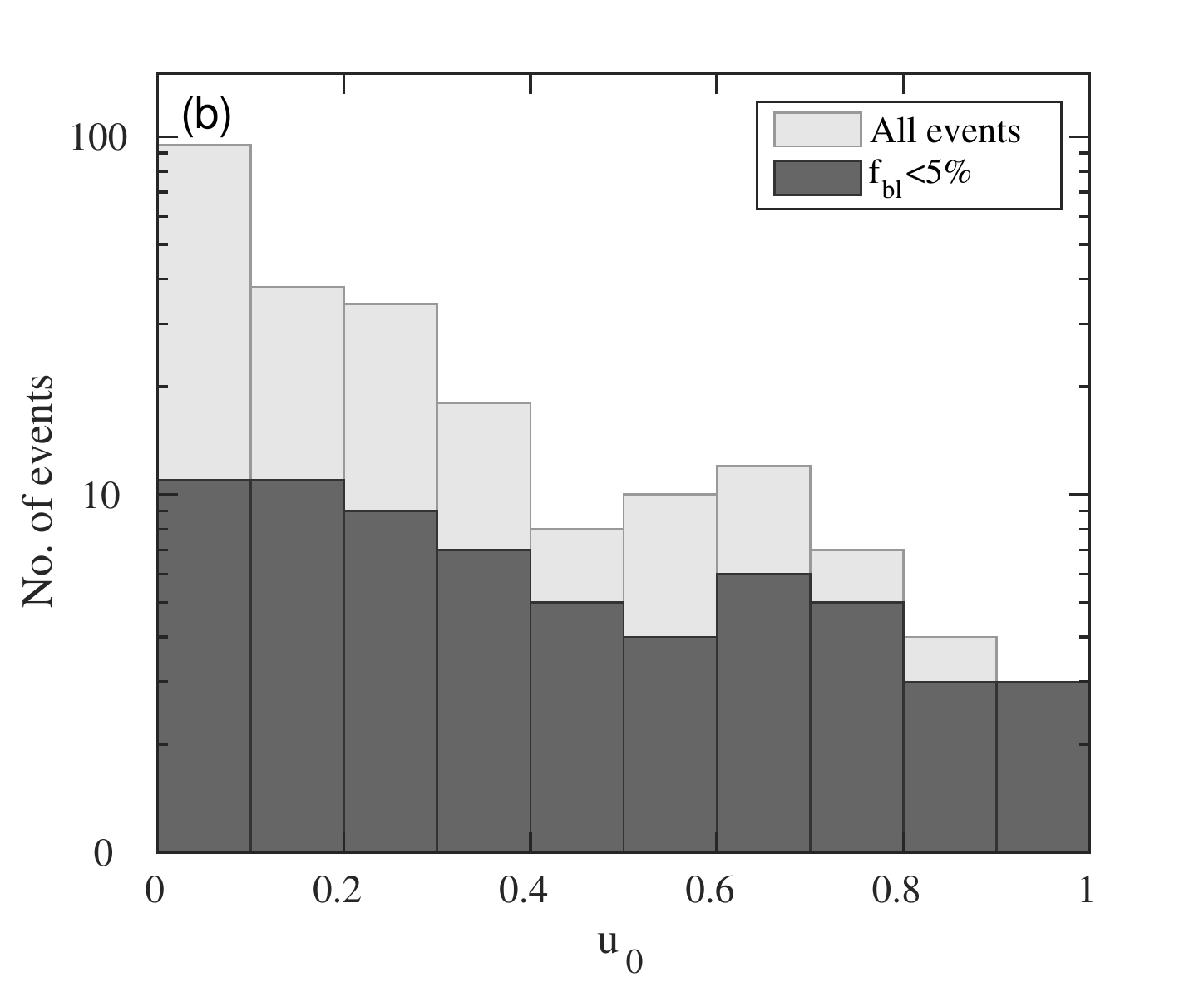}
\end{tabular}
\caption{Best-fit single-lens parameters for the sample of 224 lensing events:
(a) Distribution of event timescale, $t_{\rm E}$. The solid curve is a lognormal fit, $N\propto{{\rm exp}(-\frac{(\log(t_{\rm E})-\mu)^2}{2\sigma^2})}$, with $\mu=1.29$ (corresponding to 19.5 days)
and $\sigma=0.54$;
(b) Distribution of impact parameter, $u_0$ (light-gray). Due to selection favoring the detection of highly magnified faint events over weakly magnified faint events,
there are more events at smaller impact parameters. The dark-gray histogram shows the distribution only for unblended (i.e. relatively bright) events,
which are distributed more uniformly, as expected.\label{fig:params}}
\end{minipage}
\end{figure*}

Each microlensing event in our sample is modeled with the three Paczy{\'n}ski parameters,
the two flux inter-calibration parameters for each dataset, the finite source effects, and the orbital parallax (testing for both $u_0>0$ and $u_0<0$).
We explore this parameter space and find the best model parameters, and their uncertainties,
using a Markov-Chain Monte Carlo (MCMC) algorithm.
Outlier data points and real microlensing anomalies will obviously draw our solution away from the true one.
We therefore use an iterative algorithm for the exclusion of outliers (which naturally also applies to intervals in the light curve that include anomalies).
The photometric errors reported by DIA programs are usually underestimates.
Therefore, for each dataset we scale these errors to set $\chi^2$ per degree of freedom (DOF) equal unity, 
without the outliers and anomalies which will bias the distributions, allowing to estimate the real precision of each dataset.
Figure~\ref{fig:RMS} shows the distribution of error scaling factors for each group.
The median factors for OGLE, MOA and Wise are 1.74, 1.32 and 1.21, respectively.
Finally, we initiate the entire modeling process again without the excluded outliers and with the rescaled errors.
Figure \ref{fig:outliers} shows an example of the results of our outlier detection and exclusion algorithm, for an event with a strong parallax signal, 
where the single-lens model fits well the light curve after outlier points are excluded.

Table~\ref{table:data} in Appendix A lists the best single-lens model parameters for each event in our sample,
and their uncertainties, which were extracted from the MCMC estimation of the covariance matrix.
Microlens parallax is significantly detected ($\Delta\chi^2>100$ with respect to a model without parallax) 
in 10\% of the non-anomalous events, while finite source effects are found in 3\% of the non-anomalous events.
The values for the microlens parallax and finite source parameters for those events are marked in boldface in the table.
We do not mark the values for anomalous events, since the anomaly might affect those parameters,
and they can be reliably detected only with a full binary-lens model.

Figure~\ref{fig:params} shows, for the entire sample of events, the distributions of the best-fit values of the event timescale, $t_{\rm E}$, and the impact parameter, $u_0$.
The $t_{\rm E}$ distribution is shown fitted with a log-normal function,
$N\propto{{\rm exp}(-\frac{(\log(t_{\rm E})-\mu)^2}{2\sigma^2})}$,
with median timescale of 19.5 days.
$u_0$ should, in principle, follow a uniform probability. However, faint sources that undergo high magnification will be detected, while
similar sources undergoing a lower magnification event will be missed, due to the limiting magnitudes of the surveys, introducing a bias towards high magnification (low $u_0$) events.
The $t_{\rm E}$ and $u_0$ distributions of our sample are almost identical to the distributions of those parameters from previous years' OGLE and MOA microlensing seasons (see \citealt{Shvartzvald.2012.A}).
This shows that our sample has no obvious biases, and thus represents a typical population of lenses and sources.


\subsection{Anomaly detection filter}
\label{sec:filter}

A companion (or multiple companions) to the lens star can break the symmetry of the microlensing light curve, introducing an anomaly with respect to the single-lens model.
In order to identify microlensing anomalies in the light curves in our sample, we follow \cite{Shvartzvald.2012.A} and define an anomaly detection filter using a ''running'' $\chi^2$ estimator,
\begin{equation}
\chi^2_{\rm local} = \sum\limits_{i=1}^{N} \dfrac{(f_i-f_{\rm pl})^2}{\sigma_i^2},
\end{equation}
where $f_i$ and $\sigma_i$ are the observed flux and the re-scaled error of each epoch, respectively, and $f_{\rm pl}$ is the point-lens model flux at that time.
The local $\chi^2$ is repeatedly calculated by advancing the center of the filter one observed epoch at a time.
Since the relative duration of an anomaly is roughly proportional to the square root of the lens-companion mass ratio,
we would like the time interval of each summation to be as short as possible, in order to be sensitive to low-mass planets.
However, in order to avoid false-positive detections in short timescale events, $N$ cannot be too small, and we study the sensitivity of our filter to $N$ below
(see Section \ref{sec:false-positive} for more details).
An anomaly is considered detected if:
\begin{description}
\item[a.] $\chi^2_{\rm local}$ divided by the number of summed points, $N$, is larger than some threshold value, $P_{\rm thresh}$.
\item[b.] at least three consecutive data points have at least a 3$\sigma$ deviation from the point-lens model.
\end{description}
This latter criterion deals with the possibility that photometric outlier points will affect the local $\chi^2$ test and can lead to false detections.

\subsubsection{False-positive optimization}
\label{sec:false-positive}

In order to optimize the number of false-positive detections (i.e. to reduce their numbers without overly sacrificing true detections),
we have studied the sensitivity of our anomaly detection filter to its two parameters ($N$, $P_{\rm thresh}$).
For each of the 224 events in our sample, we use the fitted point-lens parameters to construct a magnification curve.
We then sample the light curve at the exact epochs on which the event was observed by each group. With this we account for all gaps due to weather or technical failures,
and for the different sampling cadence by each group and for each field.
The theoretical magnification curve is then randomly noised, using the observed residual distribution that we have found for each group
(see Section \ref{sec:data}), and assigned the scaled-errors from the real data, to create 500 simulated point-lens light curves for each event.
We then search for a false-positive anomaly in the simulated light curve (which has no real anomalies), varying the filter parameters.

We find that the fraction of simulated light curves that trigger the photometric outlier filter (i.e. three consecutive points with $>3\sigma$) is 0.65\%.
Examining those triggers, we find that 92\% of them have $\chi^2_{\rm local}/N<P_{\rm thresh}$, where $P_{\rm thresh}$ is in the range 7.5 to 8.
The optimal $P_{\rm thresh}$ varies from event to event, and depends linearly on the 
maximum amplification ($D_{\rm mag}$), which is the
difference between the observed baseline magnitude of
the event and its peak magnitude, with also a weak dependence on $N$.
The optimal value of $N$ is $N=40$.
For much smaller $N$, the $\chi^2_{\rm local}$ can be dominated by a single outlier point, while for much larger values, the anomalous region can be smoothed
out and not detected.
Using this $N$ and the $D_{\rm mag}$-dependent value of $P_{\rm thresh}$, 8\% of the 0.65\% of the outlier-triggered events trigger an anomaly detection, and
thus the resultant false-positive rate is at a low level of 0.05\%.
These high-threshold criteria help to reduce false-positive detections in the real data, but do not completely avoid them, due to the presence of red
noise (i.e. correlated deviant points), a noise that is not included in these randomly noised simulations.

\subsubsection{Detection efficiency}
\label{sec:efficiency}

In order to estimate the sensitivity of our microlensing survey to planets and stellar binaries, we have constructed a large sample of simulated microlensing light curves
for point-mass lenses with point-mass companions,
which were calculated with an adaptive-mesh inverse ray-shooting microlensing light curve generator (see Shvartzvald \& Maoz 2012 for details).
Specifically, for each of the 224 events in the real sample, we created a model with the same best-fit 
point-lens parameters, as described above, plus a companion to the lens star. The companion introduces three additional parameters:
$q$---the mass ratio between the companion and the lens star; 
and the two-dimensional projected position of the companion in the lens plane,
relative to the host star position, given by means of a projected angular separation, $s$, in units of the angular Einstein radius, and an angle, $\alpha$,
measured counter-clockwise from the source trajectory in the lens plane.
We explore a mass-ratio distribution, uniform in log $q$, in the range $-6 < {\rm log}~q < 0$, a uniform distribution in the scaled projected separation of $0.3 < s < 3$
(which encompasses the region microlensing is sensitive to), and all possible angles $\alpha$.
Our working assumption is that the probability for a companion is independent of an event's single-lens parameter probabilities,
which incorporate the host-star properties of mass, proper motion, and distance.
For every real event, 3000 simulated light curves, with a variety of companions, were generated. 

\begin{figure}
\centering
\begin{tabular}{c}
\includegraphics[width=0.5\textwidth]{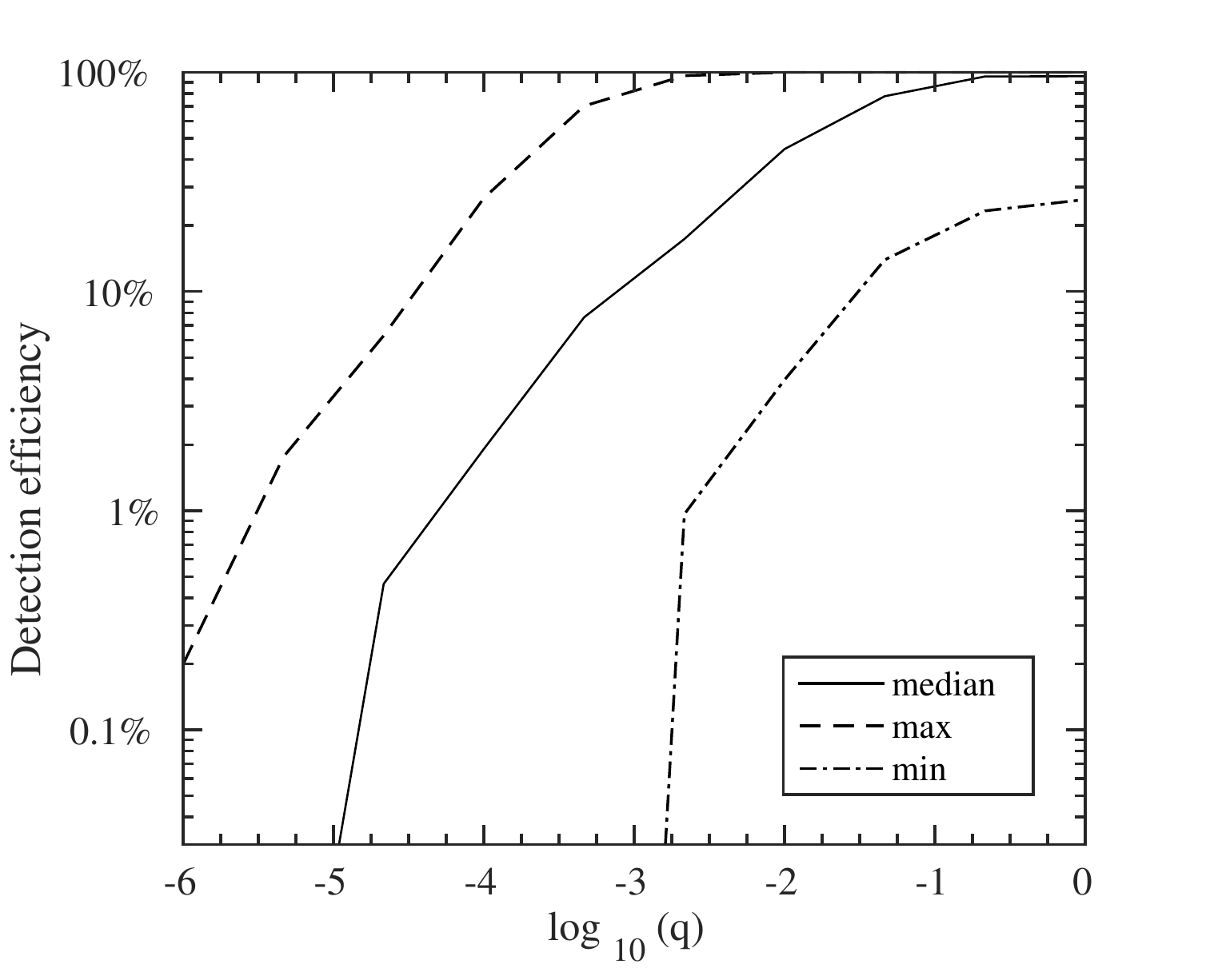}
\end{tabular}
\caption{Detection efficiency for a single companion as a function of mass ratio, $q$. The event with median efficieny is shown as the solid line.
The dashed curves represent events with $10^{\rm th}$ and $90^{\rm th}$ percentile sensitivities.\label{fig:efficiency}}
\end{figure}

The simulated magnification curves of each event are then sampled exactly like the real event and randomly noised,
as described for the point-lens models used for the false-positive optimization procedure.
We then search for anomalies in the simulated light curves with the same procedure and detection criteria that were used for the real data.
Explicitly, each light curve is modeled with a point-lens and higher order effects (including error-scaling for each observatory),
and we search for an anomaly with our detection filter, using the parameters we found for minimizing the false-positive rate.

The ``lensing zone'' is the range of separations, $s$, in which microlensing is sensitive to anomalies caused by a given mass ratio.
For stellar binaries (i.e. $q\geq10^{-2}$), the lensing zone can extend over a factor of a few in $s$. For small companions, $q\leq10^{-3}$ (likely planets), we find
that the range $0.5\leq s\leq2$ covers over 95\% of the detections, and therefore we calculate our detection efficiency for that range.
The companion detection efficiency for each event as a function of mass ratio $q$ was found
by marginalizing over $s$ and $\alpha$:
\begin{equation}
\eta (q) = \frac{\iint\Theta(q,s,\alpha) s ds d\alpha}{\iint s ds d\alpha},
\Theta(q,s,\alpha)= 
\begin{cases}
    1, \text{anomalous}\\
    0, \text{no anomaly.}
\end{cases}
\end{equation}
Figure \ref{fig:efficiency} shows the results of our detection efficiency simulation. 
The sensitivity varies by over an order of magnitude
among different events, due to the brightness of the event, its magnification and its timescale, as shown by the $90^{\rm th}$ and $10^{\rm th}$ percentile efficiency curves (dashed) in Figure \ref{fig:efficiency}.
We find that our experiment is $\sim6$ times more sensitive to $q=10^{-3}$ (corresponding to a $\sim$Jupiter/Sun mass ratio) than to $q=10^{-4}$ (``super-Neptunes'').


\subsection{Anomalous events}
\label{sec:anomlous}

\subsubsection{Heuristic detections}
\label{sec:heuristic}

Before applying to the sample our objective automated anomaly detection filter, we have identified ``by eye'' 26 anomalous events among the sample
in which deviations from the best-fit point-lens models are clear.
Figures \ref{fig:heuristic}-\ref{fig:heuristic_last} show the inter-calibrated light curves of these 26 events and their best-fit single-lens models,
obtained after excluding the anomalous regions from the fit (i.e., points detected by the anomaly detection filter).
Eight of those anomalous events have already been modeled and published, as follows.
\begin{enumerate}
\item MOA-2011-BLG-293 (\citealt{Yee.2012.A}) was the first planet discovered by our genII survey. The event was highly magnified, and so was also followed up by several groups.
However, \cite{Yee.2012.A} showed that the survey data alone were sufficient to fully constrain the planetary model.
\cite{Batista.2014.A} used Keck adaptive optics observations to further constrain the system's parameters by isolating the light from the lens star, and found that this $\sim5 M_J$ planet is actually the first microlensing planet
discovered in the habitable zone of its host star (a G-type main sequence star).
\item MOA-2011-BLG-322 (\citealt{Shvartzvald.2014.A}), a moderate-magnification event that did not trigger alerts and follow up efforts, was the first planetary microlensing event
that was detected and analyzed based solely on the genII survey data.
Using a Bayesian analysis that incorporates a Galactic structure model, Shvartzvald et al. (2014) estimated that it is a $\sim12 M_J$ planet orbiting an M-type dwarf.
\item OGLE-2011-BLG-0265 (\citealt{Skowron.2015.A}) is a Jupiter-mass planet (with two degenerate solutions of $\sim1 M_J$ and $\sim0.6 M_J$) orbiting an M dwarf.
After the main anomaly was realized from the genII data, the event was alerted. Several follow-up groups monitored the event, allowing a better characterization of the secondary anomaly. This is an example of 
the problematic, in terms of an eventual statistical interpretation, process of the ``first-generation'' mode of surveys.
The survey data alone were, again, sufficient to fully constrain the planetary model.
\item OGLE-2013-BLG-0341 (\citealt{Gould.2014.A}) is a $\sim2 M_\oplus$ planet in a $\sim$1 AU orbit around one member of a $\sim$15 AU separation binary M-dwarf system.
The event was alerted and followed up since it was highly magnified, but the genII survey data cover all three anomaly regions. Moreover, the two lower-amplitude features,
which are the ones that revealed the presence of a planet, were covered only by the genII survey data.
\item OGLE-2014-BLG-0124 (\citealt{Udalski.2015.A}) is a $\sim0.5 M_J$ planet orbiting a K-type dwarf. This was the first microlensing planet co-observed with the $Spitzer$ space mission,
and used to measure the microlens parallax from simultaneous observations from Earth and space.
\cite{Udalski.2015.A} use data only from OGLE and $Spitzer$, but the anomaly and the parallax signal are obvious also from the genII survey data alone.
\item The physical parameters of the companions in three additional published events have not been fully determined, but all of them are likely to be binary-star systems.
MOA-2011-BLG-104 (\citealt{Shin.2012.A}) is a binary with a candidate brown-dwarf companion ($q\sim 0.09$).
OGLE-2012-BLG-0456 and MOA-2012-BLG-532 (\citealt{Henderson.2014.A}) have mass ratios of $q\sim 0.9$ and $q\sim 0.5$, respectively.
\end{enumerate}

We note that the anomalies seen in OGLE-2011-BLG-0481/MOA-2011-BLG-217, which appear only in MOA data, are possibly not real,
and if not are an example of the type of false positives that can enter our sample after all.
In line with our methodology, we include this event in our sample.

\begin{figure*}
\begin{minipage}{\textwidth}
\centering
\begin{tabular}{c}
\vspace{-1.2cm}
\subfloat{\includegraphics[width=0.8\textwidth]{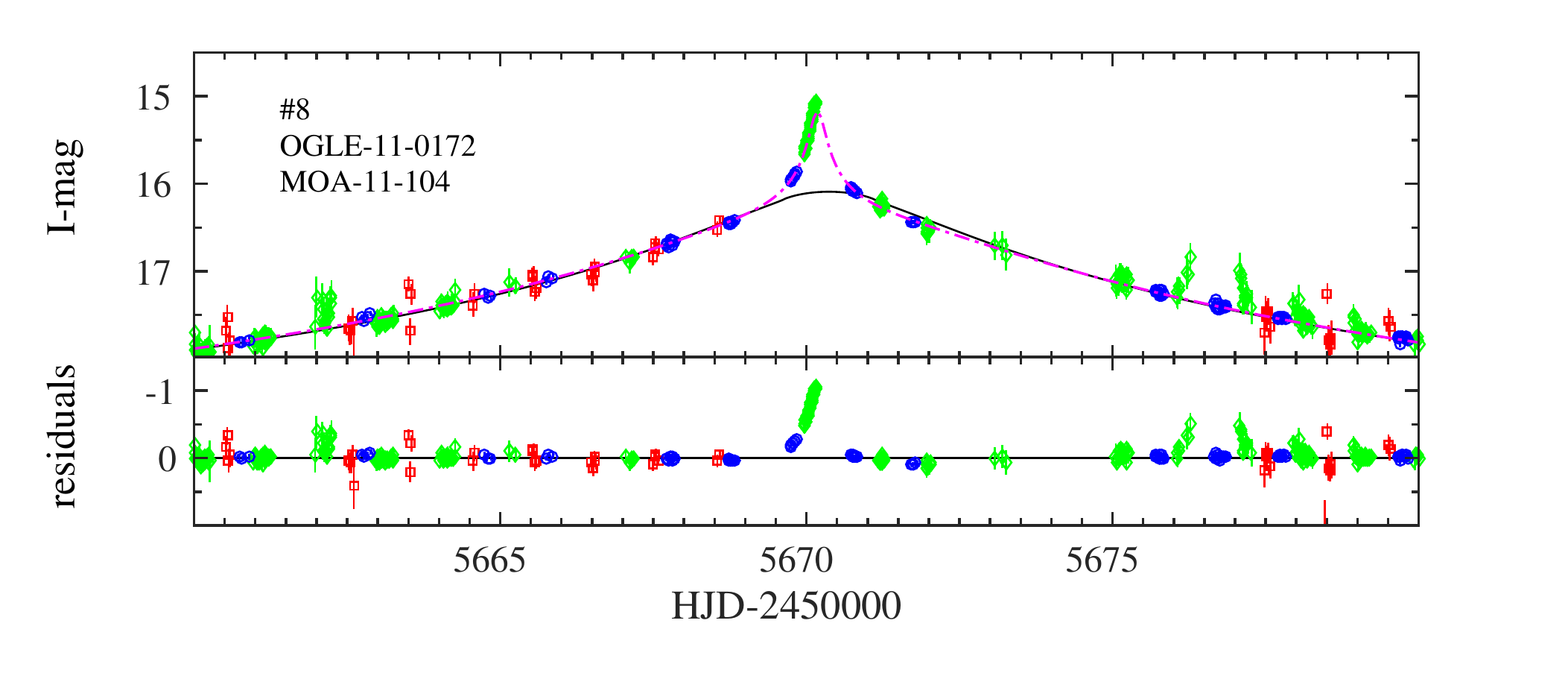}}\\
\vspace{-1.2cm}
\subfloat{\includegraphics[width=0.8\textwidth]{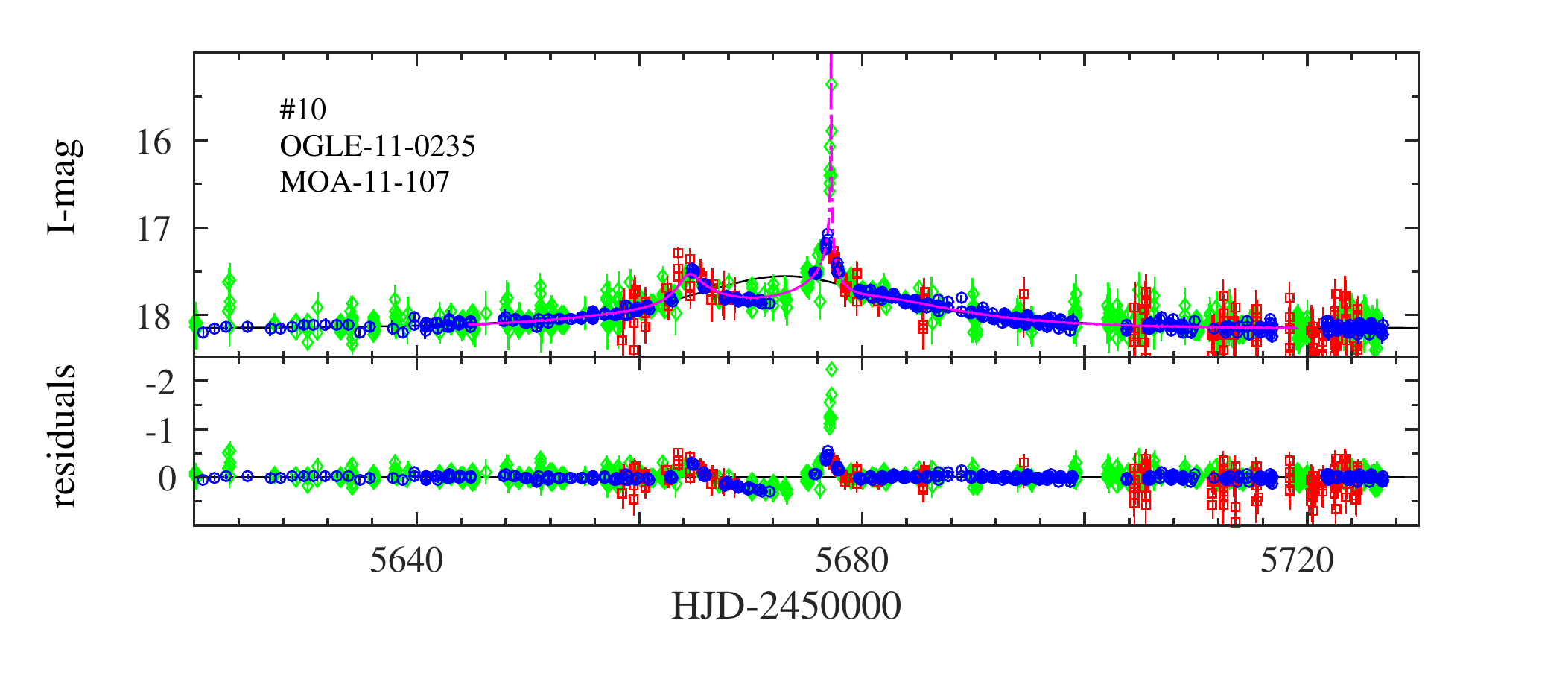}}\\
\vspace{-1.2cm}
\subfloat{\includegraphics[width=0.8\textwidth]{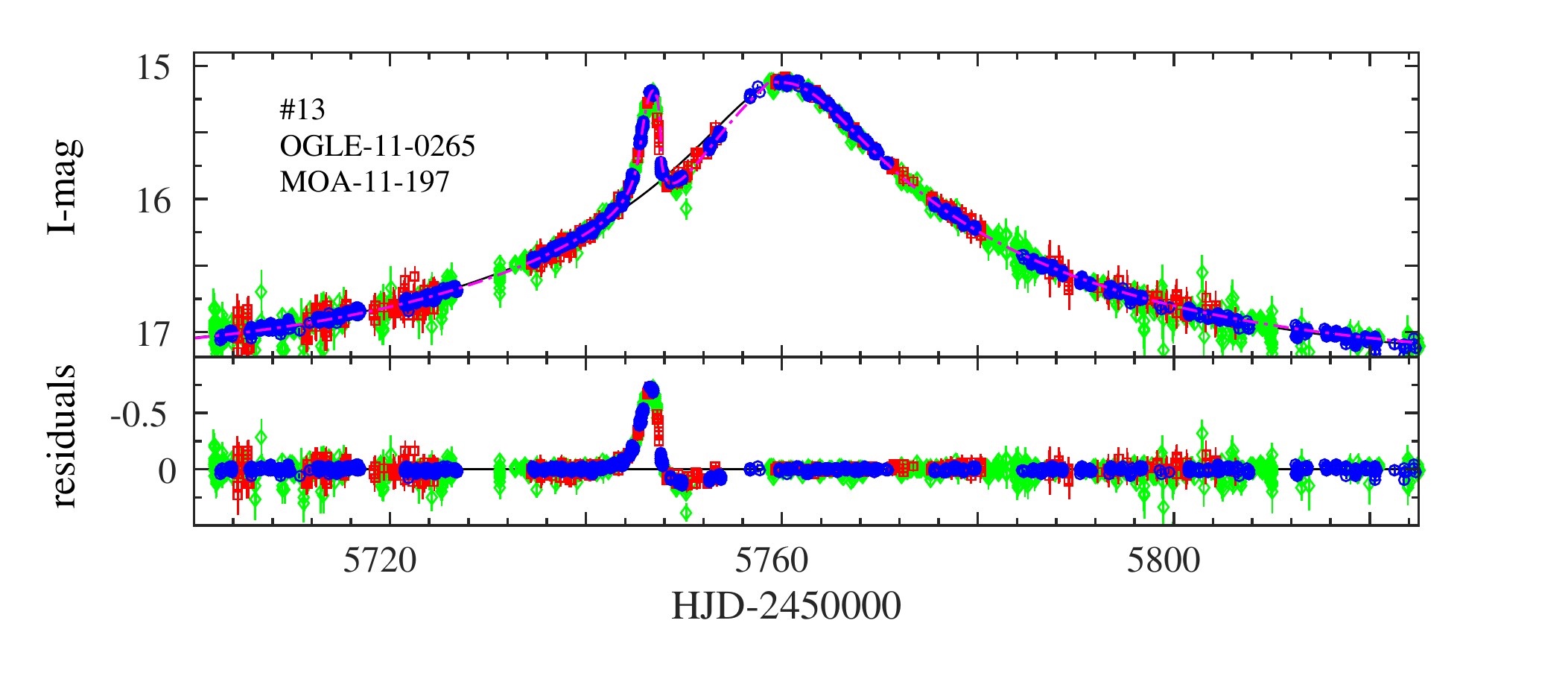}}\\
\vspace{-0.3cm}
\subfloat{\includegraphics[width=0.8\textwidth]{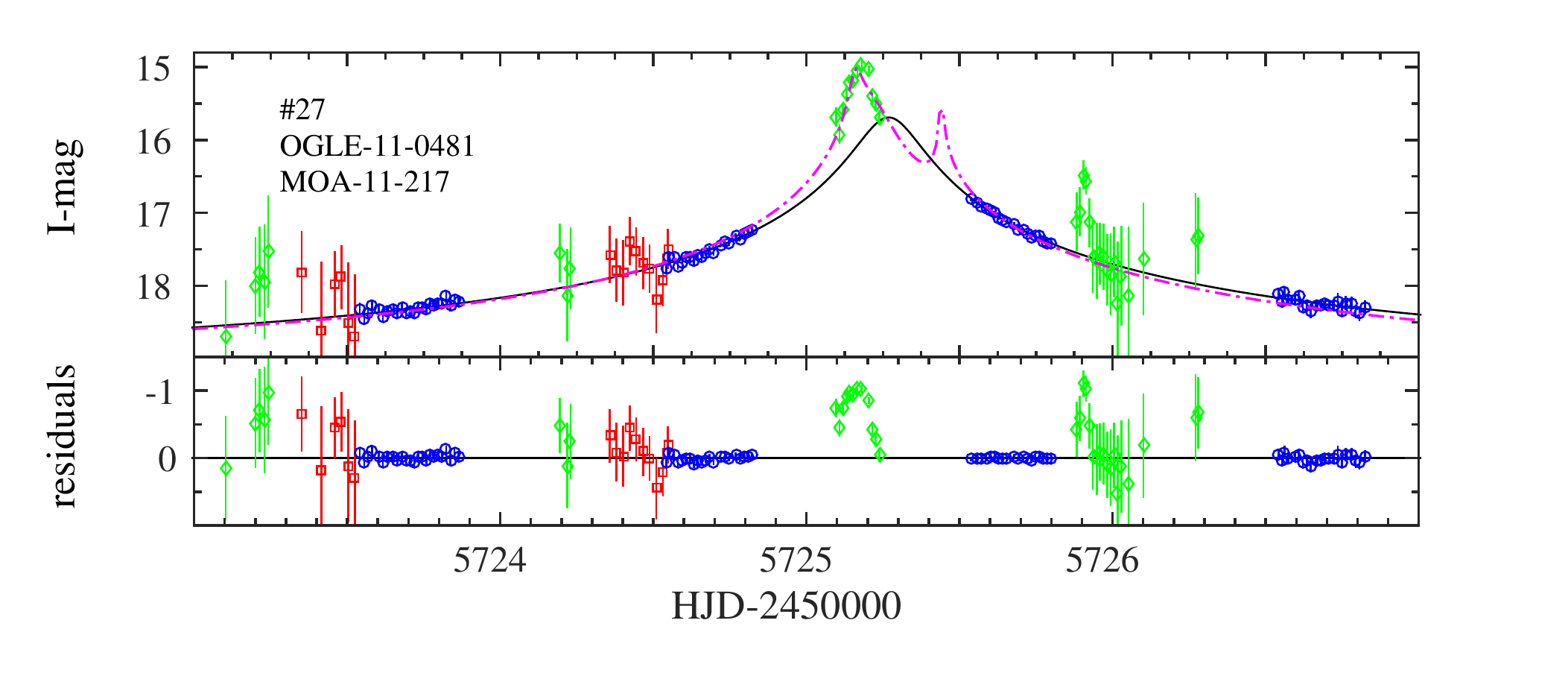}}\\
\end{tabular}
\caption{Inter-calibrated light curves of events with a clear anomaly that we identify by-eye: OGLE -- blue circles, MOA -- green diamonds, Wise -- red squares.
Magenta lines are the best-fit model, either published or from the grid search.
The residuals from the point-lens model are shown in the lower panel of each event.
The identifying number from Appendix~A is marked in the upper-left corner.\label{fig:heuristic}}
\end{minipage}
\end{figure*}

\begin{figure*}
\begin{minipage}{\textwidth}
\centering
\begin{tabular}{c}
\vspace{-1.2cm}
\subfloat{\includegraphics[width=0.8\textwidth]{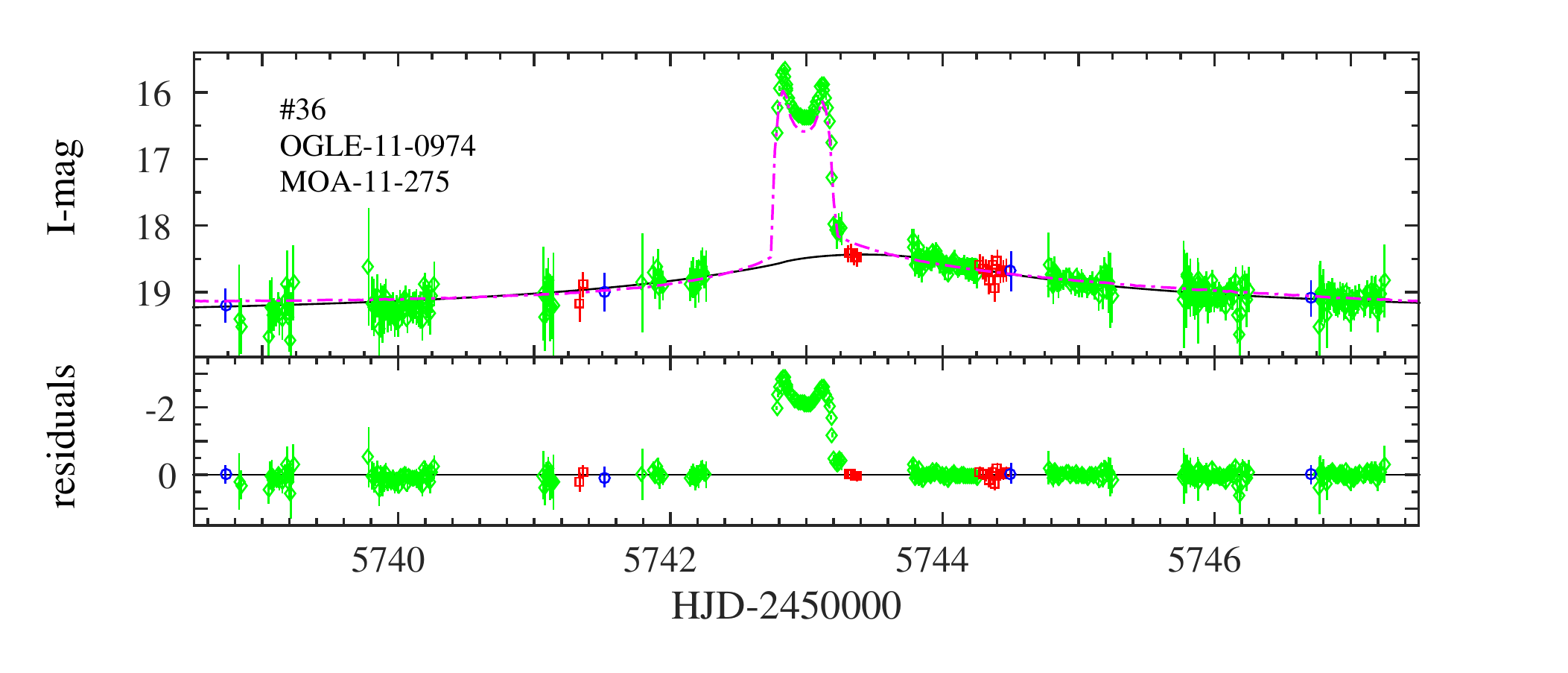}}\\
\vspace{-1.2cm}
\subfloat{\includegraphics[width=0.8\textwidth]{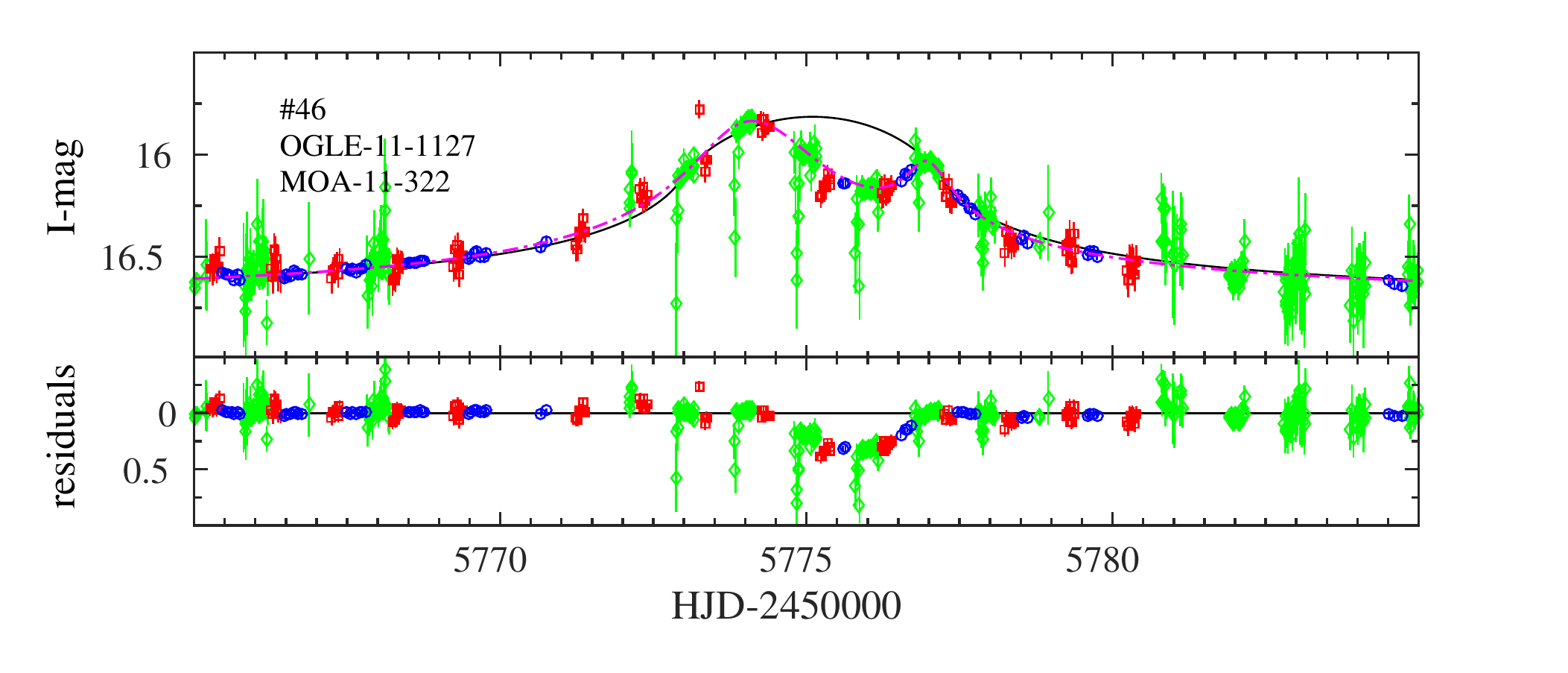}}\\
\vspace{-1.2cm}
\subfloat{\includegraphics[width=0.8\textwidth]{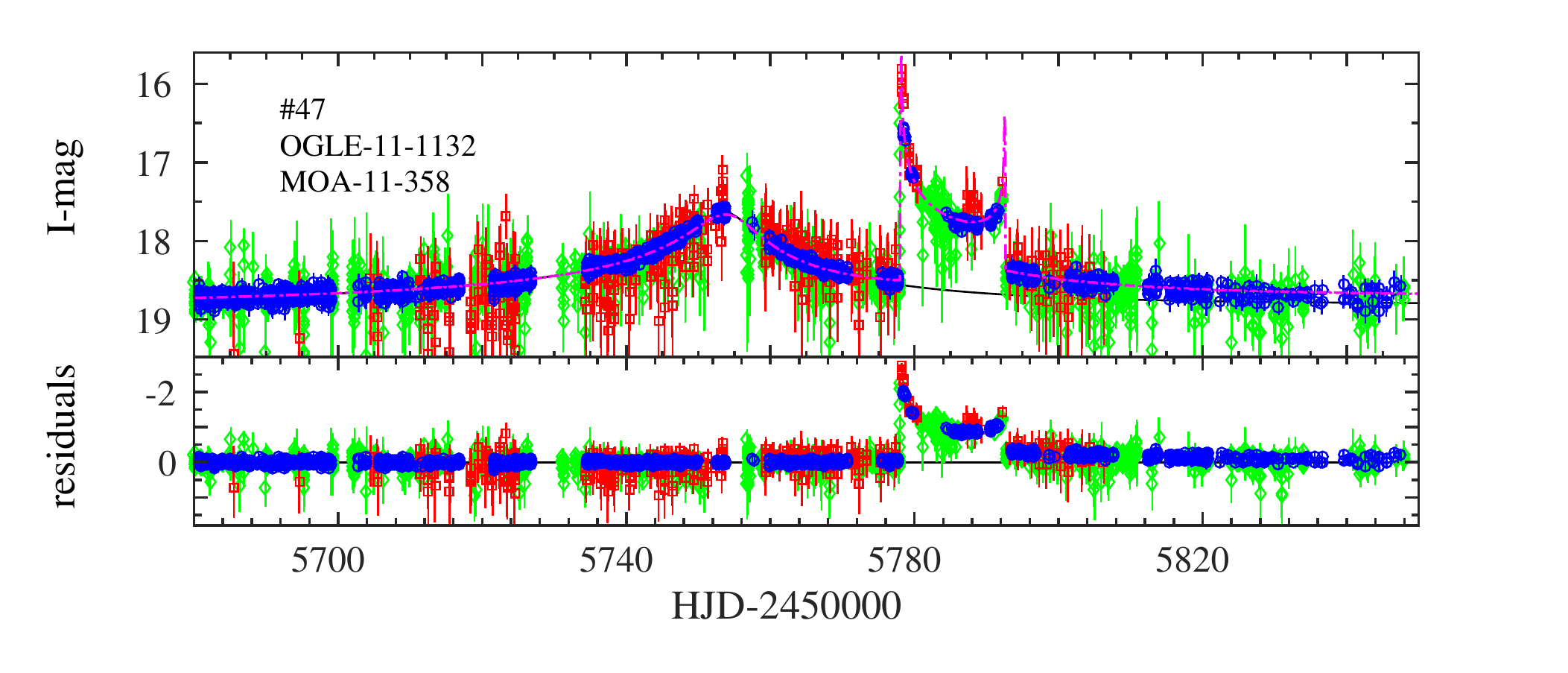}}\\
\vspace{-0.3cm}
\subfloat{\includegraphics[width=0.8\textwidth]{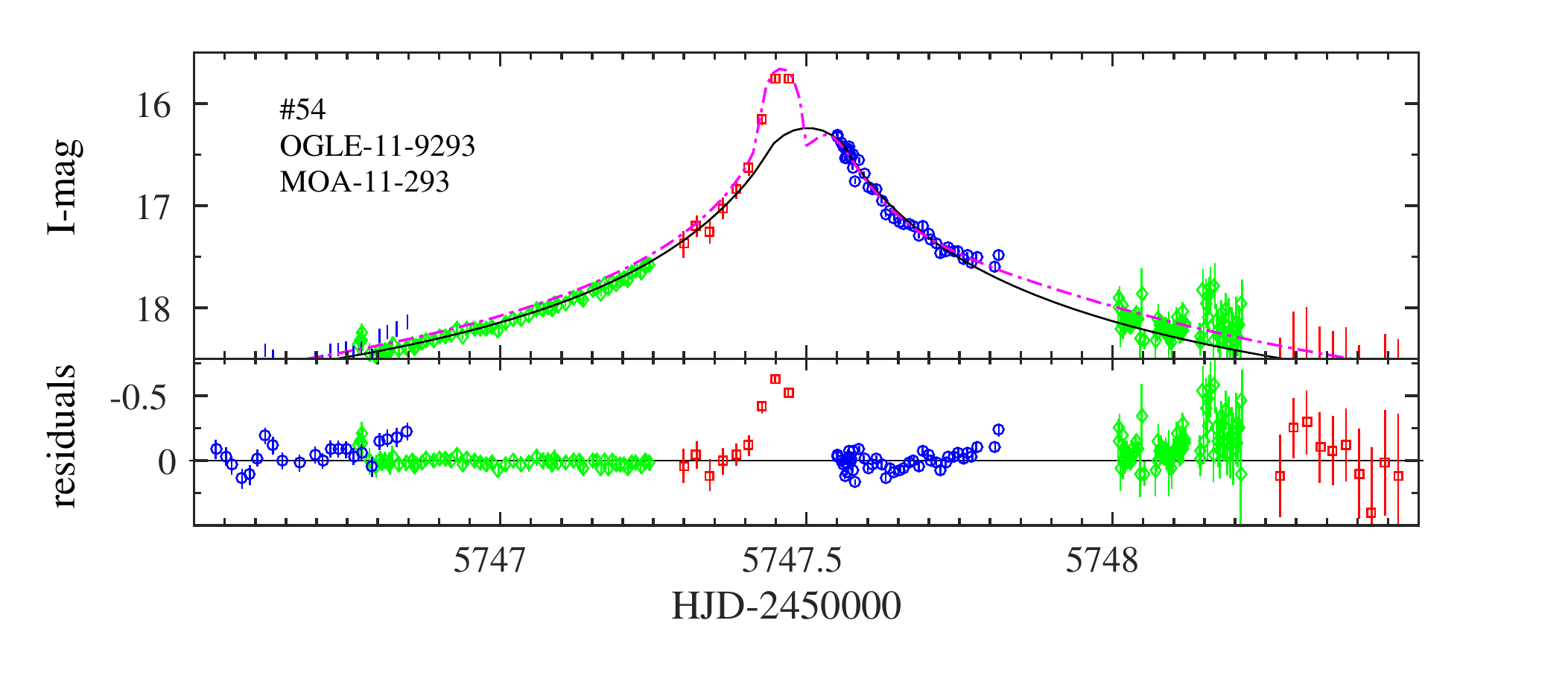}}\\
\end{tabular}
\caption{Continued from previous page.\label{fig:heuristic3}}
\end{minipage}
\end{figure*}

\begin{figure*}
\begin{minipage}{\textwidth}
\centering
\begin{tabular}{c}
\vspace{-1.2cm}
\subfloat{\includegraphics[width=0.8\textwidth]{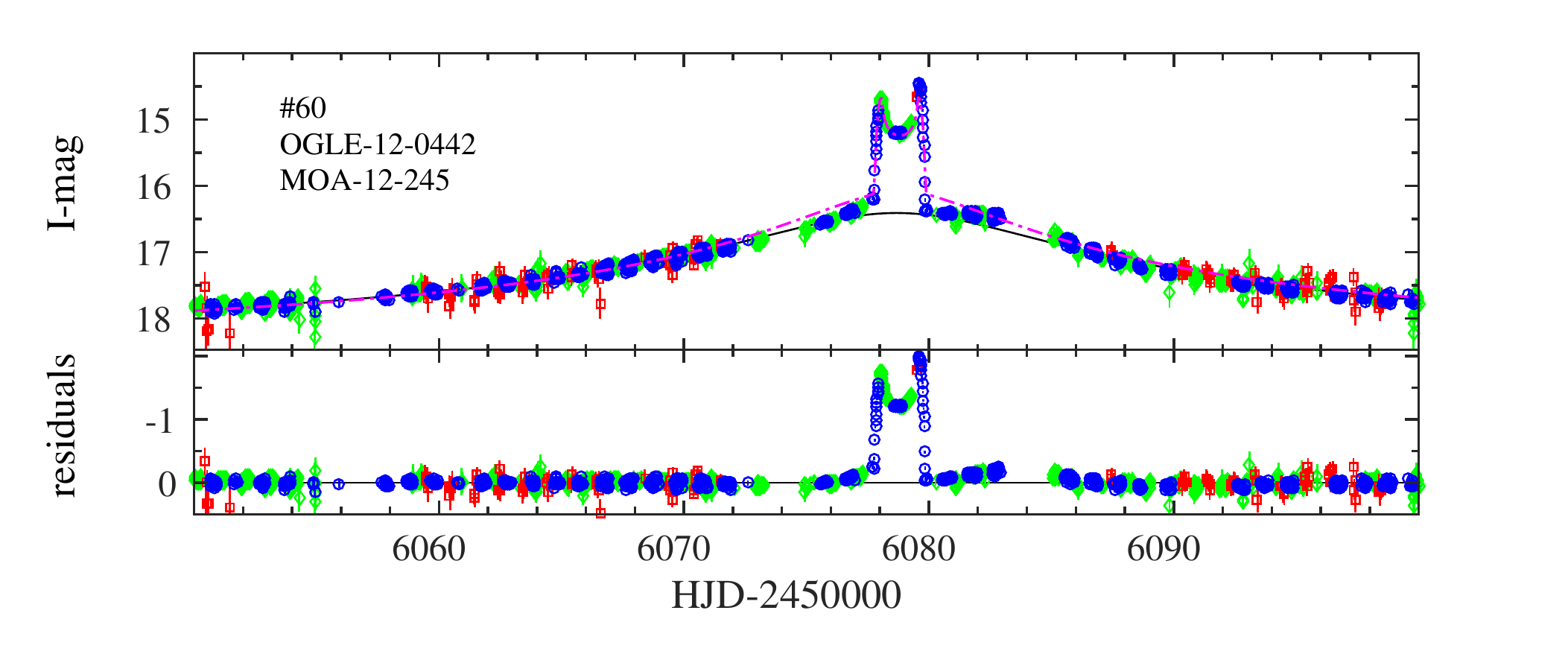}}\\
\vspace{-1.2cm}
\subfloat{\includegraphics[width=0.8\textwidth]{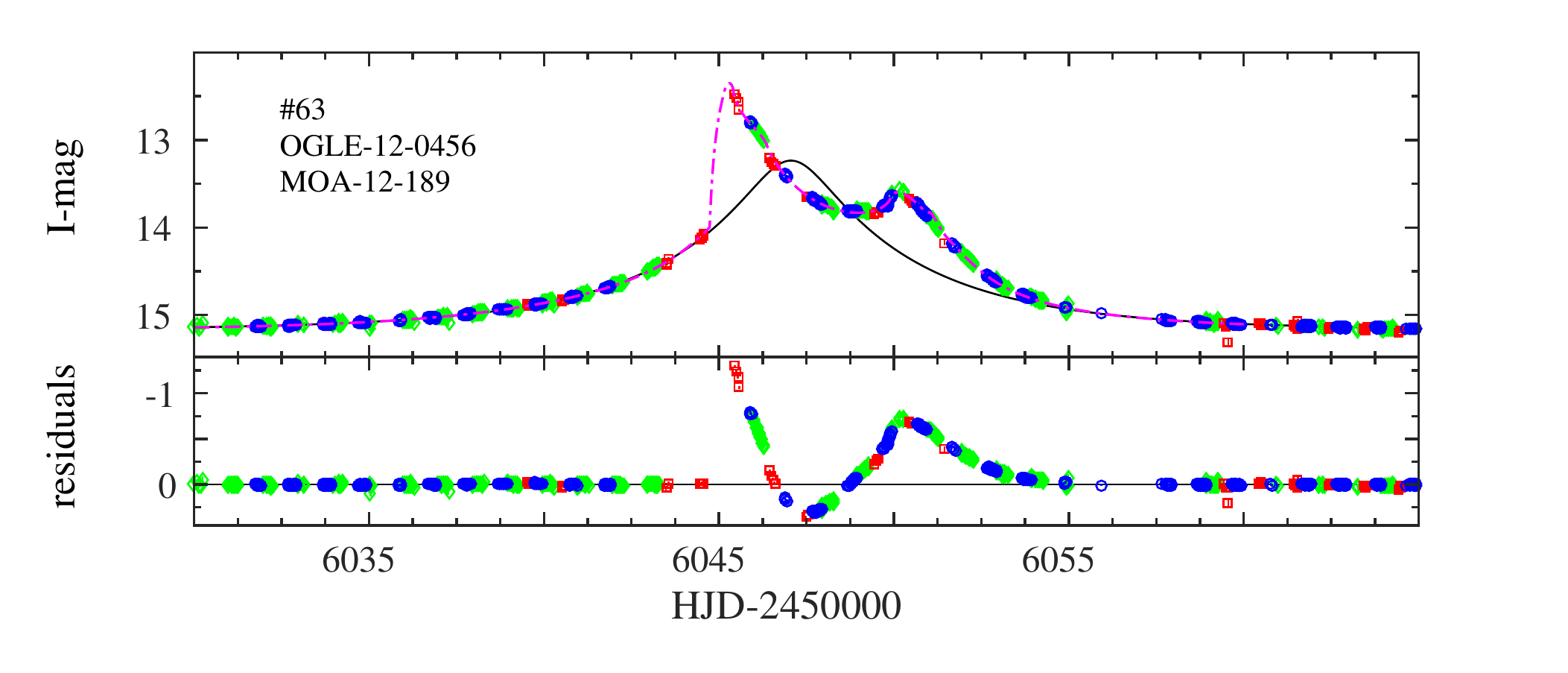}}\\
\vspace{-1.2cm}
\subfloat{\includegraphics[width=0.8\textwidth]{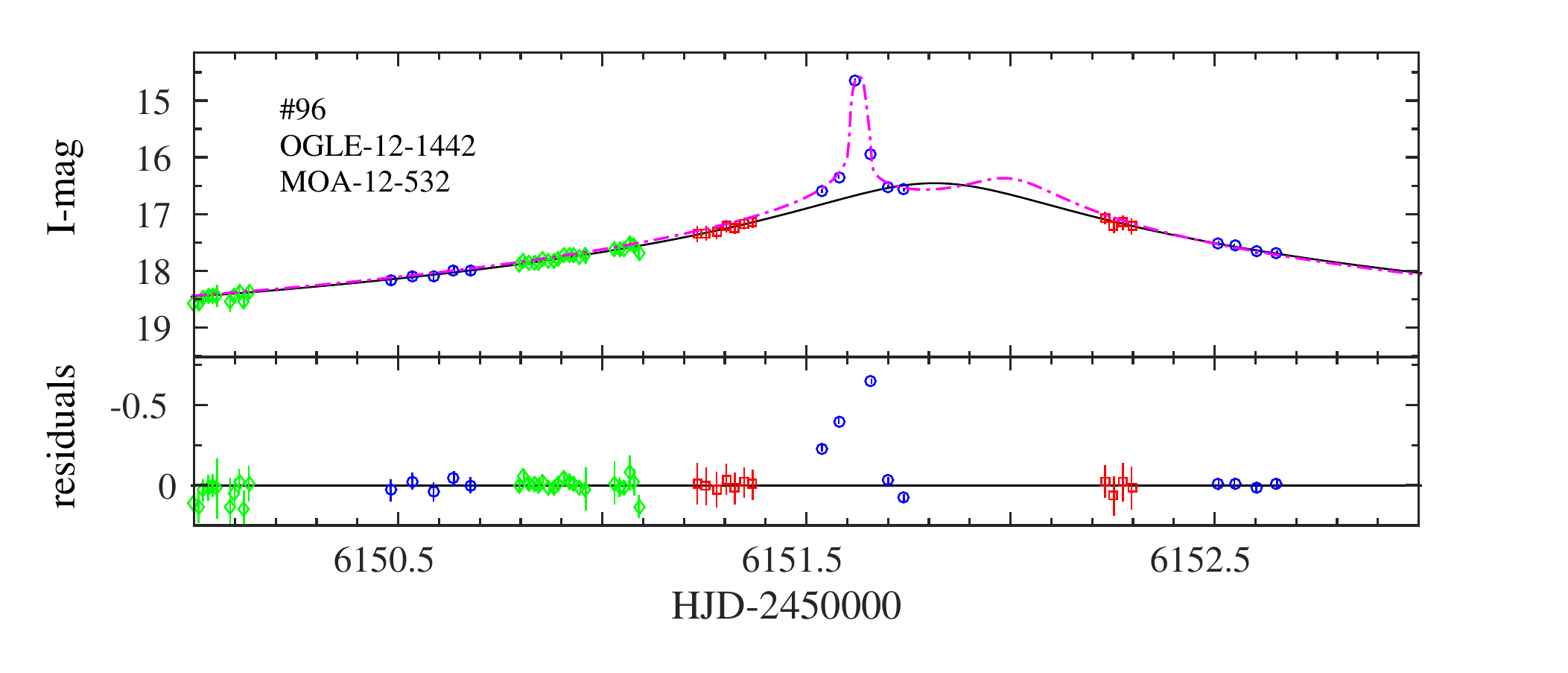}}\\
\vspace{-0.3cm}
\subfloat{\includegraphics[width=0.8\textwidth]{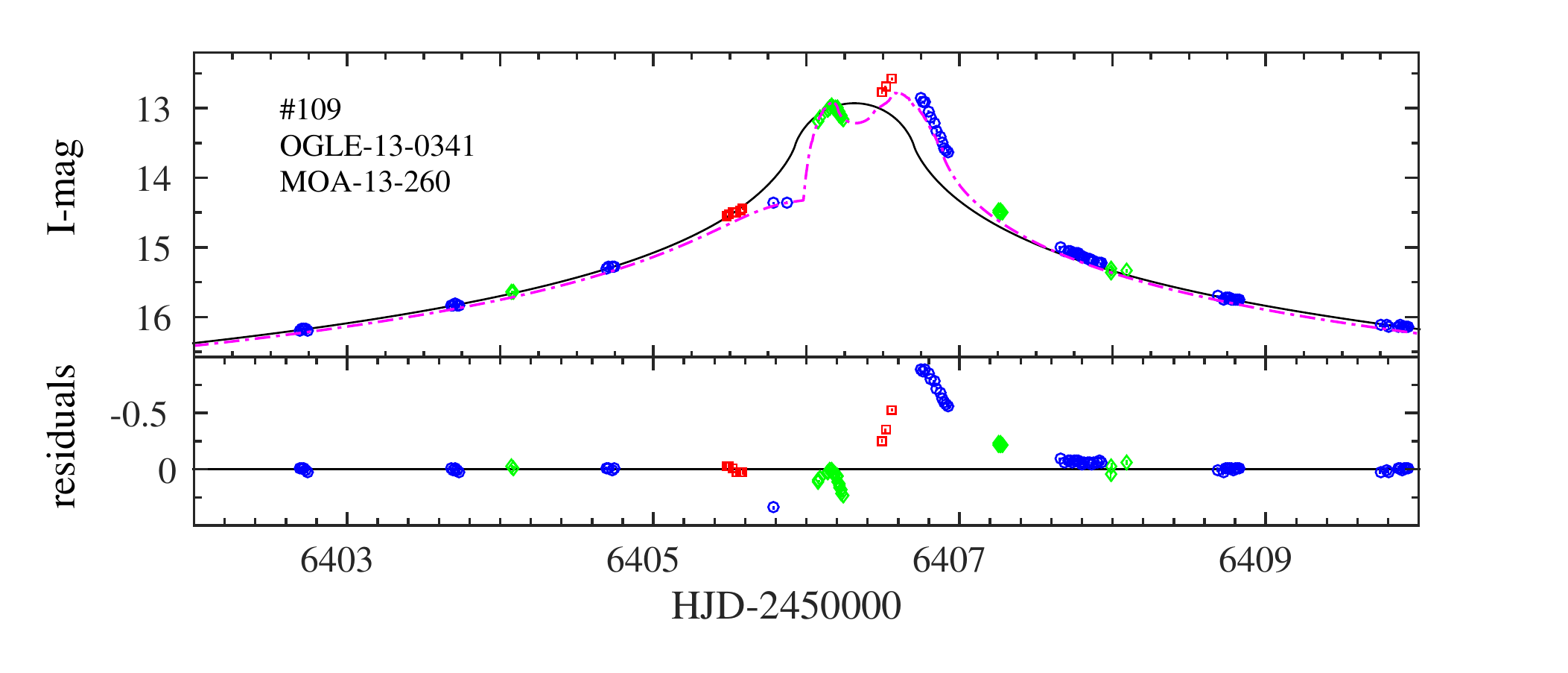}}\\
\end{tabular}
\caption{Continued from previous page.\label{fig:heuristic3}}
\end{minipage}
\end{figure*}

\begin{figure*}
\begin{minipage}{\textwidth}
\centering
\begin{tabular}{c}
\vspace{-1.2cm}
\subfloat{\includegraphics[width=0.8\textwidth]{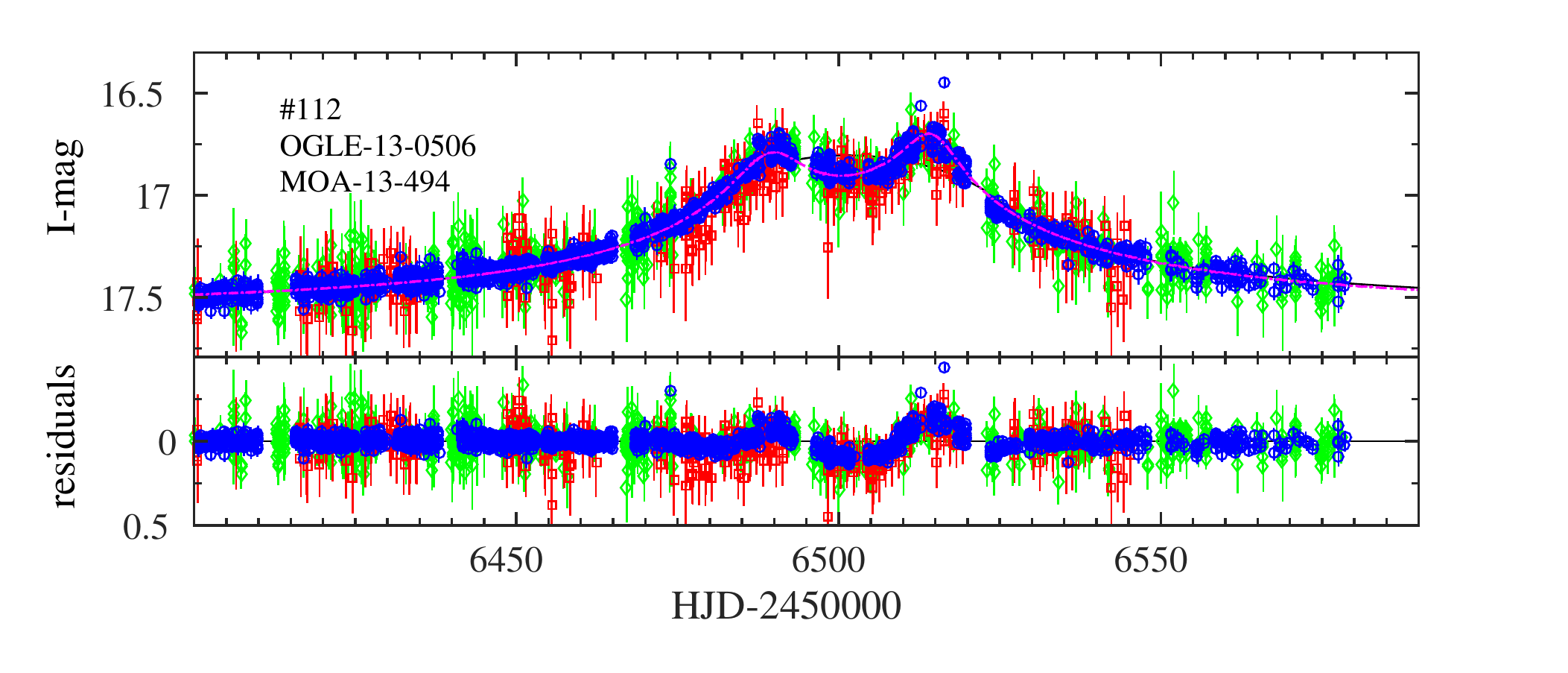}}\\
\vspace{-1.2cm}
\subfloat{\includegraphics[width=0.8\textwidth]{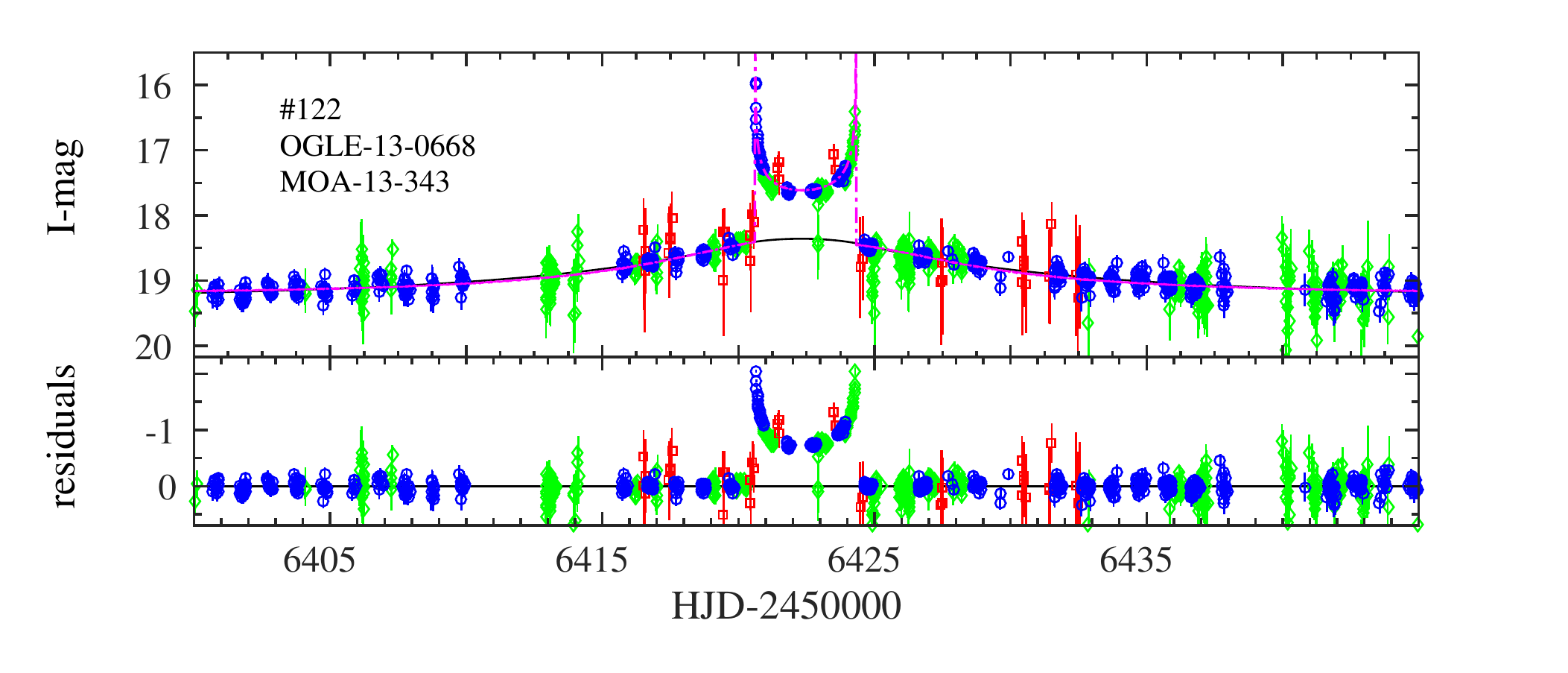}}\\
\vspace{-1.2cm}
\subfloat{\includegraphics[width=0.8\textwidth]{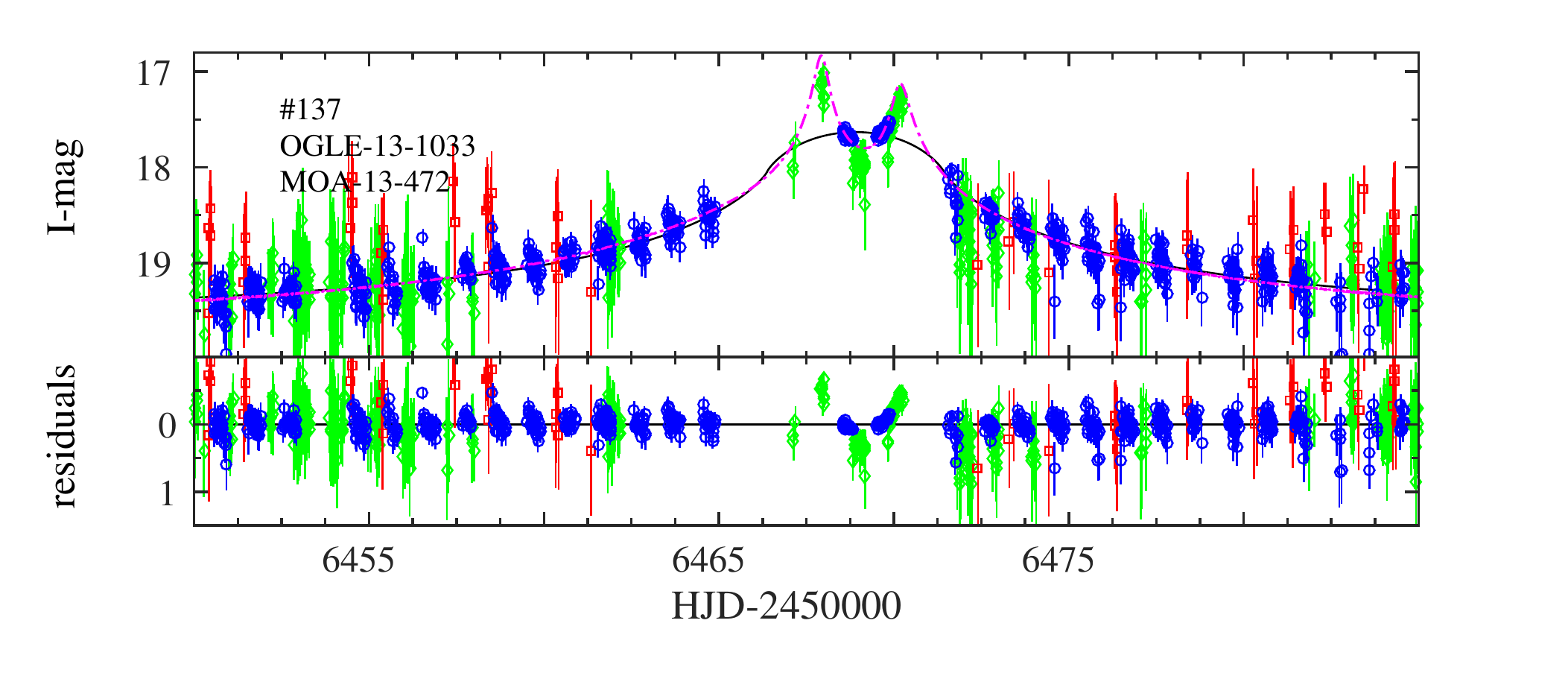}}\\
\vspace{-0.3cm}
\subfloat{\includegraphics[width=0.8\textwidth]{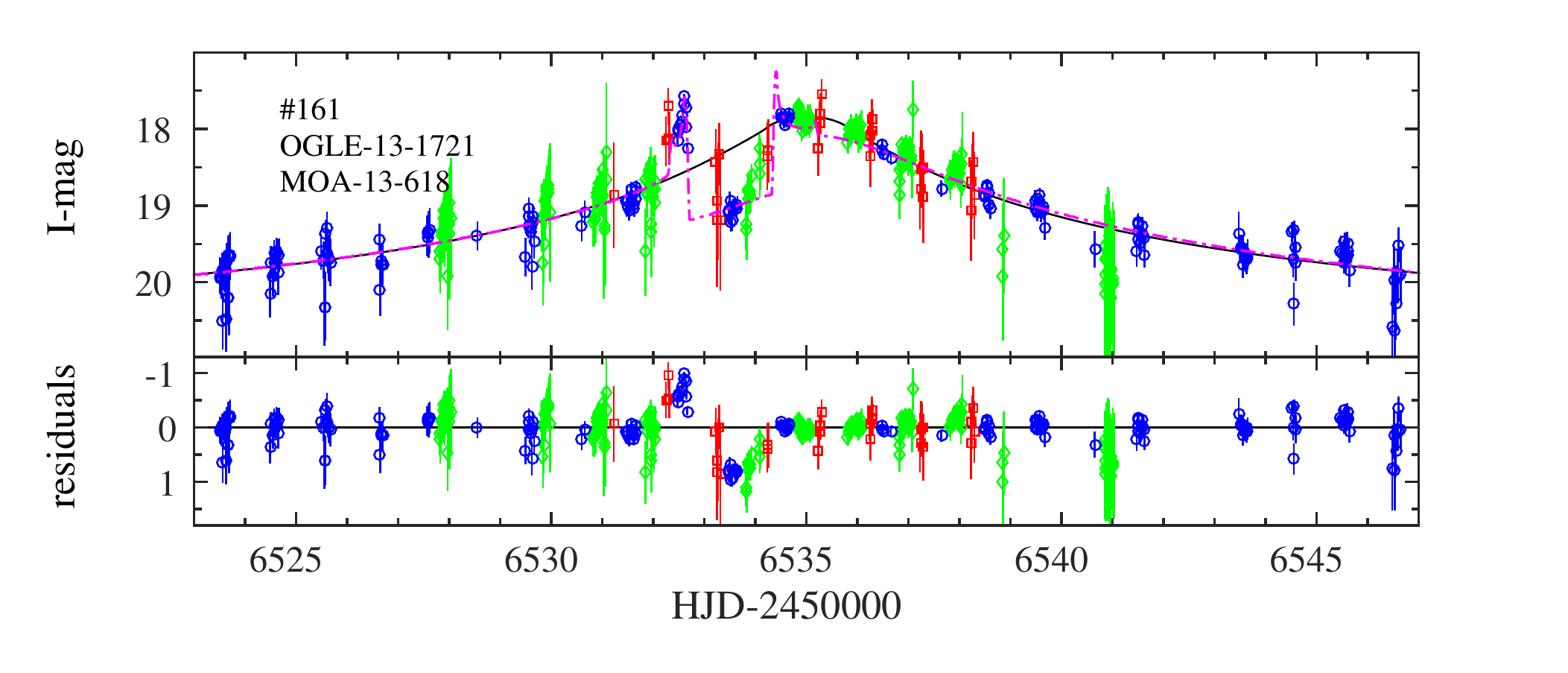}}\\
\end{tabular}
\caption{Continued from previous page.\label{fig:heuristic3}}
\end{minipage}
\end{figure*}

\begin{figure*}
\begin{minipage}{\textwidth}
\centering
\begin{tabular}{c}
\vspace{-1.2cm}
\subfloat{\includegraphics[width=0.8\textwidth]{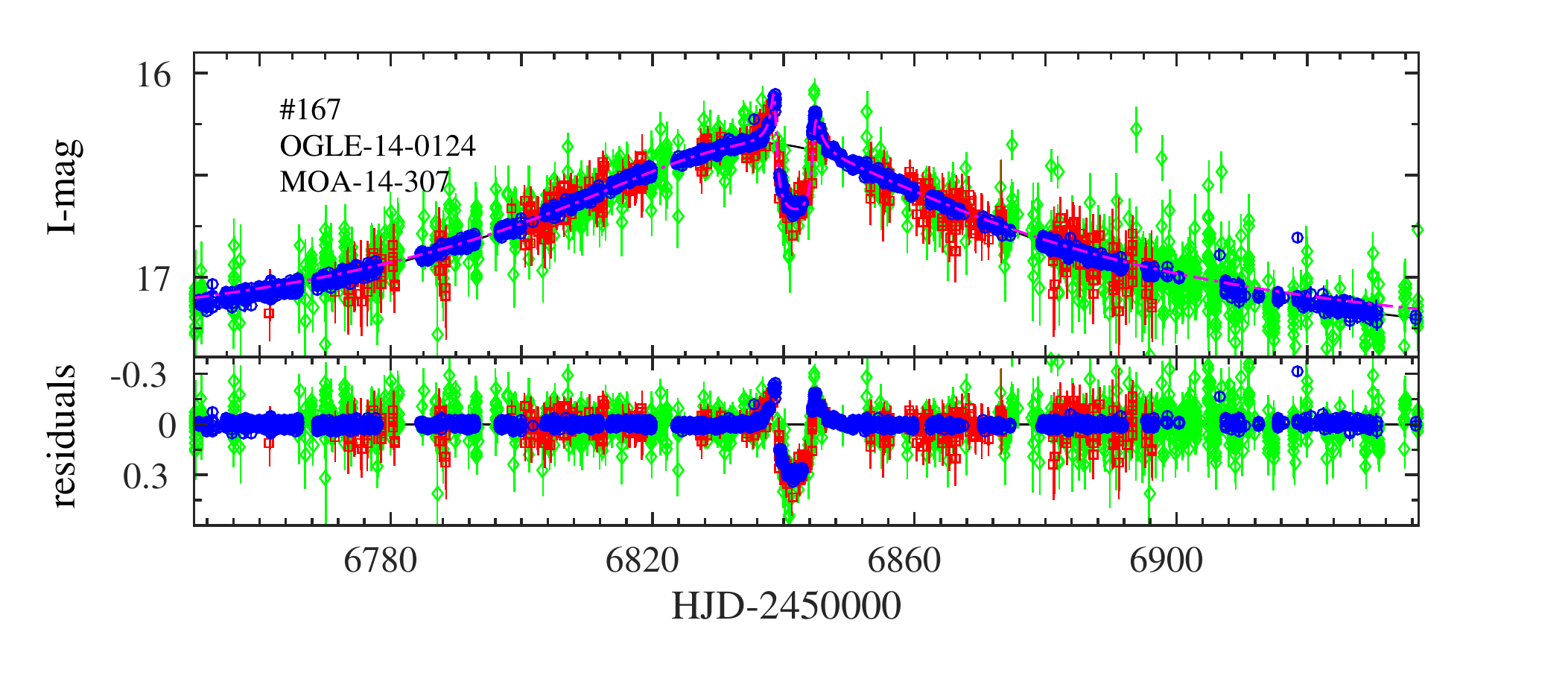}}\\
\vspace{-1.2cm}
\subfloat{\includegraphics[width=0.8\textwidth]{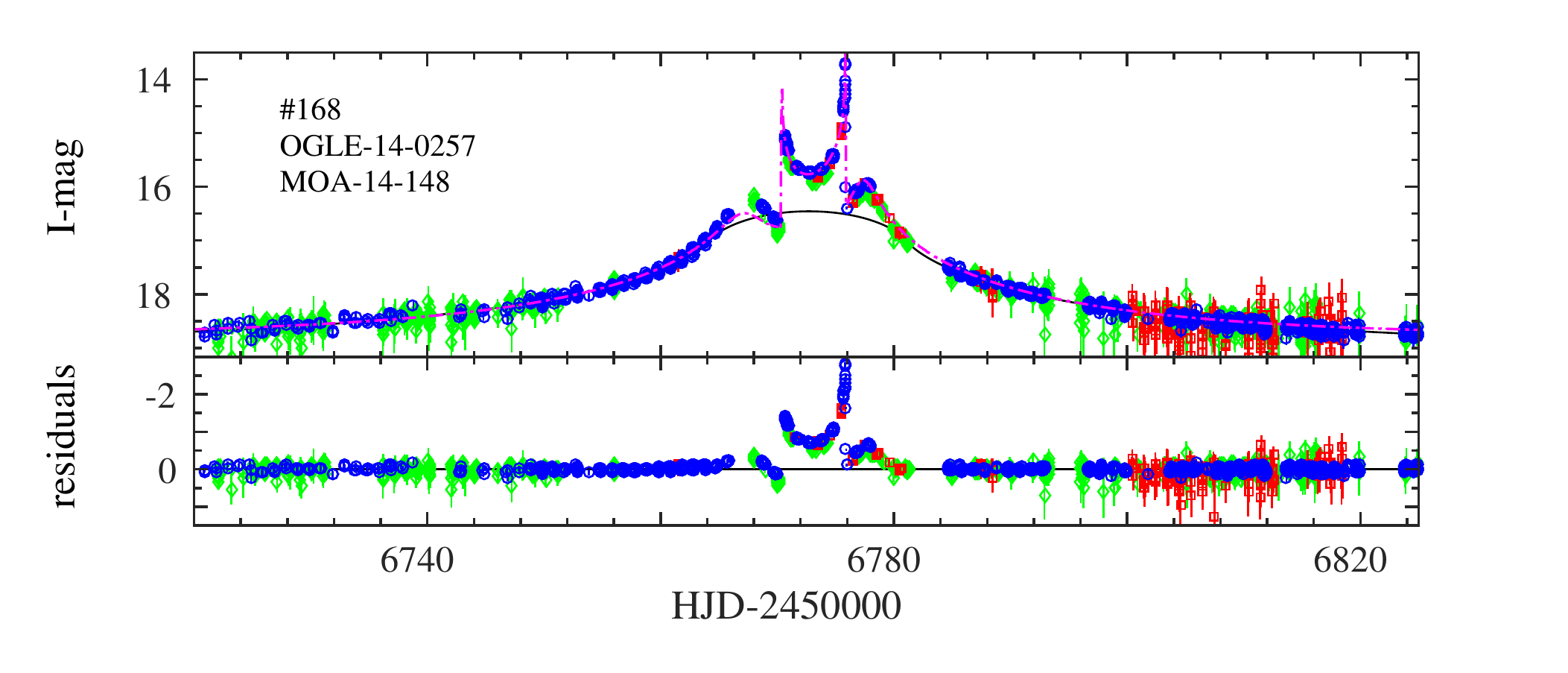}}\\
\vspace{-1.2cm}
\subfloat{\includegraphics[width=0.8\textwidth]{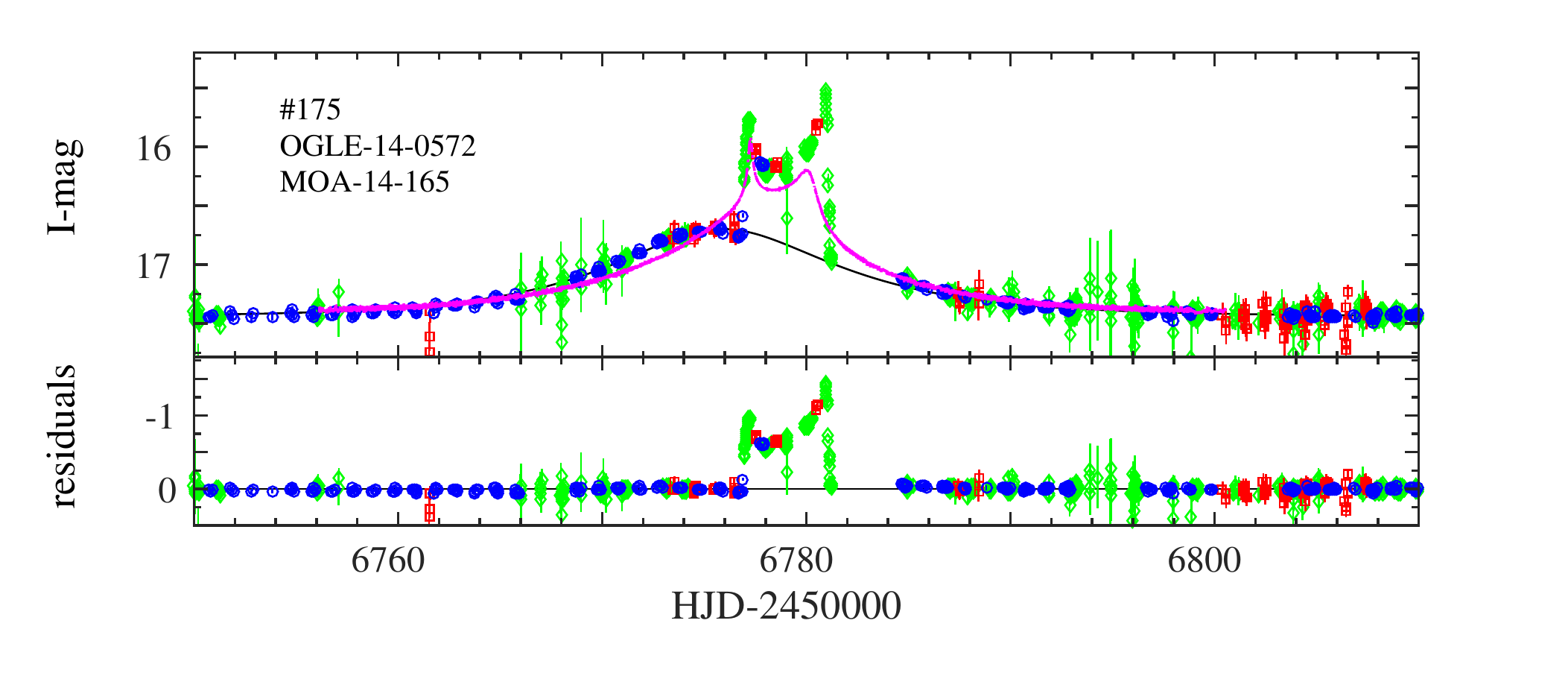}}\\
\vspace{-0.3cm}
\subfloat{\includegraphics[width=0.8\textwidth]{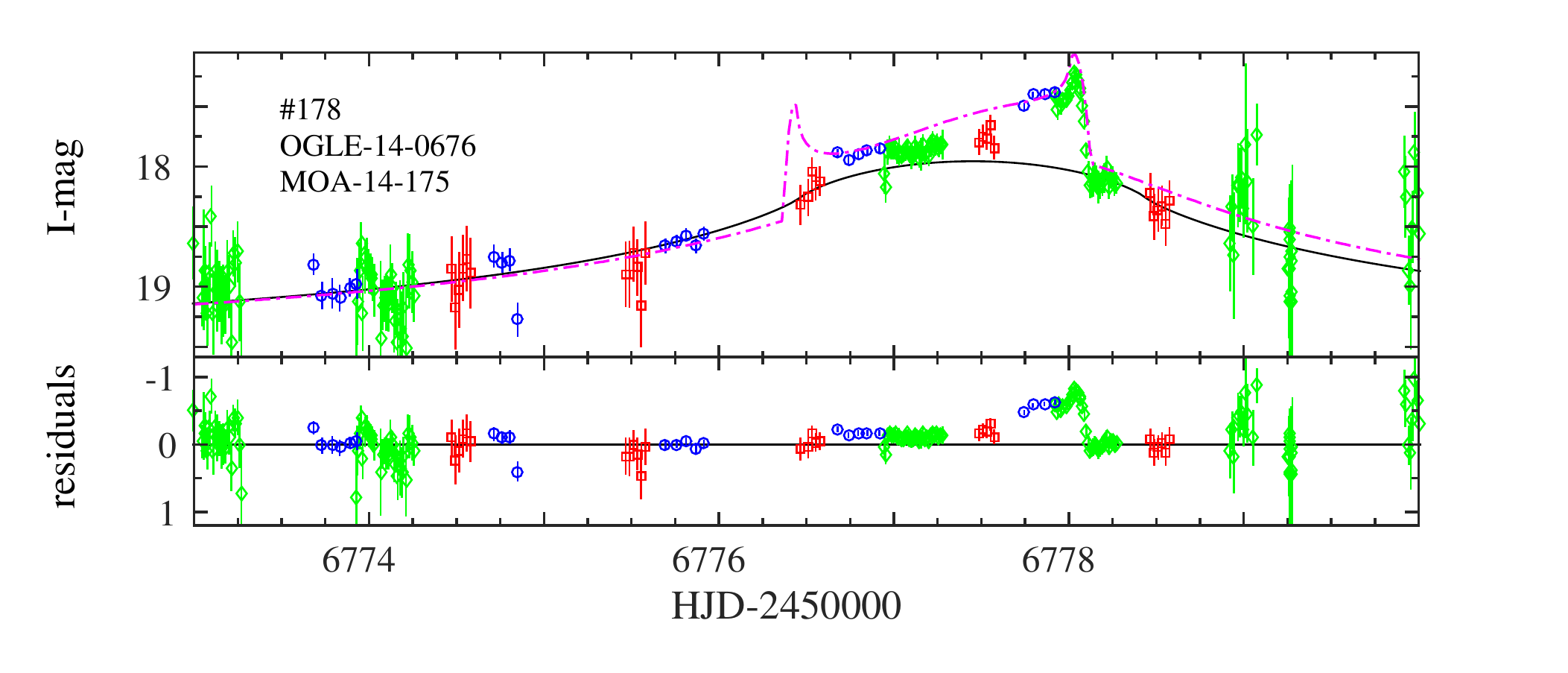}}\\
\end{tabular}
\caption{Continued from previous page.\label{fig:heuristic3}}
\end{minipage}
\end{figure*}

\begin{figure*}
\begin{minipage}{\textwidth}
\centering
\begin{tabular}{c}
\vspace{-1.2cm}
\subfloat{\includegraphics[width=0.8\textwidth]{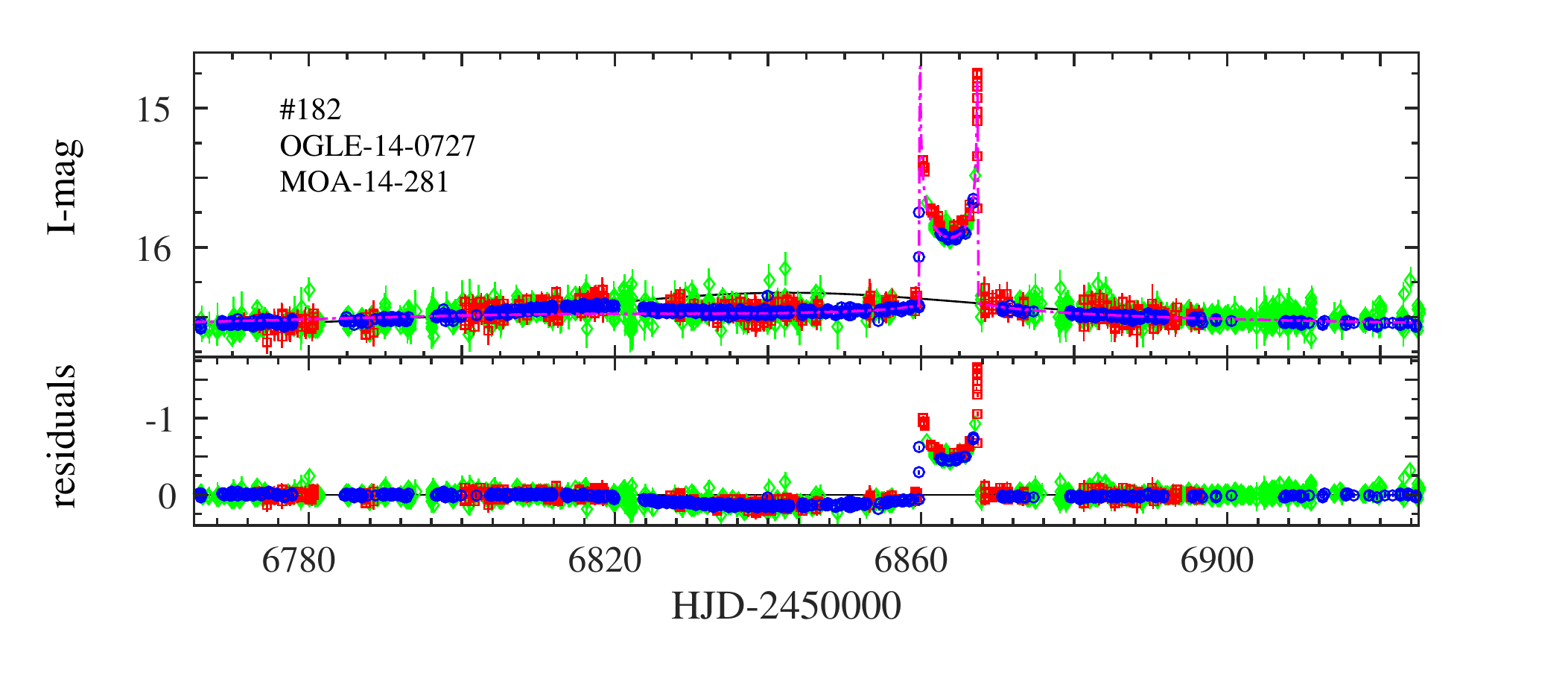}}\\
\vspace{-1.2cm}
\subfloat{\includegraphics[width=0.8\textwidth]{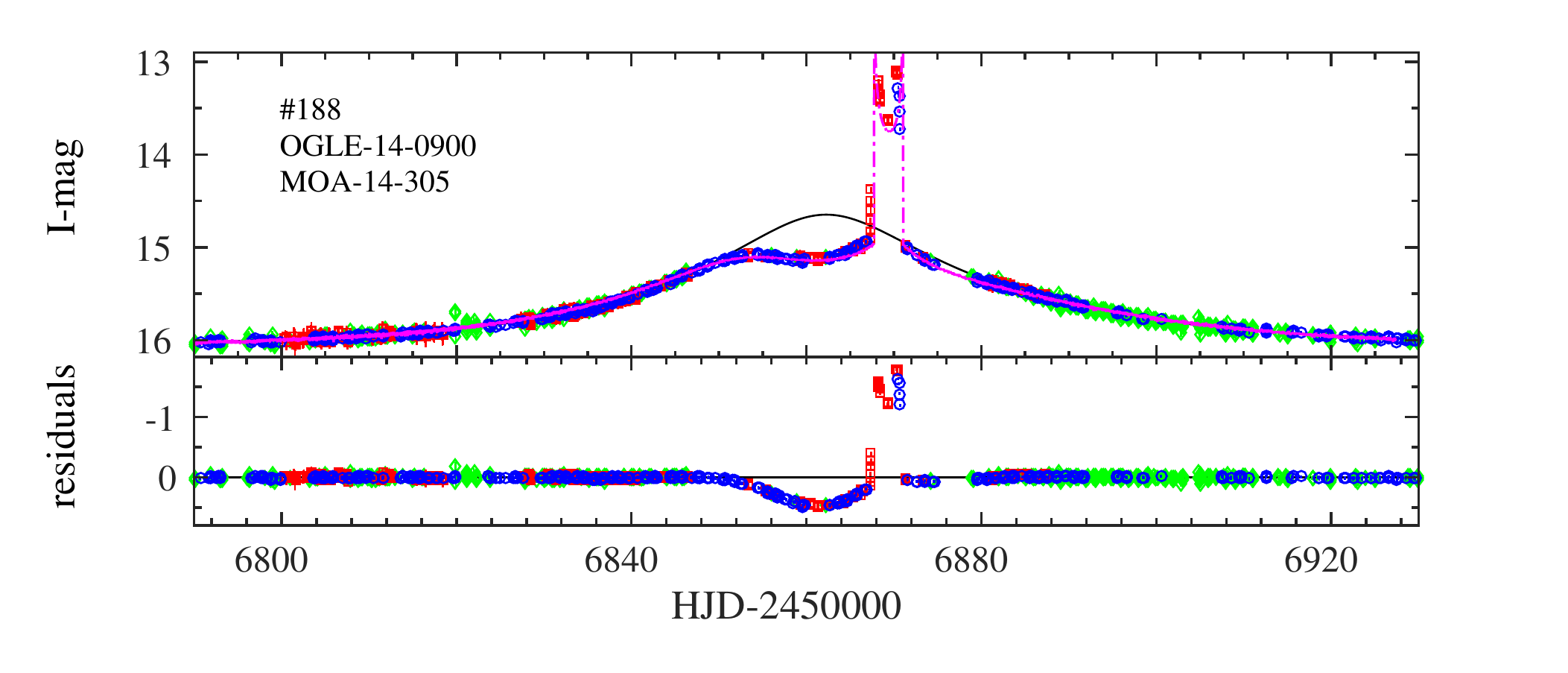}}\\
\vspace{-1.2cm}
\subfloat{\includegraphics[width=0.8\textwidth]{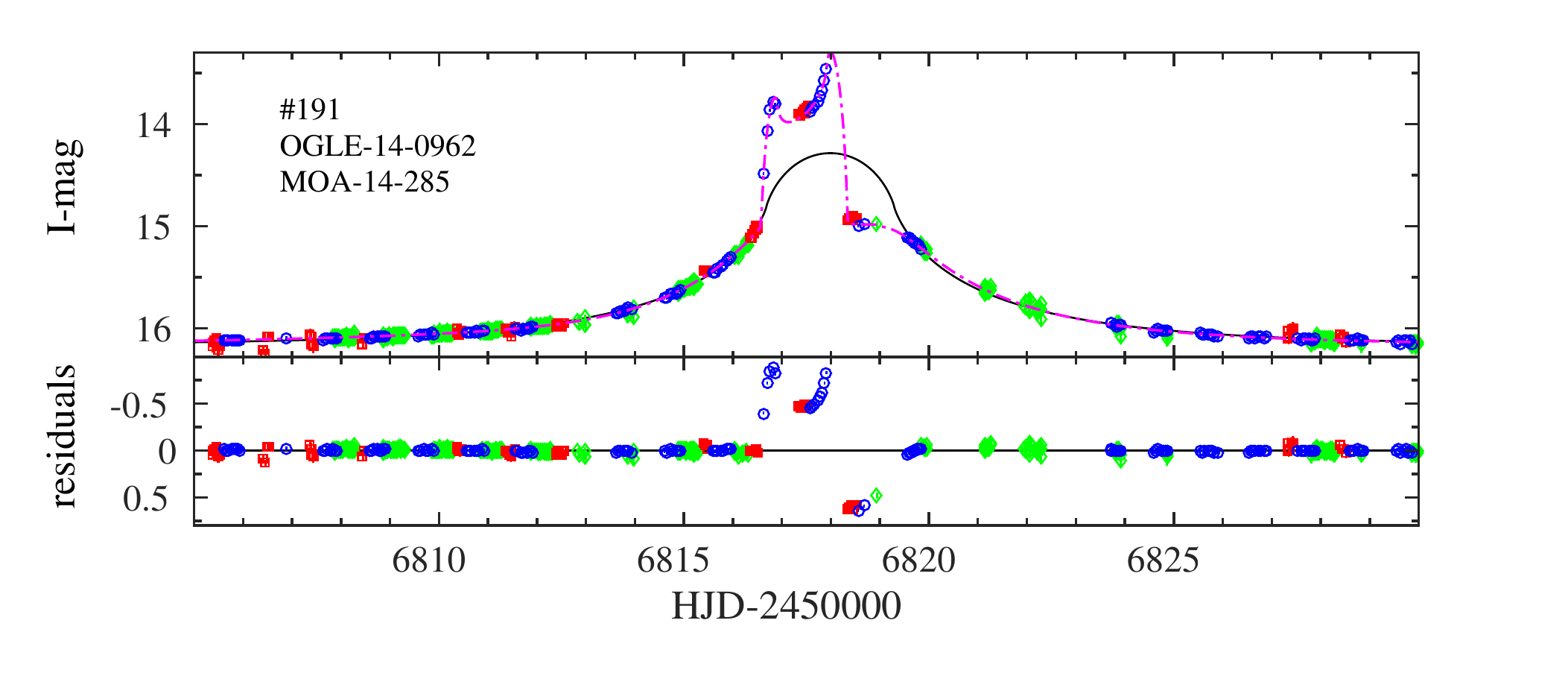}}\\
\vspace{-0.3cm}
\subfloat{\includegraphics[width=0.8\textwidth]{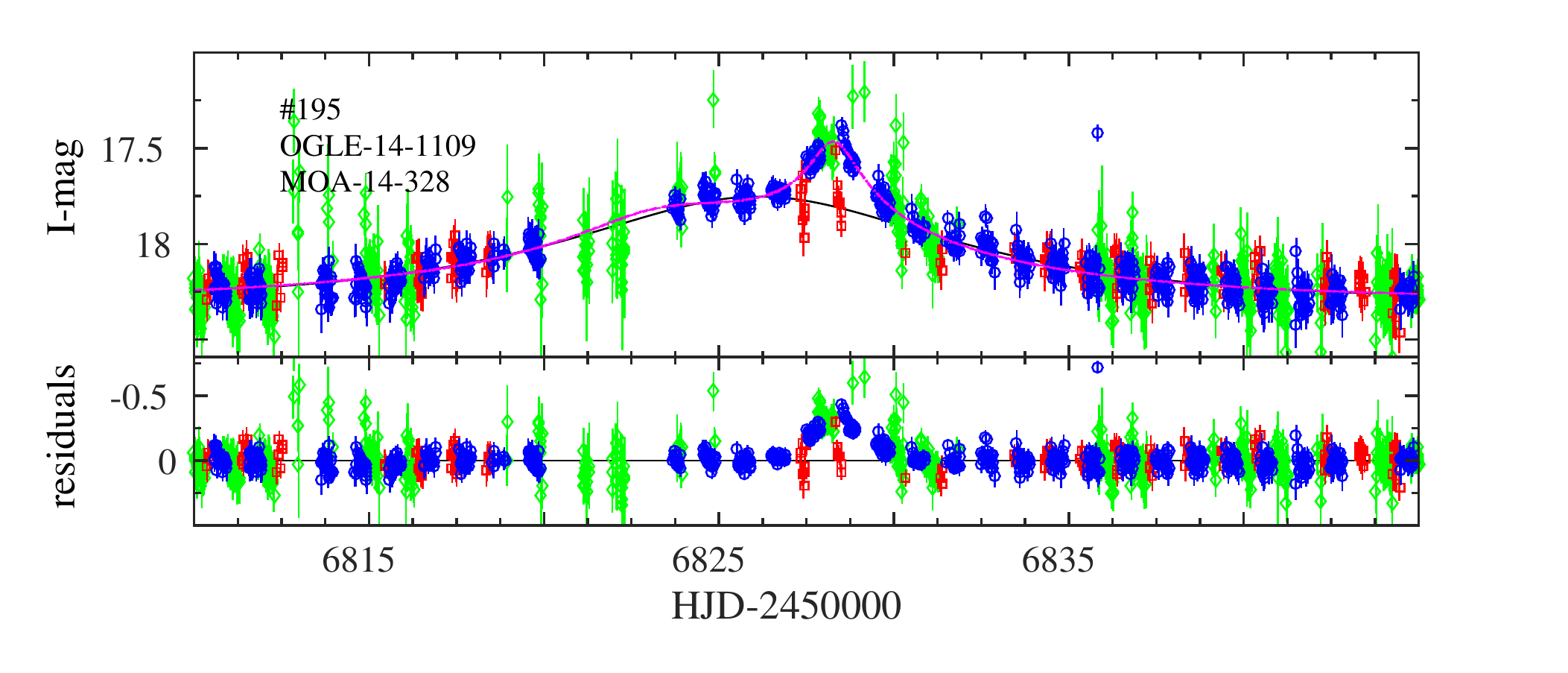}}\\
\end{tabular}
\caption{Continued from previous page.\label{fig:heuristic3}}
\end{minipage}
\end{figure*}

\begin{figure*}
\begin{minipage}{\textwidth}
\centering
\begin{tabular}{c}
\vspace{-1.2cm}
\subfloat{\includegraphics[width=0.8\textwidth]{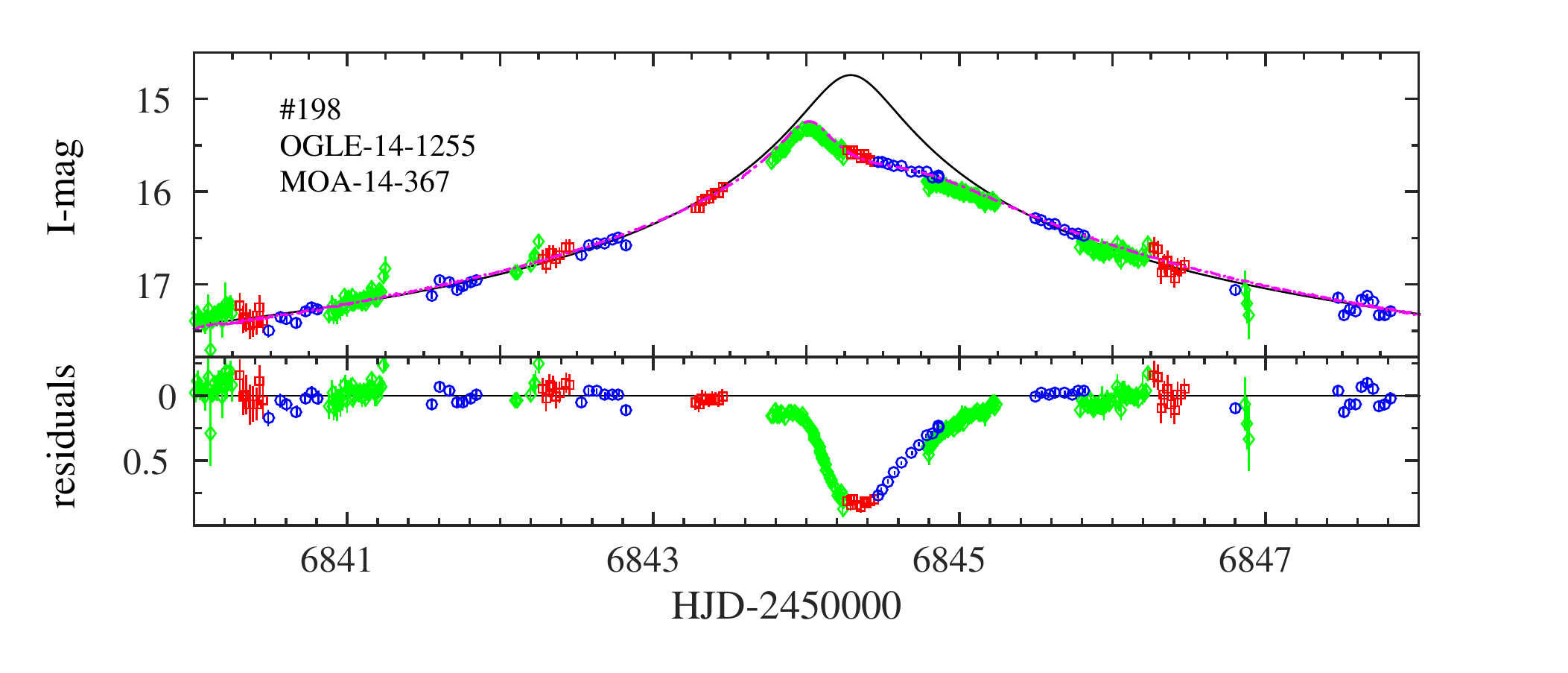}}\\
\vspace{-0.3cm}
\subfloat{\includegraphics[width=0.8\textwidth]{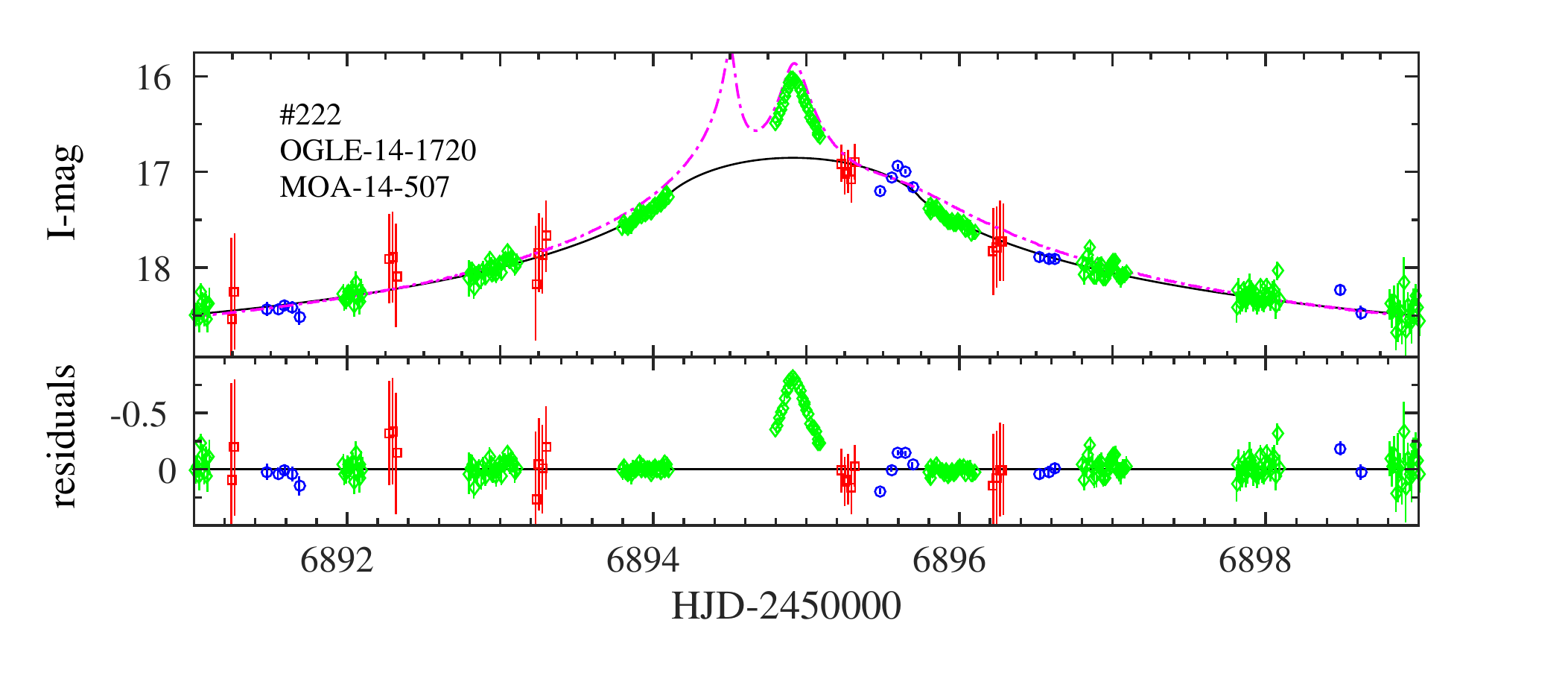}}\\
\end{tabular}
\caption{Continued from previous page.\label{fig:heuristic_last}}
\end{minipage}
\end{figure*}

\subsubsection{Objective detections}
\label{sec:Systematic}

\begin{figure*}
\begin{minipage}{\textwidth}
\centering
\begin{tabular}{c}
\vspace{-1.0cm}
\includegraphics[width=0.8\textwidth]{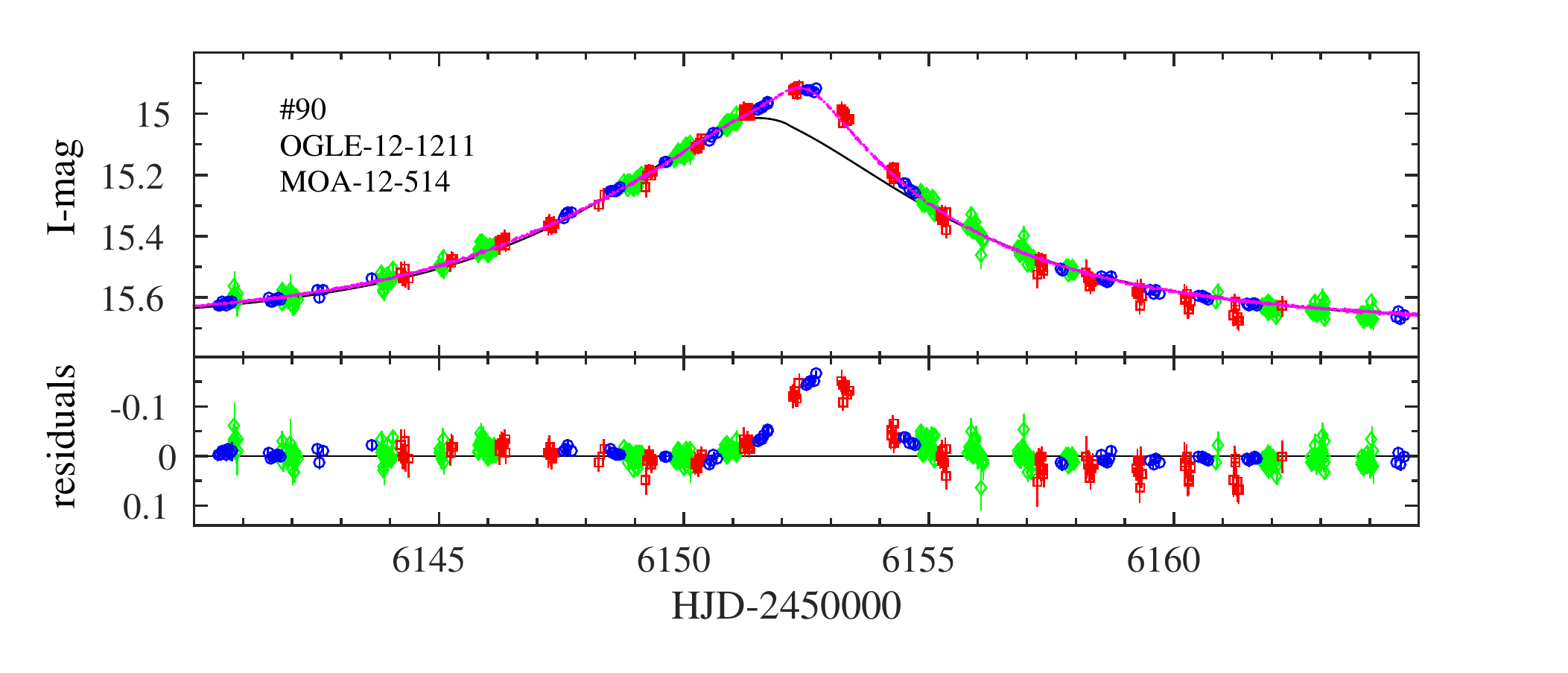}\\
\vspace{-1.0cm}
\includegraphics[width=0.8\textwidth]{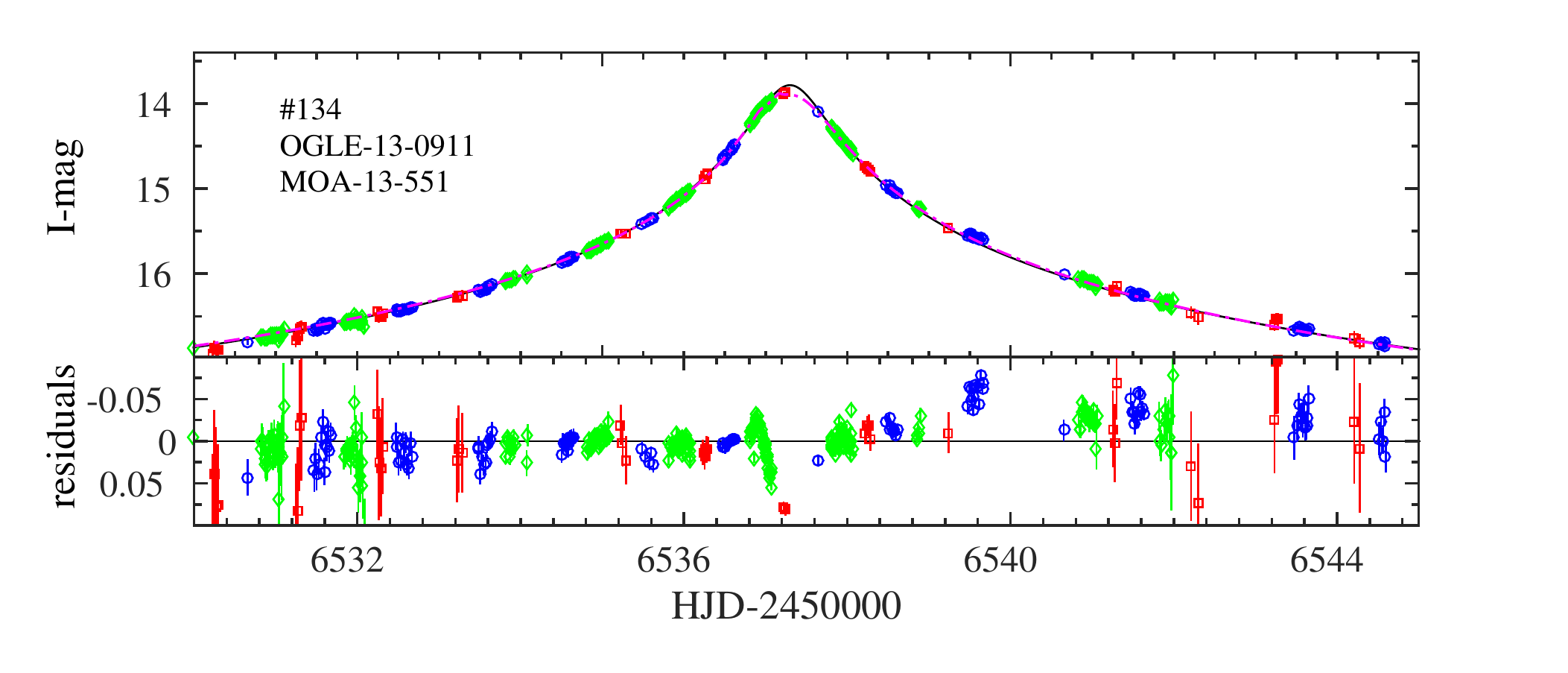}\\
\vspace{-0.3cm}
\includegraphics[width=0.8\textwidth]{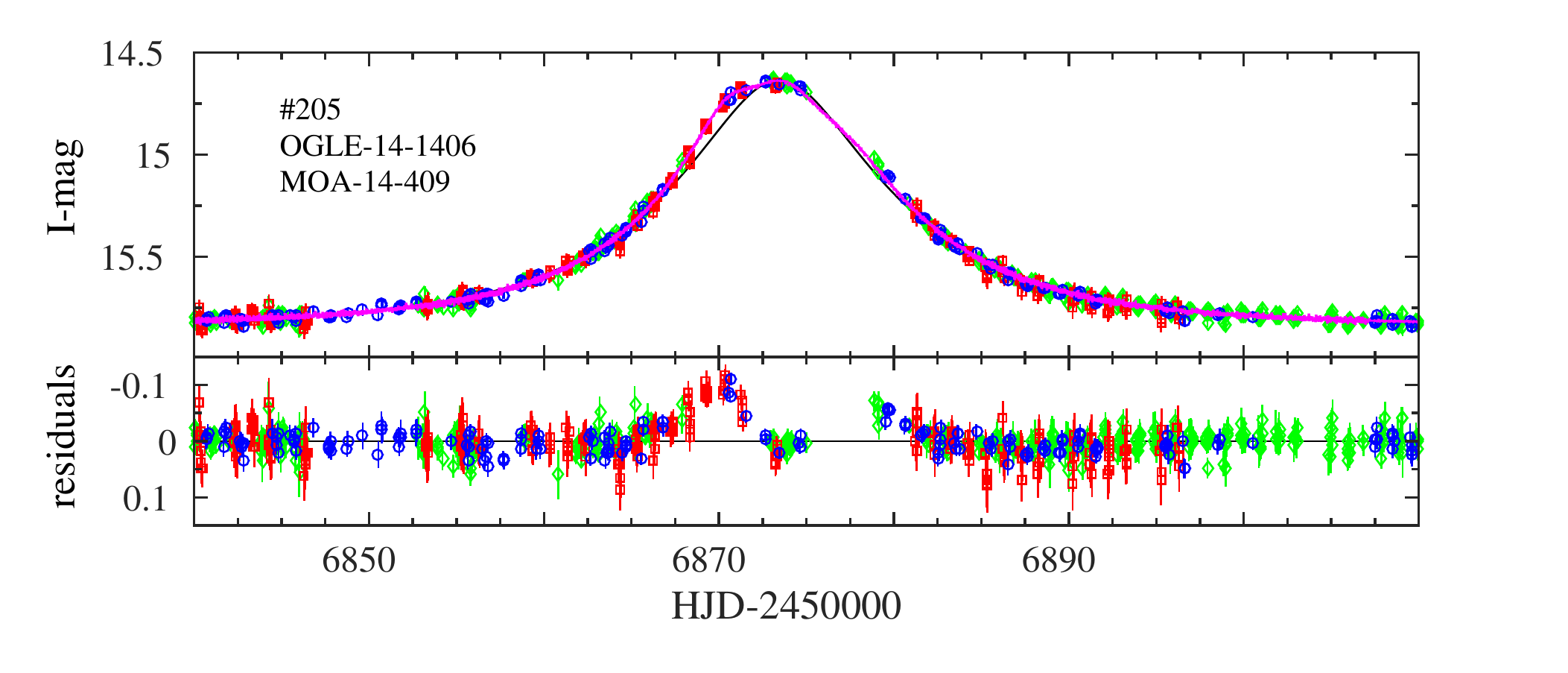}\\
\end{tabular}
\caption{Inter-calibrated light curves of three anomalous events detected by our detection filter,
in addition to those detected both by eye and by the filter, and shown in Figures \ref{fig:heuristic}-\ref{fig:heuristic_last}.
The residuals from the point-lens model are shown in the lower panel of each event, revealing the anomalous region.\label{fig:systematic}}
\end{minipage}
\end{figure*}

We next apply our automated detection filter to search for anomalies in each event in our sample, with the optimized false-positive filter parameters that we found.
The 26 anomalous events, identified by eye in Section \ref{sec:heuristic} are all easily re-detected by the automated filter.
In addition, the automated filter detects three more anomalous events.
The light curves of those events, and the residuals from their best-fit point-lens models, are shown in Figure \ref{fig:systematic}.

Table \ref{table:anomalous} summarises all anomalous events and their estimated mass ratios.
The mass ratios are either from the full binary-lens published models (eight events, see Section \ref{sec:heuristic})
or from a grid search of binary-lens models from a library we constructed, similar to the one used in Section \ref{sec:efficiency} for the detection efficiency estimates,
with 36,000 combination of $s$, $q$ and $\alpha$ (20, 50 and 36 grid points respectively).
We present the best-fit model, either published or from the grid search, as magenta lines in figures \ref{fig:heuristic}-\ref{fig:systematic}. 
From the $\chi^2$ confidence interval of the grid-based fits, we estimate a precision of $\sim0.2$ dex for the mass ratios.
A full binary-lens model will obviously set stronger constraints on the mass ratio in each system. However, for the purpose of our present statistical analysis, the above precision
is sufficient.

\begin{table*}
\begin{minipage}{\textwidth}
\center
\caption{Anomalous events.\newline The first 26 events display a clear anomaly and are detected both by eye and by the filter,
while the last three, separated by double horizontal lines, are found only by our detection filter.}
\begin{tabular}{|c|c|c|c|c|c|}
\hline
\hline
\# & OGLE No. & MOA No. & $q$ & mass  & model \\
   & 	      & 	 &     & from model & reference \\
\hline
8 & 11-0172 & 11-104 & $0.121\pm0.005$ & $0.02\pm0.01 M_\odot$ & \cite{Shin.2012.A} \\
10 & 11-0235 & 11-107 &$\sim0.77$ & & \\
13 & 11-0265 & 11-197 &  $(3.95\pm 0.06)\times 10^{-3}$ &  $0.6\pm0.2 M_{\rm{J}}$ &\cite{Skowron.2015.A} \\
27 & 11-0481 & 11-217 &  $\sim0.014$ & &\\
36 & 11-0974 & 11-275 &  $\sim0.24$& & \\ 
46 & 11-1127 & 11-322 &  $0.028\pm0.001$ & &\cite{Shvartzvald.2014.A}\\
47 & 11-1132 & 11-358 &  $\sim0.77$& & \\
54 & 11-9293 & 11-293 & $(5.3\pm0.2)\times 10^{-3}$ & & \cite{Yee.2012.A}\\
   &         &        &       & $4.8\pm0.3 M_{\rm{J}}$ &\cite{Batista.2014.A}\\
60 & 12-0442 & 12-245 &  $\sim0.093$& & \\
63 & 12-0456 & 12-189 &  $0.944\pm0.004$ & & \cite{Henderson.2014.A}\\
96 & 12-1442 & 12-532 &  $0.47\pm0.04$ & &\cite{Henderson.2014.A} \\
109 & 13-0341 & 13-260 &  $(4.8\pm0.3)\times 10^{-5}$ & $2.3\pm0.2 M_\oplus$&\cite{Gould.2014.A}\\
   &         &        &   $1.21\pm0.03$    & $0.17\pm0.01 M_\odot$ &\\
112 & 13-0506 & 13-494 &  $\sim0.50$& &\\
122 & 13-0668 & 13-343 &  $\sim0.61$& & \\
137 & 13-1033 & 13-472 &  $\sim0.4$& &\\
161 & 13-1721 & 13-618 &  $\sim1.1\times 10^{-3}$& &\\
167 & 14-0124 & 14-307 &  $(6.9\pm0.4)\times 10^{-4}$ & $0.5\pm0.1 M_{\rm{J}}$&\cite{Udalski.2015.A}\\
168 & 14-0257 & 14-148 &  $\sim0.24$& &\\
175 & 14-0572 & 14-165 &  $\sim0.15$& &\\
178 & 14-0676 & 14-175 &  $\sim1.4\times 10^{-3}$& &\\
182 & 14-0727 & 14-281 &  $\sim0.77$& &\\
188 & 14-0900 & 14-305 &  $\sim0.77$& &\\
191 & 14-0962 & 14-285 &  $\sim0.98$& &\\
195 & 14-1109 & 14-328 &  $\sim0.83$& &\\
198 & 14-1255 & 14-367 &  $\sim0.19$& &\\
222 & 14-1720 & 14-507 &  $\sim0.093$& &\\
\hline
\hline
90 & 12-1211 & 12-514 & $\sim0.19$& &\\
134 & 13-0911 & 13-551 & $\sim2.6\times 10^{-4}$& &\\
205 & 14-1406 & 14-409 & $\sim0.19$& &\\
\hline
\end{tabular}
\label{table:anomalous}
\end{minipage}
\end{table*}


\section{Planet occurrence frequency and mass-ratio distribution}
\label{sec:frequency}

The overall abundance of planets near the snowline ($0.5<s<2$) can be estimated by integrating over the frequencies of companions up to a certain mass ratio,
even without the need to fully constrain the absolute masses of the companions.
The population of lenses in our monitored genII field is dominated by old bulge stars.
The mass function of lenses toward the Galactic bulge, as modeled by \cite{Dominik.2006.A}, is a narrow distribution,
with the most probable lens mass of $\sim0.3M_\odot$.
If we define the planet mass limit at $13 M_J$ (the limiting mass for thermonuclear fusion of deuterium, assuming Solar metallicity, e.g. \citealt{Perryman.2011.A}),
then the corresponding planetary mass ratio limit
is $q=4.1\times10^{-2}$.
For the present analysis, we will therefore define planets as companions with $\log q<-1.4$.

\begin{figure}
\centering
\begin{tabular}{c}
\includegraphics[width=0.45\textwidth,angle=90]{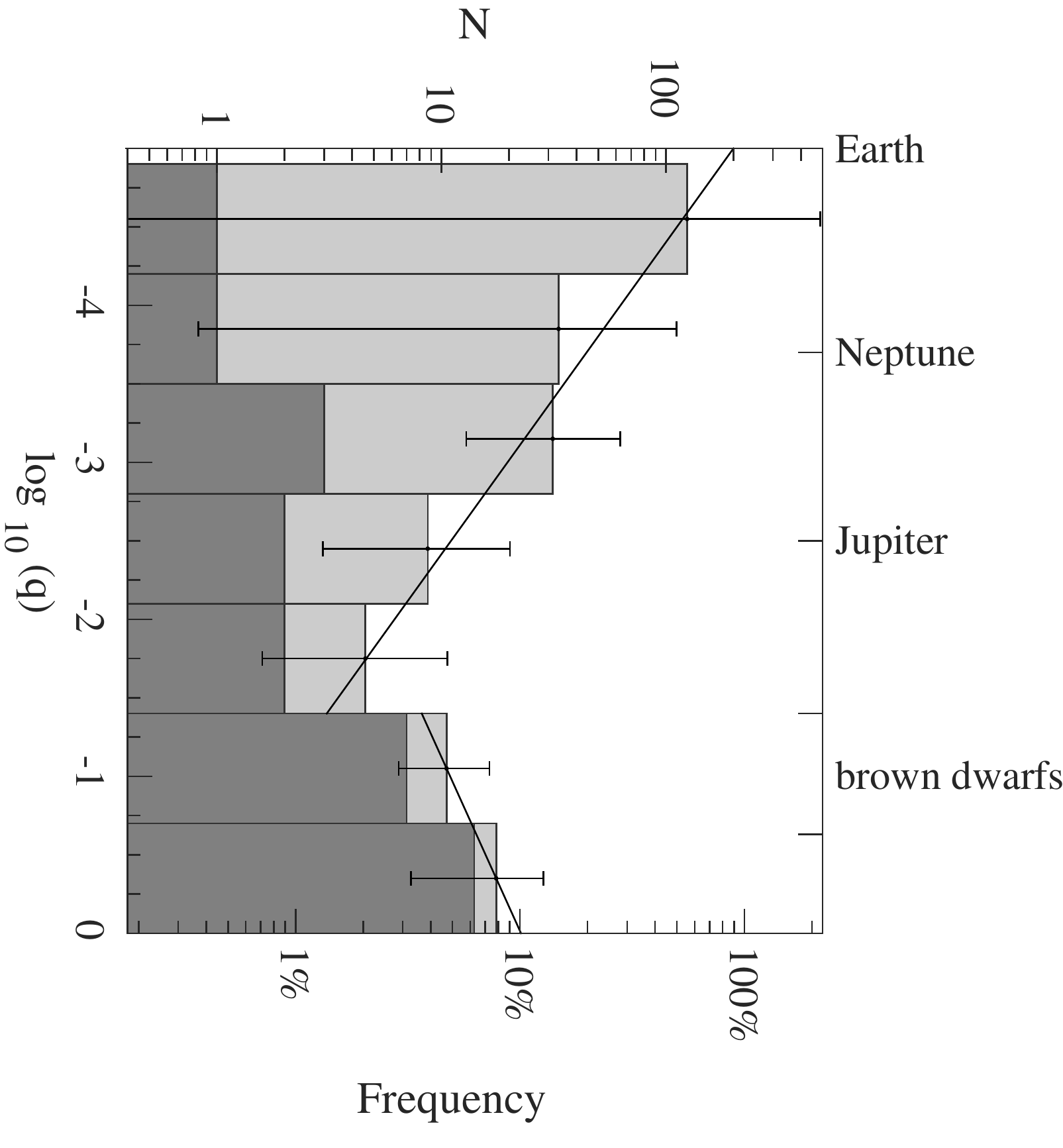}
\end{tabular}
\caption{The distribution of estimated mass ratios of our sample, before (dark gray) and after (light gray) correcting for detection efficiency.
The right-hand vertical axis shows the corresponding frequency.
The top horizontal axis indicates the masses of Earth, Neptune, Jupiter and the brown dwarf range, for a typical host mass of $0.3M_\odot$.
The distribution hints at
the existence of two populations, stellar binaries and planets, separated by a minimum that is
analogous to the brown dwarf desert found by radial velocity surveys, although here the division appears to be in the super-Jupiter range.
Solid lines are power laws \PLslope\ (in the planetary regime) and \BNslope\ (in the stellar binary regime).\label{fig:ratio}}
\end{figure}

The underlying distribution of mass ratios between companions and primary stars in the population of stars that produce microlensing events
can be inferred by correcting the observed distribution of our sample for the detection efficiencies that we derive from the simulations above.
The frequency of companions for a given range of mass ratios is simply the number of detected companions within that range, $N$,
divided by the sum of the detection efficiencies for that range, $\eta$, over all events in the sample,
\begin{equation}
f(q)=\frac{N(q)}{\sum\eta(q)}.
\end{equation}
Figure \ref{fig:ratio} shows the observed distribution of mass ratios for our sample in bins of 0.7 dex, and the recovered distribution after accounting for our detection
efficiency. The right-hand vertical axis shows the frequencies as a function of mass ratio.
We divide the planetary regime into two ranges of mass ratios, $10^{-2.8}<q<10^{-1.4}$ and $10^{-4.9}<q<10^{-2.8}$, which correspond
to Jupiters and Neptunes, respectively.
By choosing these ranges of 1.4 dex and 2.1 dex, we slightly reduce the Poisson uncertainties, though they still dominate over the detection efficiency uncertainties in each bin. 
The frequency of snowline Jupiters is
\begin{equation}
f(10^{-2.8}<q<10^{-1.4})=5.0^{+4.0}_{-2.4}\%,
\end{equation}
and the frequency of Neptunes is
\begin{equation}
f(10^{-4.9}<q<10^{-2.8})=50^{+34}_{-22}\%.
\end{equation}
Thus, the total frequency of snowline planets of these masses is
\begin{equation}
f(10^{-4.9}<q<10^{-1.4})=55^{+34}_{-22}\%.
\end{equation}
For the binary regime
we find that the frequency of brown dwarf companions is
\begin{equation}
f(10^{-1.4}<q<10^{-0.7})=4.7^{+2.6}_{-1.8}\%.
\end{equation}
and the frequency of stellar companions is
\begin{equation}
f(10^{-0.7}<q<1)=7.8^{+4.9}_{-4.6}\%.
\end{equation}

For a flat intrinsic distribution in log $q$, we expect to find a monotonically rising number of detected companions as a function of log $q$, simply because of the larger cross
section of the more massive planets, and the longer anomaly duration.
The recovered mass ratio distribution in Figure \ref{fig:ratio} shows a deficit around $q\backsimeq10^{-2}$,
and the corrected distribution can be fit with a broken---falling and rising---power law.
This result echoes previous findings, by radial velocity surveys, that have shown two distinct populations of stellar companions at orbital periods shorter than a few years---planets
and stellar binaries---likely produced via two different formation mechanisms (protoplanetary disks, and fragmentation in protostellar clouds, respectively).
In those surveys, the two populations are separated by a gap at $13-80M_J$, the ``brown-dwarf desert'' (e.g. \citealt{Grether.2006.A}).
(A recent analysis by \cite{Ranc.2015.A} compiling current knowledge of brown-dwarf companions from various discovery techniques suggests a more complex
brown-dwarf ``landscape'', with a possible period dependence of brown-dwarf occurrence frequency).
Our detection-efficiency-corrected mass-ratio distribution, assuming our typical primary has a mass of $0.3M_\odot$,
is suggestive of a similar picture, now possibly seen via microlensing.
However, the gap between the two distributions appears to occur at a lower mass, 
$\sim 3-13 M_J$, corresponding to ``super Jupiters''. This difference, if real, could be the result of the fact that microlensing probes the companions of
M stars (as opposed to FGK stars by other techniques), that microlensing probes larger separations than other techniques, or both.
Due to the small number of total events in our survey, we cannot claim a significant detection of two populations. 
Furthermore, our mass bins are large, and thus we cannot resolve the region between
the putative two populations.
Finally, full modeling of the anomalous events, together with
additional data (e.g. post-event detection of the lenses) that will help
break possible degeneracies in those models, are required in order to give
absolute masses rather than only mass ratios.
For example, if the hosts of some of the companions in the brown-dwarf bin are in reality more massive than $0.3M_\odot$, then
their companions will be low-mass M stars.
With all of these caveats in mind, the suggestion of a super-Jupiter gap in the M-star companion distribution is nevertheless intriguing.
The overall slope we find for the planetary regime ($10^{-4.9}<q<10^{-1.4}$) is \PLslope, and for the binary regime ($10^{-1.4}<q<1$) we find \BNslope.

\section{Summary and comparison to previous work}
\label{sec:Discussion_stat}

We have presented a statistical analysis of the results, to date, of the first genII microlensing survey.
This is the first controlled microlensing experiment for the abundance of snowline planets.
Among the 224 events in our sample, 29 are anomalous, revealing the presence of a companion to the lens star. 
Our results show that about half of the stars have snowline planets with mass ratios in the range $-4.9<\log (q)<-1.4$,
with Neptunes being $\sim10$ times more common than Jupiters, consistent with previous microlensing findings (e.g. \citealt{Sumi.2010.A}).

The absolute frequency we find is consistent with previous estimates from first generation microlensing surveys by \cite{Gould.2010.B},
who analyzed a sample of six detected planets from the follow-up of 13 high magnification events,
and found that the frequency for the range $-4.5<\log (q)<-2$ is $36\pm15\%$.
The frequency we find is somewhat lower than, but consistent with, the estimates by \cite{Cassan.2012.A}, who used a control sample of 196 events from the PLANET collaboration with three detected planets
(one Jupiter, one Neptune and one super-Earth), and derived a frequency of $17^{+6}_{-9}\%$ of Jupiters (in the range of $0.3-10M_J$) and $52^{+22}_{-29}\%$ Neptunes
(in the range of $10-30M_\oplus$). 
Our derived planet frequency is also consistent with results from other methods.
Radial velocity surveys find that 10.5\% of Sun-like stars host giant planets (in the range of $0.3-10M_J$) with periods $<5.5$ years (\citealt{Cumming.2008.A}).
Direct imaging surveys (\citealt{Brandt.2014.A}), which are currently sensitive only to massive planets ($>5M_J$),
find that 1.0\%-3.1\% of stars host substellar companions.

For the binary regime, we have estimated from our survey that $\sim8\%$ of lens stars have stellar companions with separations in the range $0.5<s<2$. This
is consistent with direct imaging estimates from the M-dwarfs in Multiples (MINMS) survey, of $8\pm2\%$ for companions to M stars 
with separations of 3-10 AU (\citealt{Ward-Duong.2015.A}),
and with radial-velocity surveys estimates of $\sim10\%$ for companions to FGK stars with periods $<5$ years
(\citealt{Grether.2006.A,Raghavan.2010.A}).
Although we cannot clearly identify the brown dwarf regime, we can estimate the frequency of brown-dwarf companions at $\sim5\%$.
This is significantly higher than the estimates based on radial-velocity surveys of FGK stars, of $<1\%$, by \cite{Grether.2006.A}
for companions with periods $<5$ years.
Similarly, the frequency of brown-dwarf companions to white dwarfs found by \cite{Steele.2011.A} is also low, $f_{\rm WD+BD}=0.5\%\pm0.3\%$.
Typical white dwarfs, with masses of $0.6 M_\odot$, are descended from $\sim 2 M_\odot$ stars.
It is presently unclear whether brown-dwarf companions become more common with increasing separation.
For example, \cite{Metchev.2009.A} estimate a frequency of $3.2^{+3.1}_{-2.7}\%$ of brown-dwarf companions in 28--1590 AU orbits around Sun-like stars.
This fraction is consistent with all the various brown-dwarf companion estimates above.

The companion mass ratio distribution in our sample shows a decline from companions with similar masses ($q\backsimeq1$) towards companions with $q\backsimeq10^{-2}$.
After accounting for the detection efficiency for different mass ratios, the distribution is suggestive of a rise in the planetary regime from massive planets towards less massive planets.
As discussed above, the minimum in the distribution may be the microlensing manifestation of the bimodal companion distribution found by other techniques,
but with the minimum shifted from the brown-dwarf mass range to the super-Jupiter range.

The slope we find for the planetary regime is consistent with what has been found for planetary companions to FGK stars from radial velocity surveys.
For example, \cite{Cumming.2008.A} find a a logarithmic slope of $d(\log f)/d(\log M)=-0.31\pm0.2$
for companions in the mass range 0.3--10 $M_J$ and with orbital periods of 2--2000 days.
It is also consistent with direct-imaging exoplanet survey results, e.g., \cite{Brandt.2014.A}, who find a slope of $d(\log f)/d(\log M)=-0.65\pm0.60$
for the distribution of substellar companions with masses of 5--70$M_J$ between 10 and 100 AU.
For the binary regime our slope
is shalower but, again, consistent with the $d(\log f)/d(\log M)=0.68\pm0.21$ found by radial velocity surveys (\citealt{Grether.2006.A})
for companions with masses of $0.1<(M/M_\odot)<1$,
and with the $d(\log f)/d(\log M)\sim0.5$ shown by \cite{Mazeh.2003.A} for the companion distribution in that mass range.

Full binary-lens modeling of the anomalous events in our sample is required in order to confirm these estimates.
For some of the events, the light curves show the signatures of high order effects, and thus the mass of the host star and its companion can be measured.
For the rest of the sample, high resolution imaging in the coming years could refine estimates of the masses, by isolating the light from the lens.
Together with increasing numbers of planetary detections from our ongoing genII survey, this will give a progressively clearer picture of the planetary mass distribution near the
snowlines of low-mass stars in the Galaxy.


\section*{Acknowledgments}
We thank T. Mazeh and the anonymous referee for useful comments and discussions.
This research was supported by the I-CORE program of the Planning and Budgeting Committee and the Israel Science Foundation, Grant 1829/12.
DM acknowledges support by the US-Israel Binational Science Foundation.
The OGLE project has received funding from the National Science Centre,
Poland, grant MAESTRO 2014/14/A/ST9/00121 to AU.
TS acknowledges the financial support from the JSPS, JSPS23103002,JSPS24253004 and JSPS26247023.
The MOA project is supported by the grant JSPS25103508 and 23340064.



\newpage
\onecolumn
\section*{Appendix A}

Microlens event sample: best-fit point-lens model parameters.  Significant detections of finite source effects and parallax are marked in bold.\\
At the end of the table, separated by double horizontal lines, we list the six anomalous events which
passed the limiting magnitude criteria only because of the anomaly, and are therefore excluded from the sample (see Section~\ref{sec:sample_stat} for details).

\scriptsize
\begin{longtable}{|c|c|c|c|c|c|c|c|c|c|c|c|}
\caption[Event summary]{Event summary}\label{table:data} \\

\hline \multicolumn{1}{|l|}{\textbf{\#}} & \multicolumn{1}{c|}{\textbf{OGLE}} & \multicolumn{1}{c|}{\textbf{MOA}} &
\multicolumn{1}{|c|}{\textbf{$u_0$}} & \multicolumn{1}{c|}{\textbf{$t_0-2450000$}} & \multicolumn{1}{c|}{\textbf{$t_{\rm E}$}} &
\multicolumn{1}{|c|}{\textbf{$\rho$}} & \multicolumn{1}{c|}{\textbf{$\pi_{E,E}$}} & \multicolumn{1}{c|}{\textbf{$\pi_{E,N}$}} &
\multicolumn{1}{|c|}{\textbf{$I_{\rm bl}$}} & \multicolumn{1}{|c|}{\textbf{$f_{\rm bl}$}} \\
\multicolumn{1}{|l|}{\textbf{}} & \multicolumn{1}{c|}{\textbf{no.}} & \multicolumn{1}{c|}{\textbf{no.}} &
\multicolumn{1}{|c|}{\textbf{}} & \multicolumn{1}{c|}{\textbf{[HJD]}} & \multicolumn{1}{c|}{\textbf{[Days]}} &
\multicolumn{1}{|c|}{\textbf{$10^{-3}$}} & \multicolumn{1}{c|}{\textbf{}} & \multicolumn{1}{c|}{\textbf{}} &
\multicolumn{1}{|c|}{\textbf{}} & \multicolumn{1}{|c|}{\textbf{}} \\ \hline 
\endfirsthead

\multicolumn{11}{c}%
{{\bfseries \tablename\ \thetable{} -- continued from previous page}} \\
\hline \multicolumn{1}{|l|}{\textbf{\#}} & \multicolumn{1}{c|}{\textbf{OGLE}} & \multicolumn{1}{c|}{\textbf{MOA}} &
\multicolumn{1}{|c|}{\textbf{$u_0$}} & \multicolumn{1}{c|}{\textbf{$t_0-2450000$}} & \multicolumn{1}{c|}{\textbf{$t_{\rm E}$}} &
\multicolumn{1}{|c|}{\textbf{$\rho$}} & \multicolumn{1}{c|}{\textbf{$\pi_{E,E}$}} & \multicolumn{1}{c|}{\textbf{$\pi_{E,N}$}} &
\multicolumn{1}{|c|}{\textbf{$I_{\rm bl}$}} & \multicolumn{1}{|c|}{\textbf{$f_{\rm bl}$}} \\
\multicolumn{1}{|l|}{\textbf{}} & \multicolumn{1}{c|}{\textbf{no.}} & \multicolumn{1}{c|}{\textbf{no.}} &
\multicolumn{1}{|c|}{\textbf{}} & \multicolumn{1}{c|}{\textbf{[HJD]}} & \multicolumn{1}{c|}{\textbf{[Days]}} &
\multicolumn{1}{|c|}{\textbf{$10^{-3}$}} & \multicolumn{1}{c|}{\textbf{}} & \multicolumn{1}{c|}{\textbf{}} &
\multicolumn{1}{|c|}{\textbf{}} & \multicolumn{1}{|c|}{\textbf{}} \\ \hline 
\endhead

\hline \hline
\endfoot

\hline \hline
\endlastfoot
1 & 11-0022 & 11-025 & 0.867 & 5690.508 & 56.584 & 0.000 & \bf{-0.379} & \bf{0.021} & 15.614 & 0.000 \\
  &  &  & (0.002) & (0.041) & (0.150) & (38.685) & (\bf{0.003}) & (\bf{0.007}) & (0.000) & (0.000) \\
2 & 11-0037 & 11-039 & 0.173 & 5807.481 & 128.572 & \bf{2.284} & \bf{0.263} & \bf{-0.243} & 16.155 & 0.245 \\
  &  &  & (0.001) & (0.025) & (0.565) & (\bf{14.211}) & (\bf{0.001}) & (\bf{0.002}) & (0.000) & (0.001) \\
3 & 11-0081 & 11-064 & 0.207 & 5658.523 & 106.779 & \bf{223.877} & \bf{0.437} & \bf{0.457} & 16.460 & 0.952 \\
  &  &  & (0.003) & (0.128) & (1.568) & (\bf{3.383}) & (\bf{0.014}) & (\bf{0.011}) & (0.000) & (0.000) \\
4 & 11-0082 & 11-103 & 0.295 & 5702.214 & 63.944 & 169.666 & -0.205 & -0.464 & 18.191 & 0.057 \\
  &  &  & (0.009) & (0.037) & (1.186) & (49.256) & (0.030) & (0.058) & (0.001) & (0.002) \\
5 & 11-0120 & 11-114 & 0.846 & 5670.592 & 26.837 & 0.000 & 0.015 & 1.496 & 16.964 & 0.003 \\
  &  &  & (0.031) & (0.109) & (0.698) & (71.112) & (0.040) & (0.044) & (0.001) & (0.004) \\
6 & 11-0138 & 11-082 & 0.042 & 5666.781 & 52.366 & 6.039 & 0.206 & -0.203 & 19.672 & 0.251 \\
  &  &  & (0.001) & (0.008) & (0.920) & (5.511) & (0.027) & (0.143) & (0.002) & (0.002) \\
7 & 11-0168 & 11-091 & 0.351 & 5690.133 & 24.810 & 0.000 & 0.158 & 1.498 & 16.550 & 0.215 \\
  &  &  & (0.004) & (0.014) & (0.160) & (45.810) & (0.025) & (0.104) & (0.000) & (0.001) \\
8 & 11-0172 & 11-104 & -0.050 & 5670.367 & 49.263 & 51.767 & -0.046 & -1.493 & 19.497 & 0.113 \\
  &  &  & (0.001) & (0.010) & (0.918) & (1.179) & (0.030) & (0.068) & (0.003) & (0.003) \\
9 & 11-0173 & 11-133 & -0.696 & 5689.328 & 30.654 & 0.000 & \bf{-0.160} & \bf{-1.315} & 15.828 & 0.438 \\
  &  &  & (0.052) & (0.059) & (1.142) & (147.475) & (\bf{0.029}) & (\bf{0.201}) & (0.000) & (0.001) \\
10 & 11-0235 & 11-107 & -0.669 & 5673.096 & 14.184 & 0.000 & 0.220 & 1.208 & 18.155 & 0.000 \\
  &  &  & (0.050) & (0.172) & (0.664) & (143.850) & (0.418) & (0.811) & (0.002) & (0.016) \\
11 & 11-0240 & 11-109 & 0.035 & 5659.760 & 9.381 & 44.823 & 1.424 & -1.084 & 19.000 & 0.939 \\
  &  &  & (0.018) & (0.023) & (1.604) & (24.768) & (0.802) & (0.793) & (0.004) & (0.001) \\
12 & 11-0243 & 11-151 & 0.700 & 5714.091 & 26.304 & 0.000 & 0.461 & 1.499 & 16.707 & 0.000 \\
  &  &  & (0.011) & (0.057) & (0.243) & (98.704) & (0.068) & (0.068) & (0.000) & (0.000) \\
13 & 11-0265 & 11-197 & -0.136 & 5760.259 & 51.582 & 0.000 & 0.061 & 0.628 & 17.287 & 0.000 \\
  &  &  & (0.000) & (0.009) & (0.152) & (10.814) & (0.017) & (0.110) & (0.000) & (0.000) \\
14 & 11-0266 & 11-115 & 0.664 & 5694.409 & 14.482 & 0.000 & 0.489 & 0.858 & 15.559 & 0.000 \\
  &  &  & (0.007) & (0.012) & (0.117) & (71.339) & (0.088) & (0.373) & (0.000) & (0.001) \\
15 & 11-0286 & 11-124 & 0.109 & 5680.101 & 12.338 & 76.999 & 1.494 & -0.978 & 18.636 & 0.620 \\
  &  &  & (0.032) & (0.012) & (0.496) & (28.270) & (0.340) & (0.788) & (0.003) & (0.002) \\
16 & 11-0293 & 11-190 & -0.476 & 5714.295 & 32.685 & 414.080 & -0.037 & 1.483 & 18.501 & 0.033 \\
  &  &  & (0.026) & (0.052) & (1.163) & (119.996) & (0.151) & (0.941) & (0.002) & (0.004) \\
17 & 11-0297 & 11-166 & 0.015 & 5700.670 & 136.964 & 19.059 & -0.605 & -1.446 & 18.794 & 0.907 \\
  &  &  & (0.002) & (0.021) & (14.527) & (4.812) & (0.057) & (0.096) & (0.004) & (0.001) \\
18 & 11-0318 & 11-132 & 0.503 & 5682.097 & 1.055 & 356.525 & 1.312 & -1.171 & 16.002 & 0.006 \\
  &  &  & (0.106) & (0.002) & (0.023) & (131.754) & (0.804) & (0.817) & (0.001) & (0.003) \\
19 & 11-0329 & 11-147 & -0.005 & 5691.899 & 11.928 & 5.400 & -0.663 & -0.720 & 19.194 & 0.255 \\
  &  &  & (0.000) & (0.000) & (0.231) & (0.297) & (0.541) & (0.797) & (0.004) & (0.003) \\
20 & 11-0348 & 11-189 & 0.443 & 5713.657 & 19.410 & 85.249 & 0.674 & 1.484 & 18.349 & 0.073 \\
  &  &  & (0.038) & (0.041) & (0.423) & (111.233) & (0.446) & (0.828) & (0.001) & (0.004) \\
21 & 11-0364 & 11-154 & 0.019 & 5703.622 & 74.999 & 17.449 & \bf{-0.798} & \bf{-0.242} & 16.101 & 0.986 \\
  &  &  & (0.006) & (0.009) & (6.046) & (5.385) & (\bf{0.068}) & (\bf{0.061}) & (0.000) & (0.000) \\
22 & 11-0421 & 11-160 & 0.050 & 5705.967 & 7.771 & 0.583 & -1.490 & 1.465 & 16.247 & 0.940 \\
  &  &  & (0.003) & (0.004) & (0.262) & (12.883) & (0.802) & (0.813) & (0.000) & (0.000) \\
23 & 11-0422 & 11-171 & 0.207 & 5709.443 & 17.937 & 218.483 & 1.490 & -1.496 & 18.477 & 0.537 \\
  &  &  & (0.013) & (0.026) & (0.848) & (14.174) & (0.099) & (0.336) & (0.002) & (0.002) \\
24 & 11-0424 & 11-204 & 0.314 & 5732.063 & 16.901 & 193.531 & -1.480 & 0.330 & 16.299 & 0.124 \\
  &  &  & (0.053) & (0.011) & (0.125) & (73.532) & (0.270) & (0.300) & (0.001) & (0.002) \\
25 & 11-0462 & 11-191 & 0.004 & 5763.322 & 179.829 & 2.950 & \bf{-0.068} & \bf{0.098} & 16.406 & 0.935 \\
  &  &  & (0.000) & (0.000) & (0.618) & (0.143) & (\bf{0.004}) & (\bf{0.012}) & (0.000) & (0.000) \\
26 & 11-0465 & 11-211 & 0.562 & 5781.460 & 35.139 & 109.840 & \bf{-0.127} & \bf{0.582} & 14.944 & 0.000 \\
  &  &  & (0.005) & (0.016) & (0.158) & (68.034) & (\bf{0.016}) & (\bf{0.054}) & (0.000) & (0.001) \\
27 & 11-0481 & 11-217 & -0.018 & 5725.269 & 5.607 & 2.650 & 0.873 & 0.515 & 19.111 & 0.601 \\
  &  &  & (0.016) & (0.003) & (0.389) & (17.413) & (0.808) & (0.792) & (0.003) & (0.002) \\
28 & 11-0507 & 11-287 & 0.252 & 5750.747 & 35.440 & 0.000 & \bf{-0.974} & \bf{1.482} & 17.153 & 0.867 \\
  &  &  & (0.021) & (0.055) & (2.027) & (57.304) & (\bf{0.362}) & (\bf{0.670}) & (0.001) & (0.001) \\
29 & 11-0510 & 11-247 & 0.664 & 5741.451 & 8.633 & 270.825 & -1.433 & -1.496 & 16.884 & 0.000 \\
  &  &  & (0.013) & (0.010) & (0.101) & (104.230) & (0.541) & (0.568) & (0.000) & (0.000) \\
30 & 11-0519 & 11-273 & 0.257 & 5741.040 & 10.927 & 0.000 & 1.479 & 1.499 & 15.615 & 0.943 \\
  &  &  & (0.027) & (0.019) & (0.560) & (76.924) & (0.718) & (0.799) & (0.000) & (0.000) \\
31 & 11-0521 & 11-270 & 1.025 & 5746.899 & 7.426 & 38.738 & -1.498 & -1.491 & 14.231 & 0.001 \\
  &  &  & (0.033) & (0.016) & (0.173) & (52.255) & (0.106) & (0.178) & (0.000) & (0.002) \\
32 & 11-0535 & 11-354 & -0.083 & 5768.233 & 73.757 & 82.167 & -0.038 & 0.084 & 18.321 & 0.914 \\
  &  &  & (0.008) & (0.040) & (5.729) & (23.369) & (0.045) & (0.250) & (0.001) & (0.000) \\
33 & 11-0545 & 11-290 & 0.238 & 5751.104 & 17.559 & 0.000 & -0.205 & -1.447 & 19.099 & 0.001 \\
  &  &  & (0.064) & (0.045) & (0.360) & (60.034) & (0.564) & (0.795) & (0.004) & (0.008) \\
34 & 11-0944 & 11-297 & -0.046 & 5767.394 & 12.273 & 1.624 & 0.288 & -1.499 & 16.983 & 0.000 \\
  &  &  & (0.000) & (0.001) & (0.013) & (5.479) & (0.144) & (0.654) & (0.000) & (0.000) \\
35 & 11-0966 & 11-302 & 0.077 & 5758.121 & 16.523 & 17.351 & 0.128 & -1.427 & 19.114 & 0.137 \\
  &  &  & (0.020) & (0.016) & (0.347) & (21.583) & (0.419) & (0.836) & (0.002) & (0.004) \\
36 & 11-0974 & 11-275 & 0.004 & 5743.417 & 267.266 & 4.783 & -0.981 & -1.404 & 19.485 & 0.995 \\
  &  &  & (18.946) & (0.092) & (122.304) & (16117.873) & (0.802) & (0.795) & (0.455) & (0.006) \\
37 & 11-0990 & 11-300 & -0.012 & 5758.680 & 6.788 & 16.325 & -0.975 & -0.991 & 18.480 & 0.044 \\
  &  &  & (0.003) & (0.004) & (0.108) & (4.291) & (0.736) & (0.797) & (0.002) & (0.005) \\
38 & 11-0991 & 11-312 & 0.810 & 5801.829 & 25.457 & 192.278 & 0.320 & -1.500 & 16.018 & 0.174 \\
  &  &  & (0.033) & (0.049) & (0.595) & (116.572) & (0.026) & (0.083) & (0.000) & (0.001) \\
39 & 11-0999 & 11-306 & 0.109 & 5760.856 & 5.487 & 44.185 & 1.419 & 1.178 & 19.616 & 0.197 \\
  &  &  & (0.021) & (0.006) & (0.255) & (33.615) & (0.802) & (0.800) & (0.003) & (0.005) \\
40 & 11-1003 & 11-298 & 0.071 & 5757.933 & 5.957 & 161.260 & -1.142 & -0.224 & 17.404 & 0.759 \\
  &  &  & (0.039) & (0.075) & (0.384) & (47.824) & (0.811) & (0.818) & (0.001) & (0.005) \\
41 & 11-1007 & 11-370 & -0.273 & 5789.990 & 34.525 & 121.625 & -0.045 & -1.435 & 18.798 & 0.003 \\
  &  &  & (0.008) & (0.053) & (0.836) & (56.154) & (0.053) & (0.269) & (0.002) & (0.005) \\
42 & 11-1009 & 11-335 & 0.022 & 5774.069 & 180.284 & 6.455 & 0.029 & -0.025 & 16.259 & 0.994 \\
  &  &  & (0.006) & (0.058) & (35.620) & (6.526) & (0.103) & (0.641) & (0.002) & (0.000) \\
43 & 11-1035 & 11-337 & -0.052 & 5776.564 & 32.138 & 34.412 & -0.343 & 0.032 & 19.026 & 0.796 \\
  &  &  & (0.005) & (0.010) & (1.419) & (17.005) & (0.159) & (0.521) & (0.002) & (0.001) \\
44 & 11-1036 & 11-372 & -0.198 & 5791.009 & 34.887 & 2.241 & 0.027 & 0.025 & 18.932 & 0.230 \\
  &  &  & (0.018) & (0.048) & (1.452) & (69.196) & (0.066) & (0.671) & (0.003) & (0.005) \\
45 & 11-1072 & 11-346 & 0.006 & 5775.398 & 15.875 & 5.552 & -1.496 & -0.263 & 18.143 & 0.536 \\
  &  &  & (0.000) & (0.001) & (0.293) & (0.458) & (0.140) & (0.715) & (0.002) & (0.001) \\
46 & 11-1127 & 11-322 & -0.005 & 5775.101 & 23.687 & 90.657 & 1.494 & 1.495 & 16.699 & 0.945 \\
  &  &  & (0.010) & (0.010) & (2.041) & (7.448) & (0.124) & (0.737) & (0.001) & (0.001) \\
47 & 11-1132 & 11-358 & 0.018 & 5753.381 & 298.017 & 17.452 & 0.721 & -0.529 & 18.800 & 0.970 \\
  &  &  & (0.003) & (0.073) & (73.846) & (5.238) & (0.154) & (0.224) & (0.003) & (0.000) \\
48 & 11-1162 & 11-379 & 0.693 & 5799.155 & 10.703 & 1.114 & -0.380 & -1.497 & 17.437 & 0.089 \\
  &  &  & (0.035) & (0.037) & (0.349) & (128.940) & (0.286) & (0.808) & (0.001) & (0.004) \\
49 & 11-1195 & 11-389 & -0.481 & 5804.265 & 23.892 & 459.210 & 0.055 & 1.376 & 18.218 & 0.150 \\
  &  &  & (0.092) & (0.110) & (2.200) & (132.287) & (0.141) & (0.832) & (0.009) & (0.015) \\
50 & 11-1200 & 11-381 & -0.081 & 5799.188 & 13.837 & 83.545 & -0.061 & 0.207 & 17.687 & 0.740 \\
  &  &  & (0.006) & (0.006) & (0.497) & (8.546) & (0.364) & (0.780) & (0.001) & (0.002) \\
51 & 11-1254 & 11-388 & -0.006 & 5801.312 & 29.373 & \bf{10.393} & 0.737 & 1.372 & 15.934 & 0.996 \\
  &  &  & (0.002) & (0.003) & (4.858) & (\bf{1.660}) & (0.482) & (0.803) & (0.000) & (0.000) \\
52 & 11-9112 & 11-112 & -0.310 & 5686.106 & 20.708 & 0.000 & \bf{0.427} & \bf{1.494} & 16.368 & 0.000 \\
  &  &  & (0.012) & (0.050) & (0.469) & (63.919) & (\bf{0.097}) & (\bf{0.749}) & (0.002) & (0.000) \\
53 & 11-9291 & 11-291 & 0.003 & 5747.950 & 58.011 & 1.766 & -1.494 & -1.490 & 17.420 & 0.988 \\
  &  &  & (0.001) & (0.001) & (2.370) & (0.749) & (0.100) & (0.250) & (0.001) & (0.000) \\
54 & 11-9293 & 11-293 & 0.001 & 5747.502 & 79.317 & 1.498 & -1.457 & -1.462 & 19.935 & 0.973 \\
  &  &  & (0.000) & (0.002) & (6.984) & (0.174) & (0.389) & (0.689) & (0.016) & (0.000) \\
55 & 11-9313 & 11-313 & -0.189 & 5766.639 & 5.284 & 211.722 & 1.497 & 0.798 & 18.121 & 0.640 \\
  &  &  & (0.017) & (0.008) & (0.357) & (18.439) & (0.735) & (0.789) & (0.007) & (0.013) \\
56 & 11-9367 & 11-367 & 0.052 & 5787.101 & 9.833 & 3.286 & 1.452 & 1.316 & 18.770 & 0.314 \\
  &  &  & (0.003) & (0.002) & (0.202) & (16.151) & (0.399) & (0.804) & (0.002) & (0.002) \\
57 & 11-9393 & 11-393 & 0.001 & 5803.567 & 20.698 & 5.645 & 0.414 & 0.341 & 19.005 & 0.878 \\
  &  &  & (0.002) & (0.002) & (1.499) & (1.537) & (0.457) & (0.767) & (0.004) & (0.001) \\
58 & 12-0305 & 12-182 & -0.685 & 6046.970 & 20.933 & 402.896 & -0.356 & 0.217 & 16.333 & 0.207 \\
  &  &  & (0.025) & (0.046) & (0.611) & (123.591) & (0.057) & (0.575) & (0.001) & (0.002) \\
59 & 12-0325 & 12-166 & 0.972 & 6039.774 & 15.785 & 2.207 & 0.609 & 1.485 & 16.651 & 0.000 \\
  &  &  & (0.041) & (0.092) & (0.808) & (113.530) & (0.175) & (0.886) & (0.001) & (0.000) \\
60 & 12-0442 & 12-245 & 0.174 & 6078.680 & 31.894 & 105.718 & -0.344 & 1.499 & 18.214 & 0.164 \\
  &  &  & (0.005) & (0.017) & (0.541) & (32.295) & (0.051) & (0.026) & (0.002) & (0.003) \\
61 & 12-0443 & 12-211 & 0.058 & 6046.093 & 24.627 & 41.639 & -0.342 & -0.109 & 19.515 & 0.215 \\
  &  &  & (0.002) & (0.011) & (0.705) & (14.238) & (0.164) & (0.667) & (0.005) & (0.005) \\
62 & 12-0449 & 12-216 & -0.725 & 6058.735 & 35.066 & 706.358 & -0.336 & 1.497 & 16.646 & 0.812 \\
  &  &  & (0.105) & (0.089) & (2.285) & (222.866) & (0.074) & (0.965) & (0.000) & (0.001) \\
63 & 12-0456 & 12-189 & -0.169 & 6047.058 & 7.031 & 9.316 & -1.494 & 1.422 & 15.179 & 0.000 \\
  &  &  & (0.008) & (0.013) & (0.019) & (35.654) & (0.070) & (0.573) & (0.000) & (0.000) \\
64 & 12-0462 & 12-271 & 0.055 & 6064.302 & 37.484 & 0.000 & -0.014 & 1.499 & 19.046 & 0.000 \\
  &  &  & (0.000) & (0.006) & (0.186) & (6.292) & (0.049) & (0.083) & (0.001) & (0.000) \\
65 & 12-0591 & 12-430 & 0.175 & 6170.359 & 63.396 & 78.360 & -0.013 & -0.245 & 17.075 & 0.247 \\
  &  &  & (0.003) & (0.012) & (0.491) & (24.395) & (0.006) & (0.080) & (0.001) & (0.001) \\
66 & 12-0615 & 12-277 & 0.507 & 6060.504 & 4.098 & 1.809 & -1.492 & 1.438 & 14.411 & 0.753 \\
  &  &  & (0.021) & (0.007) & (0.090) & (77.143) & (0.412) & (0.772) & (0.000) & (0.001) \\
67 & 12-0694 & 12-308 & 0.042 & 6070.255 & 12.163 & 11.372 & 1.372 & -0.673 & 19.593 & 0.744 \\
  &  &  & (0.014) & (0.007) & (1.535) & (11.920) & (0.761) & (0.773) & (0.006) & (0.003) \\
68 & 12-0722 & 12-397 & -0.173 & 6122.064 & 58.003 & 9.264 & -0.067 & -1.303 & 18.345 & 0.501 \\
  &  &  & (0.011) & (0.066) & (2.921) & (46.579) & (0.115) & (0.997) & (0.005) & (0.005) \\
69 & 12-0724 & 12-323 & 0.009 & 6071.042 & 15.176 & \bf{11.708} & -1.385 & 1.047 & 19.821 & 0.880 \\
  &  &  & (0.001) & (0.002) & (2.544) & (\bf{1.885}) & (0.756) & (0.749) & (0.005) & (0.001) \\
70 & 12-0726 & 12-351 & -0.204 & 6072.873 & 11.110 & 1.535 & 1.489 & -1.473 & 18.183 & 0.503 \\
  &  &  & (0.075) & (0.030) & (0.644) & (71.483) & (0.478) & (0.806) & (0.003) & (0.004) \\
71 & 12-0784 & 12-337 & -0.029 & 6085.181 & 88.027 & 27.140 & -0.061 & -0.027 & 16.176 & 0.988 \\
  &  &  & (0.005) & (0.018) & (24.624) & (6.748) & (0.084) & (0.252) & (0.000) & (0.000) \\
72 & 12-0825 & 12-451 & 0.352 & 6171.507 & 45.769 & 8.244 & 0.073 & 0.314 & 17.399 & 0.002 \\
  &  &  & (0.010) & (0.036) & (2.222) & (48.109) & (0.021) & (0.204) & (0.001) & (0.002) \\
73 & 12-0867 & 12-352 & 0.202 & 6110.261 & 13.949 & 0.000 & 0.128 & -1.358 & 14.260 & 0.045 \\
  &  &  & (0.001) & (0.002) & (0.035) & (18.564) & (0.242) & (0.420) & (0.000) & (0.001) \\
74 & 12-1414 & 12-205 & -0.043 & 6040.895 & 36.153 & 0.000 & -0.195 & -0.186 & 20.257 & 0.280 \\
  &  &  & (0.003) & (0.011) & (1.728) & (11.665) & (0.127) & (0.667) & (0.007) & (0.006) \\
75 & 12-1430 & 12-278 & -0.002 & 6062.104 & 18.807 & 4.130 & 0.525 & 0.941 & 20.217 & 0.001 \\
  &  &  & (0.001) & (0.001) & (0.268) & (1.150) & (0.345) & (0.835) & (0.006) & (0.006) \\
76 & 12-0941 & 12-446 & 0.245 & 6126.169 & 36.883 & 0.000 & -0.537 & 1.499 & 19.133 & 0.113 \\
  &  &  & (0.011) & (0.031) & (1.256) & (50.847) & (0.238) & (0.726) & (0.003) & (0.004) \\
77 & 12-0974 & 12-424 & -0.069 & 6110.120 & 86.425 & 91.769 & -1.477 & -0.031 & 18.802 & 0.939 \\
  &  &  & (0.019) & (0.079) & (11.899) & (9.884) & (0.143) & (0.209) & (0.005) & (0.001) \\
78 & 12-0998 & 12-447 & -0.518 & 6136.857 & 24.356 & 0.000 & 0.125 & -1.439 & 16.957 & 0.482 \\
  &  &  & (0.036) & (0.033) & (0.513) & (179.737) & (0.069) & (0.881) & (0.000) & (0.001) \\
79 & 12-1013 & 12-449 & 0.643 & 6126.889 & 21.525 & 630.572 & -0.858 & 0.248 & 18.599 & 0.209 \\
  &  &  & (0.108) & (0.078) & (2.017) & (197.902) & (0.286) & (0.627) & (0.002) & (0.006) \\
80 & 12-1049 & 12-438 & -0.338 & 6117.274 & 7.192 & 152.796 & -1.486 & -1.069 & 18.302 & 0.371 \\
  &  &  & (0.098) & (0.017) & (0.309) & (95.418) & (0.732) & (0.802) & (0.002) & (0.003) \\
81 & 12-1051 & 12-477 & 0.200 & 6132.237 & 21.209 & 0.000 & 0.025 & -0.565 & 18.494 & 0.017 \\
  &  &  & (0.026) & (0.014) & (0.273) & (37.963) & (0.115) & (0.741) & (0.002) & (0.004) \\
82 & 12-1418 & 12-435 & 0.018 & 6119.272 & 55.686 & 0.000 & 0.280 & -0.016 & 20.142 & 0.539 \\
  &  &  & (0.001) & (0.003) & (3.094) & (3.930) & (0.209) & (0.518) & (0.006) & (0.003) \\
83 & 12-1067 & 12-444 & -0.011 & 6116.566 & 7.158 & 0.396 & -1.486 & -1.430 & 18.966 & 0.848 \\
  &  &  & (0.001) & (0.000) & (0.470) & (2.402) & (0.797) & (0.816) & (0.003) & (0.001) \\
84 & 12-1069 & 12-452 & -0.220 & 6129.142 & 22.575 & 0.913 & 0.281 & -1.495 & 19.044 & 0.000 \\
  &  &  & (0.006) & (0.020) & (0.199) & (51.679) & (0.173) & (0.765) & (0.002) & (0.000) \\
85 & 12-1066 & 12-476 & -0.132 & 6148.598 & 17.562 & 80.624 & -0.086 & 0.405 & 16.670 & 0.048 \\
  &  &  & (0.002) & (0.004) & (0.096) & (19.811) & (0.050) & (0.548) & (0.001) & (0.001) \\
86 & 12-1074 & 12-519 & -0.259 & 6154.066 & 40.088 & 183.433 & 0.215 & 0.052 & 19.280 & 0.001 \\
  &  &  & (0.013) & (0.050) & (1.301) & (60.351) & (0.050) & (0.579) & (0.003) & (0.005) \\
87 & 12-1183 & 12-478 & -0.107 & 6130.447 & 7.131 & 128.561 & 0.926 & 1.082 & 19.427 & 0.593 \\
  &  &  & (0.015) & (0.028) & (0.611) & (38.220) & (0.807) & (0.818) & (0.008) & (0.008) \\
88 & 12-1193 & 12-498 & 0.106 & 6136.399 & 14.384 & 118.142 & 1.295 & 1.381 & 18.296 & 0.801 \\
  &  &  & (0.008) & (0.026) & (0.884) & (9.552) & (0.527) & (0.780) & (0.002) & (0.002) \\
89 & 12-1210 & 12-511 & -0.368 & 6143.702 & 1.881 & 1.844 & 1.499 & 1.498 & 16.507 & 0.000 \\
  &  &  & (0.010) & (0.006) & (0.015) & (82.426) & (0.796) & (0.812) & (0.000) & (0.000) \\
90 & 12-1211 & 12-514 & -0.614 & 6151.519 & 6.542 & 621.011 & 1.489 & 1.070 & 15.699 & 0.332 \\
  &  &  & (0.010) & (0.011) & (0.086) & (10.763) & (0.081) & (0.799) & (0.000) & (0.003) \\
91 & 12-1213 & 12-512 & 0.290 & 6144.519 & 18.487 & 327.066 & -1.476 & -1.420 & 19.538 & 0.000 \\
  &  &  & (0.045) & (0.057) & (0.570) & (111.028) & (0.175) & (0.744) & (0.005) & (0.000) \\
92 & 12-1250 & 12-515 & 0.060 & 6145.218 & 6.409 & 96.900 & 1.285 & 0.452 & 18.184 & 0.862 \\
  &  &  & (0.013) & (0.022) & (0.897) & (29.045) & (0.792) & (0.793) & (0.002) & (0.004) \\
93 & 12-1242 & 12-552 & -0.059 & 6160.797 & 25.921 & 1.650 & -0.380 & 0.219 & 19.306 & 0.581 \\
  &  &  & (0.003) & (0.009) & (0.863) & (13.699) & (0.130) & (0.724) & (0.002) & (0.002) \\
94 & 12-1244 & 12-555 & 0.214 & 6154.529 & 38.326 & 231.692 & 0.227 & -0.978 & 19.626 & 0.534 \\
  &  &  & (0.066) & (0.093) & (6.447) & (86.241) & (0.165) & (0.871) & (0.006) & (0.006) \\
95 & 12-1245 & 12-575 & -0.229 & 6173.239 & 41.186 & 0.000 & 0.056 & 0.125 & 19.264 & 0.320 \\
  &  &  & (0.028) & (0.137) & (4.094) & (63.580) & (0.101) & (0.660) & (0.005) & (0.007) \\
96 & 12-1442 & 12-532 & 0.025 & 6151.813 & 11.310 & 6.815 & -0.388 & -0.053 & 20.194 & 0.242 \\
  &  &  & (0.035) & (0.017) & (1.859) & (27.043) & (0.770) & (0.800) & (0.018) & (0.018) \\
97 & 12-1268 & 12-574 & 0.204 & 6171.071 & 21.705 & 0.311 & 0.259 & -1.498 & 18.668 & 0.000 \\
  &  &  & (0.004) & (0.021) & (0.294) & (43.168) & (0.086) & (0.805) & (0.002) & (0.003) \\
98 & 12-1269 & 12-560 & 0.217 & 6160.201 & 12.653 & 202.708 & -1.066 & 0.733 & 18.311 & 0.733 \\
  &  &  & (0.021) & (0.026) & (0.757) & (48.311) & (0.468) & (0.816) & (0.002) & (0.002) \\
99 & 12-1292 & 12-570 & 0.214 & 6165.816 & 5.658 & 89.066 & -0.262 & 0.688 & 16.986 & 0.082 \\
  &  &  & (0.054) & (0.003) & (0.067) & (44.433) & (0.500) & (0.795) & (0.000) & (0.002) \\
100 & 12-1581 & 12-577 & -0.149 & 6169.822 & 35.435 & 118.073 & -0.091 & -1.424 & 20.100 & 0.027 \\
  &  &  & (0.007) & (0.043) & (1.747) & (36.750) & (0.087) & (0.986) & (0.005) & (0.007) \\
101 & 12-1311 & 12-594 & 0.588 & 6170.085 & 20.773 & 0.000 & 1.056 & 0.782 & 17.841 & 0.646 \\
  &  &  & (0.127) & (0.172) & (2.486) & (216.333) & (0.260) & (0.811) & (0.001) & (0.004) \\
102 & 12-1364 & 12-580 & -0.015 & 6164.630 & 16.616 & 0.000 & 1.399 & 1.491 & 18.465 & 0.739 \\
  &  &  & (0.005) & (0.003) & (0.987) & (3.781) & (0.314) & (0.815) & (0.003) & (0.001) \\
103 & 12-9582 & 12-582 & -0.015 & 6162.944 & 1.920 & 30.483 & -0.189 & -0.202 & 18.060 & 0.929 \\
  &  &  & (0.012) & (0.002) & (4.803) & (15.692) & (0.794) & (0.798) & (0.011) & (0.027) \\
104 & 12-9591 & 12-591 & 0.006 & 6166.857 & 18.228 & 6.773 & 0.702 & -0.735 & 18.644 & 0.991 \\
  &  &  & (0.003) & (0.009) & (5.984) & (3.041) & (0.785) & (0.788) & (0.012) & (0.001) \\
105 & 12-1350 & 12-603 & -0.134 & 6175.036 & 10.578 & 136.409 & -1.241 & 0.313 & 18.846 & 0.084 \\
  &  &  & (0.011) & (0.018) & (0.474) & (37.325) & (0.534) & (0.820) & (0.006) & (0.009) \\
106 & 13-0036 & 13-399 & 0.617 & 6406.163 & 47.908 & 0.000 & 0.040 & 0.141 & 15.971 & 0.038 \\
  &  &  & (0.011) & (0.036) & (0.393) & (101.668) & (0.010) & (0.041) & (0.000) & (0.001) \\
107 & 13-0145 & 13-110 & 0.467 & 6405.939 & 29.063 & 0.065 & -0.009 & 0.562 & 16.310 & 0.000 \\
  &  &  & (0.007) & (0.019) & (0.226) & (54.737) & (0.024) & (0.233) & (0.000) & (0.002) \\
108 & 13-0183 & 13-332 & -0.652 & 6495.203 & 103.843 & 15.347 & \bf{-0.076} & \bf{0.035} & 16.888 & 0.403 \\
  &  &  & (0.023) & (0.135) & (2.225) & (91.360) & (\bf{0.006}) & (\bf{0.009}) & (0.000) & (0.001) \\
109 & 13-0341 & 13-260 & 0.000 & 6406.313 & 28.880 & 13.446 & 1.063 & -0.909 & 18.368 & 0.066 \\
  &  &  & (0.005) & (0.005) & (0.335) & (5.332) & (0.063) & (0.493) & (0.002) & (0.002) \\
110 & 13-0486 & 13-294 & -0.154 & 6420.492 & 28.978 & 0.456 & -0.171 & -0.062 & 18.247 & 0.480 \\
  &  &  & (0.004) & (0.015) & (0.389) & (33.668) & (0.055) & (0.367) & (0.001) & (0.001) \\
111 & 13-0488 & 13-355 & -0.011 & 6439.171 & 93.899 & 0.720 & -0.173 & -0.448 & 19.738 & 0.517 \\
  &  &  & (0.004) & (0.005) & (4.458) & (2.988) & (0.121) & (0.372) & (0.007) & (0.004) \\
112 & 13-0506 & 13-494 & -0.089 & 6501.345 & 154.414 & 177.820 & -0.057 & -0.011 & 17.570 & 0.898 \\
  &  &  & (0.036) & (0.246) & (13.158) & (39.844) & (0.015) & (0.099) & (0.001) & (0.001) \\
113 & 13-0513 & 13-272 & 0.136 & 6407.452 & 28.801 & 0.000 & 0.502 & 1.481 & 19.766 & 0.202 \\
  &  &  & (0.010) & (0.038) & (1.433) & (37.790) & (0.145) & (0.854) & (0.005) & (0.006) \\
114 & 13-0601 & 13-292 & -0.060 & 6414.867 & 17.778 & 80.207 & -1.158 & 0.987 & 18.756 & 0.810 \\
  &  &  & (0.015) & (0.016) & (1.414) & (22.563) & (0.390) & (0.828) & (0.003) & (0.002) \\
115 & 13-0607 & 13-376 & -0.956 & 6455.457 & 22.750 & 0.000 & -0.189 & 1.492 & 16.383 & 0.001 \\
  &  &  & (0.024) & (0.044) & (0.434) & (141.951) & (0.171) & (0.885) & (0.000) & (0.002) \\
116 & 13-0608 & 13-308 & -0.438 & 6426.720 & 14.260 & 378.693 & 0.559 & 0.521 & 17.388 & 0.023 \\
  &  &  & (0.088) & (0.023) & (0.310) & (102.202) & (0.224) & (0.736) & (0.001) & (0.003) \\
117 & 13-0610 & 13-401 & 0.253 & 6455.122 & 22.810 & 0.000 & 1.084 & 1.499 & 16.059 & 0.071 \\
  &  &  & (0.002) & (0.004) & (0.077) & (28.019) & (0.087) & (0.106) & (0.000) & (0.001) \\
118 & 13-0615 & 13-389 & -0.349 & 6448.248 & 46.213 & 0.484 & 0.248 & 1.500 & 18.110 & 0.484 \\
  &  &  & (0.036) & (0.117) & (2.467) & (113.663) & (0.129) & (0.828) & (0.002) & (0.004) \\
119 & 13-0616 & 13-331 & -0.355 & 6424.294 & 10.334 & 0.000 & -0.841 & 1.490 & 17.882 & 0.381 \\
  &  &  & (0.084) & (0.019) & (0.377) & (97.568) & (0.430) & (0.775) & (0.001) & (0.002) \\
120 & 13-0619 & 13-351 & -0.055 & 6427.437 & 21.427 & 0.000 & 0.571 & 1.497 & 20.215 & 0.169 \\
  &  &  & (0.003) & (0.007) & (0.708) & (14.559) & (0.330) & (0.839) & (0.006) & (0.006) \\
121 & 13-0621 & 13-344 & -0.101 & 6425.046 & 30.293 & 85.822 & -0.290 & 1.496 & 19.606 & 0.610 \\
  &  &  & (0.010) & (0.019) & (1.718) & (32.894) & (0.188) & (0.848) & (0.003) & (0.002) \\
122 & 13-0668 & 13-343 & 0.246 & 6422.307 & 22.570 & 266.126 & 0.806 & 0.020 & 19.287 & 0.693 \\
  &  &  & (0.068) & (0.058) & (2.768) & (82.594) & (0.348) & (0.771) & (0.003) & (0.004) \\
123 & 13-0674 & 13-346 & -0.002 & 6426.021 & 57.315 & 0.000 & -0.194 & -1.473 & 20.105 & 0.738 \\
  &  &  & (0.000) & (0.000) & (1.205) & (0.251) & (0.194) & (0.880) & (0.006) & (0.002) \\
124 & 13-0682 & 13-388 & 0.727 & 6453.990 & 21.275 & 78.987 & 1.159 & 1.488 & 16.880 & 0.002 \\
  &  &  & (0.013) & (0.045) & (0.231) & (103.167) & (0.281) & (0.256) & (0.001) & (0.003) \\
125 & 13-0689 & 13-379 & 0.237 & 6434.013 & 21.510 & 8.974 & -0.162 & 1.494 & 17.949 & 0.814 \\
  &  &  & (0.031) & (0.099) & (1.625) & (76.931) & (0.477) & (0.841) & (0.002) & (0.002) \\
126 & 13-0703 & 13-349 & -0.151 & 6429.186 & 25.032 & 81.227 & 0.129 & -0.119 & 19.014 & 0.146 \\
  &  &  & (0.044) & (0.019) & (0.740) & (28.229) & (0.170) & (0.650) & (0.003) & (0.004) \\
127 & 13-0731 & 13-377 & -0.292 & 6437.552 & 18.103 & 0.000 & -0.247 & 0.149 & 17.346 & 0.804 \\
  &  &  & (0.073) & (0.058) & (1.011) & (90.446) & (0.463) & (0.774) & (0.001) & (0.001) \\
128 & 13-0770 & 13-462 & 0.238 & 6471.819 & 26.446 & 98.029 & \bf{-1.499} & \bf{-0.862} & 17.449 & 0.342 \\
  &  &  & (0.008) & (0.023) & (0.508) & (49.735) & (\bf{0.026}) & (\bf{0.178}) & (0.001) & (0.002) \\
129 & 13-0801 & 13-386 & 0.047 & 6436.617 & 11.463 & 56.062 & -1.473 & 1.028 & 18.103 & 0.949 \\
  &  &  & (0.008) & (0.007) & (1.440) & (9.493) & (0.769) & (0.780) & (0.003) & (0.001) \\
130 & 13-0835 & 13-400 & -0.056 & 6449.971 & 9.743 & 0.000 & -1.489 & -0.011 & 18.184 & 0.204 \\
  &  &  & (0.001) & (0.001) & (0.092) & (9.652) & (0.692) & (0.781) & (0.002) & (0.002) \\
131 & 13-0850 & 13-470 & 0.959 & 6479.565 & 22.525 & 0.486 & 0.078 & -0.607 & 17.165 & 0.007 \\
  &  &  & (0.070) & (0.072) & (1.258) & (152.823) & (0.280) & (0.352) & (0.001) & (0.003) \\
132 & 13-0861 & 13-502 & 0.363 & 6516.004 & 65.655 & 0.000 & 0.037 & -0.502 & 17.402 & 0.324 \\
  &  &  & (0.017) & (0.101) & (1.497) & (87.172) & (0.020) & (0.079) & (0.001) & (0.002) \\
133 & 13-0891 & 13-433 & -0.609 & 6476.835 & 18.930 & 0.000 & 0.322 & -0.737 & 15.512 & 0.000 \\
  &  &  & (0.003) & (0.011) & (0.092) & (54.934) & (0.105) & (0.473) & (0.000) & (0.000) \\
134 & 13-0911 & 13-551 & -0.003 & 6537.299 & 126.734 & 0.533 & -0.131 & 0.218 & 19.362 & 0.444 \\
  &  &  & (0.000) & (0.000) & (1.934) & (0.720) & (0.004) & (0.077) & (0.003) & (0.002) \\
135 & 13-0925 & 13-493 & -0.358 & 6501.455 & 43.130 & 171.945 & -0.194 & 1.494 & 18.051 & 0.169 \\
  &  &  & (0.015) & (0.055) & (0.977) & (81.057) & (0.027) & (0.081) & (0.001) & (0.003) \\
136 & 13-0968 & 13-452 & -0.168 & 6460.311 & 7.046 & 21.913 & -1.489 & 1.279 & 17.675 & 0.049 \\
  &  &  & (0.004) & (0.002) & (0.056) & (32.723) & (0.783) & (0.750) & (0.001) & (0.001) \\
137 & 13-1033 & 13-472 & -0.139 & 6468.808 & 21.183 & 196.985 & -0.371 & 1.377 & 19.571 & 0.550 \\
  &  &  & (0.098) & (0.050) & (3.002) & (82.722) & (0.777) & (0.760) & (0.005) & (0.007) \\
138 & 13-1037 & 13-483 & 0.048 & 6473.098 & 36.512 & 1.531 & -1.500 & -0.754 & 20.185 & 0.437 \\
  &  &  & (0.015) & (0.019) & (3.706) & (16.175) & (0.511) & (0.697) & (0.013) & (0.009) \\
139 & 13-1078 & 13-460 & 0.000 & 6462.061 & 27.261 & \bf{0.990} & -1.210 & 1.338 & 20.259 & 0.929 \\
  &  &  & (0.000) & (0.000) & (1.700) & (\bf{0.072}) & (0.798) & (0.806) & (0.015) & (0.002) \\
140 & 13-1080 & 13-456 & 0.004 & 6462.038 & 22.224 & 0.697 & 1.380 & -1.492 & 20.388 & 0.904 \\
  &  &  & (0.002) & (0.001) & (4.915) & (1.221) & (0.808) & (0.818) & (0.007) & (0.001) \\
141 & 13-9455 & 13-455 & 0.058 & 6456.795 & 7.492 & 110.370 & 0.588 & -1.359 & 18.519 & 0.684 \\
  &  &  & (0.030) & (0.023) & (1.567) & (32.662) & (0.811) & (0.812) & (0.012) & (0.019) \\
142 & 13-1114 & 13-485 & 0.008 & 6474.942 & 68.566 & 0.000 & -0.707 & -0.383 & 20.374 & 0.292 \\
  &  &  & (0.001) & (0.003) & (5.160) & (2.113) & (0.147) & (0.300) & (0.018) & (0.012) \\
143 & 13-1124 & 13-500 & 0.269 & 6489.987 & 22.335 & 0.727 & 0.671 & -1.478 & 18.292 & 0.593 \\
  &  &  & (0.047) & (0.045) & (1.035) & (96.262) & (0.263) & (0.806) & (0.001) & (0.002) \\
144 & 13-1145 & 13-523 & -0.053 & 6503.507 & 36.562 & 8.814 & -0.173 & 1.443 & 18.536 & 0.654 \\
  &  &  & (0.011) & (0.009) & (1.204) & (13.386) & (0.131) & (0.886) & (0.002) & (0.001) \\
145 & 13-1157 & 13-490 & -0.287 & 6504.274 & 13.587 & 140.112 & 0.116 & -1.493 & 15.260 & 0.005 \\
  &  &  & (0.003) & (0.004) & (0.049) & (39.911) & (0.065) & (0.581) & (0.000) & (0.001) \\
146 & 13-1227 & 13-545 & -0.052 & 6515.517 & 93.445 & 13.861 & -0.165 & 0.803 & 17.576 & 0.924 \\
  &  &  & (0.003) & (0.027) & (3.798) & (10.893) & (0.022) & (0.121) & (0.001) & (0.000) \\
147 & 13-1250 & 13-516 & 0.552 & 6491.580 & 2.588 & 582.467 & -0.264 & 0.240 & 15.668 & 0.925 \\
  &  &  & (0.117) & (0.019) & (0.302) & (208.685) & (0.801) & (0.789) & (0.000) & (0.001) \\
148 & 13-1253 & 13-519 & -0.110 & 6497.594 & 7.984 & 18.925 & 1.469 & 1.476 & 17.138 & 0.786 \\
  &  &  & (0.006) & (0.008) & (0.260) & (27.287) & (0.593) & (0.803) & (0.001) & (0.001) \\
149 & 13-1257 & 13-556 & 0.253 & 6514.259 & 23.877 & 128.485 & 0.994 & -1.475 & 18.625 & 0.513 \\
  &  &  & (0.068) & (0.038) & (0.839) & (85.036) & (0.157) & (0.771) & (0.001) & (0.002) \\
150 & 13-1279 & 13-518 & 0.082 & 6499.373 & 3.808 & \bf{139.541} & 0.000 & 0.000 & 15.799 & 0.000 \\
  &  &  & (0.001) & (0.001) & (0.010) & (\bf{0.859}) & (0.000) & (0.000) & (0.000) & (0.000) \\
151 & 13-1287 & 13-537 & 0.576 & 6503.415 & 13.465 & 0.000 & 0.163 & -1.446 & 17.502 & 0.314 \\
  &  &  & (0.051) & (0.046) & (0.710) & (155.043) & (0.314) & (0.839) & (0.001) & (0.003) \\
152 & 13-1492 & 13-548 & -0.243 & 6504.597 & 13.462 & 254.502 & -1.064 & 0.286 & 17.274 & 0.911 \\
  &  &  & (0.074) & (0.054) & (2.230) & (111.644) & (0.757) & (0.795) & (0.002) & (0.002) \\
153 & 13-1498 & 13-568 & -0.232 & 6517.917 & 18.762 & 184.671 & 0.737 & -1.410 & 18.842 & 0.000 \\
  &  &  & (0.010) & (0.031) & (0.403) & (59.591) & (0.178) & (0.873) & (0.002) & (0.000) \\
154 & 13-1506 & 13-541 & -0.110 & 6512.931 & 12.343 & 105.627 & -0.742 & 0.612 & 16.009 & 0.017 \\
  &  &  & (0.004) & (0.002) & (0.125) & (8.549) & (0.135) & (0.816) & (0.001) & (0.002) \\
155 & 13-1551 & 13-572 & -0.155 & 6517.452 & 15.373 & 163.872 & -0.963 & -0.276 & 19.404 & 0.506 \\
  &  &  & (0.022) & (0.024) & (1.063) & (53.582) & (0.505) & (0.833) & (0.005) & (0.005) \\
156 & 13-1598 & 13-570 & -0.754 & 6532.931 & 10.952 & 4.218 & \bf{-1.500} & \bf{-1.461} & 15.584 & 0.133 \\
  &  &  & (0.035) & (0.024) & (0.325) & (69.121) & (\bf{0.019}) & (\bf{0.858}) & (0.000) & (0.003) \\
157 & 13-1617 & 13-560 & -0.200 & 6515.385 & 11.019 & 163.536 & -0.865 & -0.955 & 17.642 & 0.152 \\
  &  &  & (0.007) & (0.007) & (0.225) & (41.817) & (0.296) & (0.777) & (0.001) & (0.003) \\
158 & 13-1669 & 13-585 & -0.007 & 6527.163 & 52.476 & 9.301 & 0.163 & 1.496 & 20.463 & 0.013 \\
  &  &  & (0.000) & (0.003) & (2.138) & (0.471) & (0.111) & (1.054) & (0.019) & (0.019) \\
159 & 13-1678 & 13-595 & -0.204 & 6531.602 & 7.545 & 0.000 & 1.494 & 1.379 & 18.312 & 0.413 \\
  &  &  & (0.028) & (0.018) & (0.479) & (77.941) & (0.469) & (0.811) & (0.003) & (0.006) \\
160 & 13-1689 & 13-593 & -0.081 & 6534.384 & 5.177 & 47.757 & 0.931 & -0.432 & 17.420 & 0.086 \\
  &  &  & (0.003) & (0.001) & (0.062) & (20.507) & (0.646) & (0.794) & (0.001) & (0.003) \\
161 & 13-1721 & 13-618 & -0.086 & 6535.020 & 22.085 & 92.966 & -0.336 & -1.119 & 20.502 & 0.254 \\
  &  &  & (0.009) & (0.042) & (2.239) & (26.987) & (0.447) & (0.825) & (0.010) & (0.011) \\
162 & 13-1730 & 13-602 & -0.214 & 6538.838 & 17.116 & 3.528 & -0.216 & -0.140 & 18.848 & 0.330 \\
  &  &  & (0.021) & (0.041) & (0.940) & (64.806) & (0.414) & (0.803) & (0.004) & (0.007) \\
163 & 13-1738 & 13-628 & 0.286 & 6542.673 & 12.642 & 8.011 & -1.265 & 0.079 & 18.951 & 0.328 \\
  &  &  & (0.075) & (0.042) & (0.809) & (87.152) & (0.427) & (0.776) & (0.002) & (0.005) \\
164 & 13-1775 & 13-614 & -0.008 & 6538.850 & 46.183 & 7.671 & -0.060 & -0.392 & 20.126 & 0.750 \\
  &  &  & (0.001) & (0.004) & (7.734) & (1.873) & (0.300) & (0.750) & (0.012) & (0.003) \\
165 & 14-0099 & 14-109 & -0.343 & 6876.998 & 123.255 & 263.093 & \bf{0.194} & \bf{0.291} & 16.683 & 0.338 \\
  &  &  & (0.005) & (0.087) & (1.073) & (71.902) & (\bf{0.004}) & (\bf{0.020}) & (0.000) & (0.001) \\
166 & 14-0115 & 14-273 & -0.348 & 6859.582 & 97.229 & 307.209 & \bf{0.110} & \bf{0.039} & 17.117 & 0.010 \\
  &  &  & (0.009) & (0.040) & (0.858) & (21.747) & (\bf{0.005}) & (\bf{0.030}) & (0.001) & (0.001) \\
167 & 14-0124 & 14-307 & -0.160 & 6836.531 & 167.579 & 89.568 & 0.129 & -0.074 & 17.382 & 0.712 \\
  &  &  & (0.004) & (0.047) & (5.816) & (38.330) & (0.014) & (0.027) & (0.001) & (0.001) \\
168 & 14-0257 & 14-148 & 0.001 & 6772.704 & 61.720 & 139.227 & -0.140 & -0.078 & 19.027 & 0.333 \\
  &  &  & (0.006) & (0.028) & (1.309) & (2.858) & (0.015) & (0.153) & (0.003) & (0.003) \\
169 & 14-0337 & 14-194 & -0.543 & 6822.955 & 45.276 & 207.333 & \bf{0.146} & \bf{1.416} & 16.768 & 0.000 \\
  &  &  & (0.006) & (0.023) & (0.187) & (78.241) & (\bf{0.042}) & (\bf{0.075}) & (0.000) & (0.000) \\
170 & 14-0383 & 14-147 & 0.013 & 6761.393 & 15.139 & 10.691 & -0.041 & -0.549 & 18.032 & 0.583 \\
  &  &  & (0.004) & (0.001) & (0.293) & (3.899) & (0.171) & (0.753) & (0.002) & (0.001) \\
171 & 14-0419 & 14-283 & -0.248 & 6822.954 & 47.010 & 0.000 & -0.551 & -0.861 & 17.918 & 0.194 \\
  &  &  & (0.012) & (0.024) & (1.166) & (57.567) & (0.175) & (0.333) & (0.002) & (0.002) \\
172 & 14-0512 & 14-157 & 0.675 & 6782.126 & 11.354 & 0.000 & \bf{0.914} & \bf{1.492} & 15.679 & 0.000 \\
  &  &  & (0.003) & (0.010) & (0.038) & (40.879) & (\bf{0.078}) & (\bf{0.259}) & (0.000) & (0.000) \\
173 & 14-0530 & 14-220 & -0.117 & 6791.737 & 20.530 & 51.944 & -0.034 & -0.372 & 17.684 & 0.093 \\
  &  &  & (0.002) & (0.005) & (0.162) & (18.433) & (0.103) & (0.650) & (0.001) & (0.001) \\
174 & 14-0531 & 14-229 & 0.186 & 6784.771 & 41.753 & 1.559 & 0.151 & -1.499 & 18.045 & 0.802 \\
  &  &  & (0.029) & (0.071) & (3.073) & (56.852) & (0.173) & (0.792) & (0.001) & (0.001) \\
175 & 14-0572 & 14-165 & -0.561 & 6775.903 & 9.009 & 3.733 & 1.498 & -1.480 & 17.445 & 0.000 \\
  &  &  & (0.007) & (0.036) & (0.083) & (65.306) & (0.052) & (0.555) & (0.001) & (0.000) \\
176 & 14-0582 & 14-151 & 0.005 & 6766.182 & 2.074 & 84.547 & -1.127 & 0.463 & 19.220 & 0.453 \\
  &  &  & (0.028) & (0.026) & (0.245) & (32.638) & (0.793) & (0.805) & (0.004) & (0.006) \\
177 & 14-0617 & 14-215 & 0.141 & 6771.464 & 1.565 & 236.811 & 0.469 & -0.516 & 18.756 & 0.718 \\
  &  &  & (0.085) & (0.007) & (0.162) & (67.919) & (0.802) & (0.778) & (0.003) & (0.004) \\
178 & 14-0676 & 14-175 & 0.004 & 6778.228 & 8.015 & 110.012 & -0.910 & 0.487 & 19.537 & 0.456 \\
  &  &  & (0.077) & (0.226) & (5.435) & (50.335) & (0.828) & (0.789) & (0.016) & (0.026) \\
179 & 14-0692 & 14-214 & 0.255 & 6778.824 & 7.763 & 274.031 & 1.199 & -0.476 & 18.853 & 0.762 \\
  &  &  & (0.057) & (0.037) & (0.737) & (74.266) & (0.738) & (0.769) & (0.002) & (0.004) \\
180 & 14-0696 & 14-199 & 0.405 & 6788.774 & 6.832 & 3.167 & \bf{-1.393} & \bf{1.496} & 14.183 & 0.000 \\
  &  &  & (0.001) & (0.003) & (0.011) & (24.469) & (\bf{0.098}) & (\bf{0.512}) & (0.000) & (0.000) \\
181 & 14-0734 & 14-243 & -0.753 & 6791.432 & 5.851 & 424.005 & 1.483 & -0.740 & 17.725 & 0.019 \\
  &  &  & (0.046) & (0.027) & (0.240) & (176.220) & (0.707) & (0.813) & (0.001) & (0.005) \\
182 & 14-0727 & 14-281 & 1.043 & 6842.546 & 38.110 & 0.000 & -0.819 & -0.075 & 16.607 & 0.056 \\
  &  &  & (0.162) & (2.346) & (4.552) & (306.481) & (0.142) & (0.127) & (0.001) & (0.013) \\
183 & 14-0729 & 14-262 & -0.663 & 6799.616 & 6.557 & 4.023 & -1.226 & 1.430 & 16.463 & 0.000 \\
  &  &  & (0.008) & (0.017) & (0.045) & (85.907) & (0.486) & (0.778) & (0.000) & (0.000) \\
184 & 14-0783 & 14-253 & -0.017 & 6791.411 & 21.003 & \bf{19.919} & \bf{0.072} & \bf{1.187} & 20.117 & 0.532 \\
  &  &  & (0.001) & (0.002) & (1.410) & (\bf{1.993}) & (\bf{0.488}) & (\bf{0.794}) & (0.005) & (0.003) \\
185 & 14-0874 & 14-302 & -0.199 & 6845.679 & 24.877 & 130.938 & -0.030 & 1.482 & 15.887 & 0.000 \\
  &  &  & (0.004) & (0.006) & (0.096) & (41.271) & (0.083) & (0.351) & (0.000) & (0.000) \\
186 & 14-0893 & 14-296 & -0.024 & 6809.366 & 74.901 & 0.000 & 0.000 & 0.000 & 19.660 & 0.857 \\
  &  &  & (0.007) & (0.014) & (8.718) & (6.110) & (0.243) & (0.551) & (0.007) & (0.002) \\
187 & 14-0894 & 14-333 & -0.458 & 6846.403 & 28.577 & 0.000 & 0.213 & -1.499 & 17.530 & 0.031 \\
  &  &  & (0.010) & (0.026) & (0.491) & (71.085) & (0.074) & (0.870) & (0.001) & (0.003) \\
188 & 14-0900 & 14-305 & 0.280 & 6862.267 & 33.421 & 0.000 & -0.202 & -1.499 & 16.060 & 0.000 \\
  &  &  & (0.010) & (0.041) & (0.238) & (73.755) & (0.022) & (0.017) & (0.001) & (0.000) \\
189 & 14-0928 & 14-361 & -0.387 & 6877.343 & 55.656 & 106.812 & -0.076 & -1.402 & 16.570 & 0.181 \\
  &  &  & (0.014) & (0.043) & (1.576) & (66.610) & (0.023) & (0.068) & (0.000) & (0.001) \\
190 & 14-9958 & 14-275 & 0.090 & 6817.268 & 79.072 & 34.569 & 0.120 & 0.884 & 19.074 & 0.000 \\
  &  &  & (0.001) & (0.031) & (0.290) & (13.950) & (0.022) & (0.049) & (0.002) & (0.000) \\
191 & 14-0962 & 14-285 & -0.002 & 6817.988 & 6.540 & 199.230 & -1.476 & 1.493 & 16.171 & 0.524 \\
  &  &  & (0.027) & (0.014) & (0.085) & (10.398) & (0.237) & (0.622) & (0.000) & (0.002) \\
192 & 14-0996 & 14-310 & -0.064 & 6812.742 & 16.899 & 1.083 & 1.257 & 1.409 & 20.275 & 0.000 \\
  &  &  & (0.002) & (0.007) & (0.150) & (15.487) & (0.650) & (0.812) & (0.003) & (0.000) \\
193 & 14-1029 & 14-312 & 0.002 & 6811.241 & 125.789 & 0.950 & -1.425 & 0.101 & 20.353 & 0.984 \\
  &  &  & (0.001) & (0.005) & (53.289) & (0.855) & (0.282) & (0.381) & (0.006) & (0.000) \\
194 & 14-1101 & 14-362 & 0.713 & 6855.431 & 20.393 & 358.065 & -0.897 & 1.372 & 16.892 & 0.000 \\
  &  &  & (0.035) & (0.045) & (0.499) & (167.500) & (0.319) & (0.745) & (0.000) & (0.000) \\
195 & 14-1109 & 14-328 & 0.785 & 6826.141 & 7.855 & 608.533 & -1.414 & 1.499 & 18.304 & 0.044 \\
  &  &  & (0.085) & (0.076) & (0.681) & (205.985) & (0.805) & (0.747) & (0.002) & (0.011) \\
196 & 14-1111 & 14-401 & -0.181 & 6861.902 & 32.831 & 2.518 & 0.077 & 0.676 & 18.740 & 0.006 \\
  &  &  & (0.005) & (0.026) & (0.587) & (36.410) & (0.089) & (0.706) & (0.003) & (0.004) \\
197 & 14-1153 & 14-389 & -0.366 & 6853.536 & 16.342 & 0.000 & 0.081 & -1.485 & 17.928 & 0.001 \\
  &  &  & (0.008) & (0.018) & (0.174) & (68.999) & (0.216) & (0.837) & (0.001) & (0.002) \\
198 & 14-1255 & 14-367 & 0.011 & 6844.289 & 25.208 & 0.000 & 0.050 & 1.400 & 18.551 & 0.636 \\
  &  &  & (0.013) & (0.006) & (1.188) & (11.132) & (0.589) & (0.770) & (0.004) & (0.002) \\
199 & 14-1284 & 14-392 & 0.037 & 6852.991 & 40.898 & 0.000 & 0.551 & -0.992 & 19.230 & 0.687 \\
  &  &  & (0.009) & (0.010) & (1.745) & (7.153) & (0.217) & (0.785) & (0.003) & (0.001) \\
200 & 14-1285 & 14-370 & -0.567 & 6848.987 & 6.316 & 0.000 & 1.500 & 1.180 & 16.076 & 0.000 \\
  &  &  & (0.007) & (0.008) & (0.033) & (72.928) & (0.464) & (0.812) & (0.000) & (0.000) \\
201 & 14-1308 & 14-357 & 0.493 & 6846.555 & 10.199 & 0.000 & -0.894 & -1.372 & 17.794 & 0.061 \\
  &  &  & (0.103) & (0.040) & (0.233) & (142.274) & (0.707) & (0.792) & (0.003) & (0.014) \\
202 & 14-1317 & 14-364 & -0.045 & 6840.933 & 40.873 & 0.000 & -1.484 & -0.200 & 20.254 & 0.651 \\
  &  &  & (0.007) & (0.032) & (4.059) & (16.625) & (0.708) & (0.744) & (0.014) & (0.006) \\
203 & 14-1340 & 14-372 & -0.084 & 6845.154 & 6.697 & 0.085 & 1.491 & 1.497 & 19.393 & 0.000 \\
  &  &  & (0.003) & (0.004) & (0.173) & (18.720) & (0.811) & (0.819) & (0.004) & (0.000) \\
204 & 14-1392 & 14-398 & 0.097 & 6854.863 & 13.448 & 97.901 & -0.434 & -0.910 & 20.017 & 0.191 \\
  &  &  & (0.007) & (0.010) & (0.571) & (22.785) & (0.593) & (0.795) & (0.004) & (0.005) \\
205 & 14-1406 & 14-409 & -0.344 & 6873.449 & 12.544 & 0.000 & 0.100 & -1.497 & 15.840 & 0.000 \\
  &  &  & (0.003) & (0.026) & (0.065) & (49.425) & (0.195) & (0.827) & (0.001) & (0.000) \\
206 & 14-1409 & 14-393 & -0.029 & 6853.793 & 11.255 & 14.686 & -1.192 & 0.144 & 18.945 & 0.890 \\
  &  &  & (0.006) & (0.010) & (1.519) & (12.319) & (0.794) & (0.810) & (0.002) & (0.001) \\
207 & 14-1410 & 14-424 & 0.199 & 6868.477 & 21.526 & 195.626 & -0.268 & -1.486 & 18.435 & 0.434 \\
  &  &  & (0.023) & (0.027) & (0.883) & (71.835) & (0.230) & (0.844) & (0.002) & (0.004) \\
208 & 14-1437 & 14-407 & -0.056 & 6858.442 & 7.216 & 0.000 & 1.463 & 1.132 & 18.466 & 0.426 \\
  &  &  & (0.020) & (0.006) & (0.354) & (15.973) & (0.819) & (0.793) & (0.005) & (0.004) \\
209 & 14-1472 & 14-433 & 0.147 & 6864.195 & 12.511 & 40.003 & 0.426 & 0.926 & 19.705 & 0.007 \\
  &  &  & (0.006) & (0.022) & (0.413) & (34.658) & (0.705) & (0.814) & (0.006) & (0.009) \\
210 & 14-1483 & 14-506 & 0.063 & 6895.944 & 54.295 & 57.645 & -0.009 & 0.261 & 18.862 & 0.762 \\
  &  &  & (0.003) & (0.029) & (2.656) & (14.109) & (0.060) & (0.394) & (0.002) & (0.001) \\
211 & 14-1486 & 14-423 & -0.107 & 6864.709 & 5.525 & 82.538 & 1.428 & 1.499 & 16.596 & 0.565 \\
  &  &  & (0.004) & (0.001) & (0.055) & (26.708) & (0.535) & (0.813) & (0.000) & (0.000) \\
212 & 14-1498 & 14-456 & -0.150 & 6881.422 & 25.862 & 0.000 & 0.000 & 0.037 & 19.411 & 0.336 \\
  &  &  & (0.006) & (0.032) & (0.753) & (33.263) & (0.112) & (0.659) & (0.002) & (0.003) \\
213 & 14-1501 & 14-426 & 0.007 & 6859.360 & 43.698 & 0.000 & 0.004 & 0.093 & 20.328 & 0.099 \\
  &  &  & (0.001) & (0.002) & (2.755) & (2.010) & (0.215) & (0.701) & (0.012) & (0.011) \\
214 & 14-1570 & 14-451 & -0.697 & 6887.094 & 12.102 & 451.023 & 0.236 & 0.752 & 16.121 & 0.169 \\
  &  &  & (0.025) & (0.017) & (0.303) & (130.543) & (0.110) & (0.737) & (0.000) & (0.001) \\
215 & 14-1812 & 14-444 & 0.017 & 6873.149 & 68.707 & 26.821 & \bf{1.459} & \bf{1.003} & 21.368 & 0.243 \\
  &  &  & (0.002) & (0.017) & (14.538) & (3.729) & (\bf{0.187}) & (\bf{0.797}) & (0.072) & (0.053) \\
216 & 14-1596 & 14-445 & -0.206 & 6874.945 & 16.352 & 210.268 & -1.476 & -0.359 & 18.457 & 0.575 \\
  &  &  & (0.014) & (0.057) & (0.760) & (53.485) & (0.343) & (0.771) & (0.003) & (0.005) \\
217 & 14-1598 & 14-448 & -0.096 & 6874.447 & 3.484 & 0.844 & -1.488 & -1.155 & 19.253 & 0.121 \\
  &  &  & (0.031) & (0.008) & (0.238) & (28.613) & (0.823) & (0.801) & (0.004) & (0.008) \\
218 & 14-1600 & 14-494 & -0.285 & 6889.198 & 27.047 & 0.000 & -0.198 & 1.486 & 19.070 & 0.277 \\
  &  &  & (0.039) & (0.058) & (1.592) & (113.552) & (0.115) & (0.812) & (0.003) & (0.005) \\
219 & 14-1647 & 14-489 & 0.164 & 6892.046 & 16.237 & 0.000 & -0.086 & -1.482 & 17.716 & 0.840 \\
  &  &  & (0.012) & (0.018) & (0.580) & (49.741) & (0.310) & (0.842) & (0.001) & (0.001) \\
220 & 14-1701 & 14-488 & -0.124 & 6891.664 & 6.797 & 122.430 & -1.459 & -1.397 & 17.559 & 0.668 \\
  &  &  & (0.005) & (0.005) & (0.162) & (9.349) & (0.652) & (0.790) & (0.001) & (0.001) \\
221 & 14-1706 & 14-502 & -0.807 & 6892.816 & 3.298 & 0.000 & -1.500 & -0.994 & 17.221 & 0.000 \\
  &  &  & (0.038) & (0.024) & (0.104) & (141.679) & (0.729) & (0.815) & (0.001) & (0.010) \\
222 & 14-1720 & 14-507 & -0.045 & 6894.912 & 13.556 & 75.223 & -1.370 & -0.662 & 19.072 & 0.718 \\
  &  &  & (0.004) & (0.006) & (0.817) & (5.115) & (0.631) & (0.788) & (0.005) & (0.004) \\
223 & 14-1748 & 14-496 & -0.001 & 6891.179 & 6.973 & 0.000 & -1.349 & -1.282 & 19.827 & 0.664 \\
  &  &  & (0.002) & (0.001) & (0.976) & (1.931) & (0.796) & (0.834) & (0.011) & (0.006) \\
224 & 14-1816 & 14-498 & 0.065 & 6893.400 & 13.480 & 18.307 & -1.496 & -1.411 & 19.947 & 0.000 \\
  &  &  & (0.003) & (0.008) & (0.293) & (15.881) & (0.140) & (0.765) & (0.005) & (0.000) \\
\hline \hline 
X1 & 11-0975 & 11-305 & 0.408 & 5760.427 & 14.646 & 1145.523 & -0.473 & 0.120 & 19.344 & 0.009 \\
  &  &  & (0.252) & (0.238) & (2.751) & (162.081) & (0.738) & (0.795) & (0.007) & (0.043) \\
X2 & 12-0937 & 12-405 & -0.063 & 6102.288 & 9.248 & 725.970 & 1.226 & 1.221 & 20.328 & 0.071 \\
  &  &  & (0.131) & (0.099) & (0.523) & (37.411) & (0.806) & (0.800) & (0.008) & (0.019) \\
X3 & 12-1502 & 12-567 & 0.064 & 6138.430 & 393.584 & 71.659 & -0.810 & 0.053 & 20.365 & 0.933 \\
  &  &  & (0.041) & (2.725) & (79.223) & (80.405) & (0.310) & (0.229) & (0.012) & (0.002) \\
X4 & 13-0966 & 13-428 & -0.352 & 6450.141 & 8.388 & 223.542 & 1.180 & -0.863 & 20.108 & 0.404 \\
  &  &  & (0.236) & (0.143) & (11.326) & (221.974) & (0.796) & (0.788) & (0.013) & (0.026) \\
X5 & 14-0289 & 14-092 & 0.572 & 6852.483 & 230.254 & 837.450 & -0.260 & 1.061 & 18.813 & 0.243 \\
  &  &  & (0.170) & (28.456) & (113.809) & (395.201) & (0.456) & (0.200) & (0.031) & (0.082) \\
X6 & 14-0921 & 14-270 & -0.003 & 6803.854 & 374.898 & 3.852 & 0.490 & 0.229 & 19.724 & 0.996 \\
  &  &  & (0.234) & (0.145) & (90.046) & (151.397) & (0.788) & (0.819) & (0.021) & (0.000) \\
\end{longtable}
\normalsize

\end{document}